%% file: main.tex
\documentclass[floatfix,a4paper,aps,prd,twocolumn,preprintnumbers,amsfonts,amsmath]{revtex4}
\usepackage{geometry}
\DeclareMathAlphabet{\scr}{U}{rsfs}{m}{n}

\usepackage[dvips]{color}
\usepackage{graphicx}

\newcommand{\newc}{\newcommand}

\newc{\mcal}{\mathcal}
\newc{\nn}{\nonumber}

\newc{\ra}{\rightarrow}
\newc{\lra}{\leftrightarrow}
\newc{\wtilde}{\widetilde}

\newc{\beq}{\begin{equation}}
\newc{\eeq}{\end{equation}}
\newc{\beqn}{\begin{eqnarray}}
\newc{\eeqn}{\end{eqnarray}}
\renewcommand{\eqref}[1]{Eq.~(\ref{#1})}
\newc{\eqsref}[2]{Eqs.~(\ref{#1}),\,(\ref{#2})}
\newc{\figref}[1]{Fig.~\ref{#1}}

\newc{\eps}{\epsilon}
\newc{\lam}{\lambda}
\newc{\lamp}{\lambda^{\prime}}
\newc{\Lam}{\Lambda}

\newc{\lan}{\mathcal{L}}
\newc{\abs}[1]{\lvert#1\rvert}
\newc{\oas}{\mathcal{O}(\alpha_s^2)}
\newc{\del}{\partial}
\newc{\rb}[1]{\raisebox{1.5ex}[-1.5ex]{#1}}

\newc{\ie}{{\it i.e. }}
\newc{\eg}{{\it e.g. }}
\newc{\cf}{{\it cf. }}

\newc{\rpv}{\mathrm{B}_3}

\def\lsim{\raise0.3ex\hbox{$\;<$\kern-0.75em\raise-1.1ex\hbox{$\sim\;$}}}
\def\gsim{\raise0.3ex\hbox{$\;>$\kern-0.75em\raise-1.1ex\hbox{$\sim\;$}}}

\newc{\rRPV}{\stackrel{\lamp}{\longrightarrow}}
\newc{\rRPVlam}{\stackrel{\lam}{\longrightarrow}}
\newc{\hr}{\hookrightarrow}
\newc{\hrRPV}{\stackrel{\lamp}{\hookrightarrow}}
\newc{\hrRPVlam}{\stackrel{\lam}{\hookrightarrow}}
\newc{\met}{\not{\!\!E}_T}

\newcommand{\Ye}{\big(\mathbf{Y}_E\big)}
\newcommand{\Yu}{\big(\mathbf{Y}_U\big)}
\newcommand{\Yd}{\big(\mathbf{Y}_D\big)}
\newcommand{\Ydsim}{\mathbf{Y}_D}

\newc{\neut}[1]{\tilde{\chi}_{#1}^0}
\newc{\stau}{\tilde{\tau}}

\begin{document}

\title{$\tilde\tau_1$ LSP Phenomenology: Two- Versus Four-Body Decay Modes \\
and Resonant Single Slepton Production at the LHC as an Example}

\author{H.~K.~Dreiner}
\email[]{dreiner@th.physik.uni-bonn.de}
\author{S.~Grab}
\email[]{sgrab@th.physik.uni-bonn.de}

\affiliation{Bethe Center for Theoretical Physics \& Physikalisches Institut der Universit\"at Bonn, 
Nu\ss allee 12, 53115 Bonn, Germany}

\author{M.~K.~Trenkel}
 \email[]{trenkel@mppmu.mpg.de}
\affiliation{Max-Planck-Institut f\"ur Physik, F\"ohringer Ring 6, 
80805 M\"unchen, Germany}

% \pacs{12.60.Jv, 04.65.+e, 14.80.Ly}

%%%%%%%%%%%%%%%%%%%%%%%%%%%

\begin{abstract}
\noindent We investigate B$_3$ mSUGRA models, where the lightest stau,
$\tilde\tau_1$, is the LSP. B$_3$ models allow for lepton number and
R-parity violation; the LSP can thus decay. We assume one non-zero
B$_3$~coupling $\lambda'_{ijk}$ at $M_{\rm GUT}$, which
generates further B$_3$~couplings at $M_Z$. We study the
RGEs and give numerical examples. The new couplings lead to additional
$\tilde\tau_1$ decays, providing distinct collider signatures. We classify
the $\tilde\tau_1$ decays and describe their dependence on the mSUGRA
parameters.
We exploit our results for single slepton production at the LHC. As an
explicit numerical example, we investigate single smuon production,
focussing on like-sign dimuons in the final state. Also considered are
final states with three or four muons.
\end{abstract}

\preprint{BONN-TH-2008-08}
\preprint{MPP-2008-92}

\maketitle

%%%%%%%%%%%%%%%%%%%%%%%%%%%

\input{Chapter1-Introduction}

\input{Chapter2-Model}

\input{Chapter3-LSPdecays}

\input{Chapter4-SingleSlepton}

\input{Chapter5-ExampleCalc}

\section{Conclusion}
\label{conclusion}

$\rpv$ interactions allow for LSP decays and thus reopen large regions
in the SUSY parameter space, where the LSP is charged. We have 
investigated for the first time in detail the phenomenology of $\rpv
$ mSUGRA models with a $\stau_1$~LSP. We have hereby assumed only one
non-vanishing $\rpv$~coupling $\lambda'_{ijk}$ at $M_{\rm GUT}$.

\medskip

An essential feature of the $\rpv$ mSUGRA signatures is the decay of
the $\stau_1$~LSP.  Given only one $\rpv$~coupling at $M_{\rm GUT}$, we would
expect either a 4-body or 2-body decay of the $\stau_1$~LSP depending on
whether it couples directly to the dominant $\rpv$ operator or
not. However, in $\rpv$ mSUGRA models the RGEs are highly coupled and
further couplings are generated at the weak scale. These are of course
suppressed relative to the dominant coupling but may lead to 2-body decays,
which have larger phase space and do not involve heavy propagators.

\medskip

We have here numerically investigated the generation of $\lambda_{i33}$~couplings
via dominant $\lamp_{ijk}$~couplings. 
The generated couplings are typically smaller by at least two orders of magnitude;
see Figs.~\ref{fig_runningcouplingsD} and \ref{fig_runningcouplingsU}. 
We have then performed a first detailed analysis of the parameter dependence 
of the $\stau_1$~LSP decay modes. It turned out that in large regions of parameter space 
the 2-body decay dominates over the 4-body decay, 
see Figs.~\ref{tanb_dmixing}-\ref{M0_dependence}.

\medskip

In the second part of the paper, we applied our results to resonant
single slepton production at the LHC, which is possible in $\rpv$
scenarios with a non-zero $\lamp_{ijk}$~coupling. We first studied
the general decay signatures. From the experimental point of view, the
final states with two like-sign or even more charged leptons are of
special interest. Each event is also accompanied by at least one tau. 

\medskip

We further investigated numerically single smuon production for 
$\lamp_{2jk} \neq 0$ within two representative $\stau_1$~LSP
scenarios, \ie for two sets of $\rpv$ mSUGRA parameters.  
We include the 2-body $\stau_1$~LSP decays via the generated $\lam_{233}$
couplings in our analysis.
The cross sections for like-sign dimuon final states are given in
Tab.~\ref{tab_numbersA} and Tab.~\ref{tab_numbersB} and those for
final states with three or four muons in
Tab.~\ref{tab_mehrals2muonenA} and Tab.~\ref{tab_mehrals2muonenB}. For
example, we found resulting cross sections for exclusive like-sign dimuon events of
$\mathcal{O}(100\, \text{fb})$ for $\lambda'_{2jk}|_{\rm GUT}=0.01$.
Additional three- and four-muon events can occur with the same rate.
This is a novel discovery mechanism for the LHC and should be investigated 
in more detail, also by the LHC experimental groups.

% % % % % % % % % % % % % % % % % % % % % % % % % % d

\begin{acknowledgments}
  We thank Benjamin Allanach and Markus Bernhardt for valuable
  help with the as-yet unpublished B$_3$ version of {\tt Softsusy}.
  SG thanks the theory groups of Fermilab National Accelerator, Argonne
  National Laboratory and UC Santa Cruz for helpful discussions and warm
  hospitality. SG also thanks the `Deutsche Telekom Stiftung' and
  the `Bonn-Cologne Graduate School of Physics and Astronomy' for
  financial support. This work was partially supported by BMBF grant
  05 HT6PDA and by the Helmholtz Alliance \mbox{HA-101} `Physics at the Terascale'.
\end{acknowledgments}

\appendix

\input{Chapter6-Appendices}

\bibliographystyle{h-physrev}
\bibliography{references}

\end{document}

%% file: Chapter1-Introduction.tex
\section{Introduction}

Supersymmetry
\cite{Wess:1974tw,Drees:1996ca,Nilles:1983ge,Martin:1997ns}
(SUSY) is one of the most promising extensions of the Standard Model
of particle physics (SM) \cite{Glashow:1961tr, Weinberg:1967tq}. In
its simplest form, we obtain the supersymmetric Standard Model (SSM),
with a doubling of the SM particle content and one extra Higgs
doublet. The SSM solves the hierarchy problem of the SM if SUSY is
broken at a mass scale $\lsim{\cal O} (10\,\mathrm{TeV})$.  Therefore,
SUSY should be testable at the LHC \cite{Armstrong:1994it,CMS:1996fp},
which will start taking data this year. 

\medskip

If they exist, supersymmetric particles are typically much heavier
than their SM partners and at colliders will mostly decay
rapidly. This leads to cascade decay chains in the detector to the
lightest supersymmetric particle (LSP). The nature of the LSP and its
possible decay modes is thus an essential feature for all
supersymmetric signatures. It is the purpose of this paper to study a
novel supersymmetric phenomenology, namely with the lightest scalar
tau (stau) $\stau_1$ as the LSP
\cite{Allanach:2006st,Allanach:2003eb}. In particular we analyze in
detail the potential $\stau_1$ decays in baryon-triality, B$_3$, models
\cite{Ibanez:1991hv,Ibanez:1991pr,Grossman:1998py,Dreiner:2005rd,Dreiner:2006xw}. 
We then study the discovery potential of a specific signature in this
framework, namely resonant single slepton production at the LHC,
resulting in multiple muons in the final state.

\subsection{The B$_3$ Framework}

The most general renormalizable superpotential of the SSM is
\cite{Sakai:1981pk, Weinberg:1981wj}
\beqn
W_{\mathrm{SSM}}&=& W_{\text{P}_6}+W_{\not \text{P}_6}\,,
\label{superpot} 
\eeqn
\beqn
W_{\text{P}_6}&=&\eps_{ab}\left[(\mathbf{Y}_E)_{ij}L_i^aH_d^b
  \bar{E}_j + (\mathbf{Y}_D)_{ij}Q_i^{ax}H_d^b\bar{D}_{jx}
\right.  \notag  \\ & & 
\left.+(\mathbf{Y}_U)_{ij}Q_i^{ax}H_u^b\bar{U}_{jx} + \mu
  H_d^aH_u^b\right],
\label{P6-superpot} 
\\[1.5mm]
%\eeqn
%\beqn
W_{\not \text{P}_6} & = & \eps_{ab}\left[\frac{1}{2} \lam_{ijk} L_i^aL_j^b
\bar{E}_k + \lam'_{ijk}L_i^aQ_j^{bx}\bar{D}_{kx}\right]\notag 
\\&&
+\epsilon_{ab}\kappa^i  L_i^aH_u^b
+\frac{1}{2}\eps_{xyz} \lam''_{ijk}
\bar{U}_i^{\,x} \bar{D}_j^{\,y} \bar{D}_k^{\,z} \,.
\label{notP6-superpot} 
\eeqn
Here we use the standard notation of Ref.~\cite{Allanach:1999ic}. 

\medskip

The superpotential (\ref{superpot}) consists of two different parts.
$W_{\text{P}_6}$, involves the lepton $\mathbf{Y}_E$, down-quark 
$\mathbf{Y}_D$, and up-quark ${\bf Y}_U$ Yukawa matrices, which give mass to the
leptons and quarks after electroweak symmetry breaking.

\medskip 

$W_{\not \text{P}_6}$, consists of lepton and baryon number violating
operators, which together can lead to rapid proton decay
\cite{Smirnov:1996bg,Bhattacharyya:1998bx,Barbier:2004ez,Shiozawa:1998si}.
The SSM thus requires an additional symmetry
\cite{Dreiner:2005rd,Ibanez:1991hv,Ibanez:1991pr} to stabilize the 
proton. The most widely assumed symmetry is R-parity which prohibits
$W_{\not\text{P}_6}$, leading to the MSSM. But R-parity allows
dangerous dimension-five proton decay operators such as $QQQL$
\cite{Dimopoulos:1981dw}, thus proton-hexality, $\text{P}_6$, is preferred
\cite{Dreiner:2005rd}. Here, we consider a third possibility,
baryon-triality, $\rpv$. $\rpv$ is a discrete $\boldsymbol{Z}_3
$-symmetry which prohibits only the $\bar{U}\bar{D}\bar{D}$ operators in
\eqref{notP6-superpot} but also the dangerous dimension five
operators. See for example Refs.~\cite{Lee:2007fw, Lee:2007qx,
Lee:2008pc} for $\rpv$ models that provide a dark matter candidate.

\medskip

The $\text{B}_3$ SSM has some distinguishing features compared to the 
MSSM \cite{Dreiner:1997uz,Barbier:2004ez}, which can have a strong 
impact on (hadron) collider phenomenology 
\cite{Dreiner:1991pe,Allanach:1999bf}:
\begin{enumerate}
\item Lepton flavor and lepton number are violated.
\item The renormalization group equations (RGEs) get 
additional contributions
\cite{Martin:1993zk,Allanach:1999mh,Allanach:2003eb}.
\item Neutrino masses can be generated as experimentally
observed \cite{Hall:1983id, Hempfling:1995wj, Borzumati:1996hd,
Hirsch:2000ef, Allanach:2007qc, Dreiner:2007uj}.
\item The LSP is not stable.
\item Supersymmetric particles can be produced singly, possibly on resonance. 
\end{enumerate}

\medskip

Since the LSP is not stable, we are not restricted to the lightest
neutralino $\tilde{\chi}_1^0$ as the LSP \cite{Ellis:1983ew}. The most
general B$_3$ SSM has more than 200 parameters and in principle any
SUSY particle can be the LSP. Within the MSSM, the most widely studied
constrained model is minimal supergravity (mSUGRA) with conserved
$\text{P}_6$ and radiative electroweak symmetry breaking
\cite{Chamseddine:1982jx,Barbieri:1982eh,Hall:1983iz,Soni:1983rm,Ibanez:1982fr}.
The 124 free parameters of the MSSM are reduced to only five
\begin{equation}
M_0, \, M_{1/2}, \, A_0, \, \tan \beta, \, \text{sgn}(\mu)\, ,
\label{mSUGRA}
\end{equation}
which are fixed at the grand unification (GUT) scale, $M_{\rm GUT}$. We
have a universal scalar mass $M_0$, a universal gaugino mass $M_{1/2}
$, a universal trilinear scalar coupling $A_0$, the ratio of the Higgs
vacuum expectation values (vev's) $\tan\beta$, and the sign of the
Higgs mixing parameter $\text {sgn}(\mu)$. For a wide range of these
parameters a $\tilde{\chi}_1^0$~LSP is in fact obtained at the weak
scale, $M_Z$ \cite{Ibanez:1984vq}. There are also wide ranges of
parameter space with a $\tilde\tau_1$~LSP, but these are
cosmologically excluded in the MSSM or mSUGRA \cite{Ellis:1983ew}.

\medskip

In the B$_3$ mSUGRA model we consider here
\cite{Allanach:2003eb,Allanach:2006st}, we have six parameters at the
GUT scale
\begin{align}
M_0,\,  M_{1/2},\,A_0,\, \tan\beta,\, \textnormal{sgn}(\mu),\;\; 
\text{and}\;\; {\mathbf \Lambda'},
\label{eq_rpvmsugra}
\end{align}
where ${\mathbf \Lambda'}$ stands for one non-vanishing coupling
$\lambda'_{ijk}$. A first investigation of the parameter space has
shown, that there are extensive regions with a neutralino, a stau or a
sneutrino LSP
\cite{Allanach:2003eb,Allanach:2006st}. We shall focus here on a $\stau_1$
LSP. $\stau_1$~LSP scenarios have been studied in the
 literature~\cite{Akeroyd:1997iq,deGouvea:1998yp,Akeroyd:2001pm,Bartl:2003uq,Allanach:2003eb,Allanach:2006st,Allanach:2007vi,Dreiner:2007uj,Bernhardt:2008mz,steve}.
As we now discuss, we go beyond this work in several aspects.

\subsection{New Phenomenology and Outline}
 
The $\stau_1$~LSP might decay via the dominant $L_i Q_j\bar{D}_k$
operator, \eqref{notP6-superpot}; for example via a 4-body decay
in the presence of a non-vanishing $\lambda'_{211}$
\begin{equation}
	\tilde{\tau}^-_1 \overset{\lam'_{211}}{\longrightarrow} 
	\tau^- \mu^- u \bar d \, .
\label{eq_4bdy}
\end{equation} 
An important feature of B$_3$ mSUGRA models is that additional $\rpv$
couplings are generated via the RGE running. These new couplings can
lead to 2-body decays of the $\stau_1$~LSP. For example, $\lambda'_
{211}$ will generate $\lambda_{233}$ which allows for the decay
\begin{equation}
	\tilde{\tau}_1^- \overset{\lam_{233}}{\longrightarrow} 
	\mu^- \nu_\tau \, .
\label{eq_2bdy}
\end{equation}

Even though $\lam_{233}\ll\lam_{211}^\prime$, this might be the
dominant decay mode. The decay (\ref{eq_4bdy}) is suppressed by phase
space and  heavy propagators.

\medskip

We analyze in detail the conditions for a dominance of the 2-body
decay over the 4-body decay. We provide for the first time an
extensive study of $\rpv$ $\stau_1$~LSP decays and extend and specify
thus the results of \cite{steve}, where a first estimate has been
performed. This is useful when studying both pair produced and singly
produced SUSY particles within the $\rpv$ mSUGRA model. Typically all
heavy SUSY particle decay to the ($\stau_1$)~LSP.

\medskip

In the second half of our paper, we consider the $\rpv$ mSUGRA model
with a $\stau_1$~LSP and focus on resonant single (left-handed)
charged slepton $\tilde{\ell}_{Li}$ and sneutrino $\tilde{\nu}_i$
production at hadron colliders, which proceeds via a dominant $L_i Q_j
\bar{D}_k$ operator:
\beqn
\bar u_j d_k &\overset{\lam'_{ijk}}{\longrightarrow}& \tilde{\ell}^-_{Li} \, , \\
\bar d_j d_k &\overset{\lam'_{ijk}}{\longrightarrow}& \tilde{\nu}_{i}.
\eeqn
Here, $u_j$ ($d_k$) is an up-type (down-type) quark of generation $j$
($k$). 

\medskip

Single slepton production allows us also to study two $\rpv$~couplings
at a time, depending on the scenario. The slepton is always produced
via a $\lambda'$ whereas the decay of the $\stau_1$~LSP in the decay
chain of the slepton might proceed via a generated $\lambda$, \cf \eqref{eq_2bdy}.

\medskip 

Single slepton production within a $\tilde{\chi}_1^0$~LSP scenario
leads to like-sign dileptons in the final state and has thus a very
promising signature for experimental studies, see
Refs.~\cite{Dreiner:2000vf,Dreiner:2000qf,Dreiner:1998gz,Moreau:2000bs,Deliot:2000mf}.
Here we show that for a $\stau_1$~LSP, we also obtain like-sign
dilepton events and additionally events with three or four leptons in
the final state.  We give event rates for the LHC for two
representative sets of $\rpv$ mSUGRA parameters. We also discuss the
background, although a detailed signal over background analysis is
beyond the scope of this paper. This is the first study of single
slepton production in $\stau_1$~LSP scenarios.

\medskip

We assume in the following that only one non-vanishing $\lambda'_{ijk}
$ is present at $M_{\mathrm{GUT}}$, similar to the dominant top Yukawa
in the SM. Allowing for more than one coupling leads to stricter bounds
\cite{Dreiner:1997uz,Barbier:2004ez,Chemtob:2004xr,Dreiner:2006gu,Allanach:1999ic,Agashe:1995qm}. 
The bounds for a single $\lambda'_{ijk}$ lie between $\mathcal{O}(1)$
and $\mathcal{O}(10^{-4})$ depending on the flavor indices and
sparticle masses.  These bounds can be up to four orders of magnitude
stronger at $M_{\rm GUT}$ if one includes the generation of neutrino masses
\cite{Allanach:1999ic,Allanach:2003eb}. We therefore assume below that
$\lambda'_{ijk}\lsim \mathcal{O}(10^{-2})$ and require it to be
consistent with the observed neutrino masses.

\medskip

Resonant slepton production at hadron colliders via the $L_iQ_j\bar{D}
_k$ operator was first investigated in
\cite{Dimopoulos:1988jw,Dimopoulos:1988fr}, using tree-level production
cross sections. Three-lepton final states and like-sign dilepton events
were investigated in
Ref.~\cite{Dreiner:2000vf,Dreiner:2000qf,Dreiner:1998gz,Moreau:2000bs,Deliot:2000mf}.
Ref.~\cite{Allanach:2003wz} considered scenarios with a gravitino LSP.
Experimental studies by the D0 collaboration at the Tevatron were
performed in Refs.~\cite{Abazov:2002es,Abazov:2006ii} assuming a $
\tilde{\chi}_1 ^0$~LSP and a non-vanishing $\lamp_{211}$. The NLO QCD
corrections to the cross section were computed in
\cite{Choudhury:2002au,Yang:2005ts,Dreiner:2006sv,Chen:2006ep}. The 
SUSY-QCD corrections were included by \cite{Dreiner:2006sv}. The
latter can modify the NLO QCD prediction by up to 35\%. In
Refs.~\cite{Bernhardt:2008mz,Borzumati:1999th,Accomando:2006ga,Belyaev:2004qp}
single slepton production in association with a single top quark was
considered.

\medskip

The outline of our paper is as follows. In Sect.~\ref{sec_model} we
review the $\rpv$ mSUGRA model and approximate formul{\ae} for
sparticle masses. We define two $\rpv$ mSUGRA scenarios with a
$\stau_1$~LSP, as a reference for phenomenological studies. We then
derive approximate equations for the RGE generation of $\lambda$ from
$\lambda'$. In Sect.~\ref{stau_LSP_decays}, we classify the different
decay modes of the $\stau_1$~LSP and investigate the conditions for a
dominance of the 2-body decay over the 4-body decay and \textit{vice versa}.
In Sect.~\ref{single_slep_signatures}, we classify all possible
signatures for resonant single slepton production in $\rpv$ mSUGRA
models with a $\stau_1$~LSP. In Sect.~\ref{numerical_example} we
calculate event rates for like-sign dimuon events as well as for
three- and four-muon events, at the LHC. We also discuss backgrounds
and cuts for like-sign dimuon events. We conclude in
Sect.~\ref{conclusion}.

%% file: Chapter2-Model.tex
\section{The low energy spectrum of the $\rpv$ mSUGRA Model with a $\stau_1$~LSP}
\label{sec_model}

We have defined the $\rpv$ mSUGRA model in Eq.~(\ref{eq_rpvmsugra})
via six input parameters at the GUT scale
\cite{Allanach:2003eb,Allanach:2006st}. We now discuss the low energy 
spectrum.  Sparticle masses and couplings are obtained by running the
respective RGEs down to the weak scale. Due to the mixing of different
quark flavors, described by the Cabibbo-Kobayashi-Maskawa (CKM)
matrix, the RGEs of the $\rpv$~couplings are not independent, but
highly coupled.  Therefore, a single non-zero $\lamp_{ijk}$ at the GUT
scale generates a set of other non-zero $\rpv$~couplings at lower
scales.  Assuming a diagonal charged lepton Yukawa matrix
$\mathbf{Y}_E$, only those couplings can be generated which violate
the same lepton number as $\lamp_{ijk}$, {\it i.e.} $\lamp_{imn}$ and
$\lam_{ill}$.  No additional source of lepton number violation is
introduced.  Phenomenologically particularly relevant is the
generation of $\lam_{i33}$, which we discuss in detail in
Sect.~\ref{sec_rges}.

\subsection{Sparticle Spectra}

The low energy SUSY particle masses depend strongly on the universal
mSUGRA parameters (\ref{mSUGRA}) and only weakly on $\lambda' \lsim 
\mathcal{O}(10^{-2})$ \cite{Allanach:2006st}. For later use, we cite
here approximate expressions for the relevant SUSY particle masses in
terms of the mSUGRA parameters as given in \cite{Drees:1995hj}, \cf
also the original work in Ref.~\cite{Ibanez:1984vq}. The masses of the
sleptons of the first and second generation are
\begin{align}
\begin{split}
m_{\tilde{\ell}_R}^2 & = M_0^2 + 0.15 M_{1/2}^2 - \sin^2\theta_W M_Z^2 
\cos 2\beta, \\
m_{\tilde{\ell}_L}^2 & = M_0^2 + 0.52 M_{1/2}^2 - (0.5 - \sin^2\theta_W) 
M_Z^2 \cos 2\beta, \\
m_{\tilde{\nu}}^2 & = M_0^2 + 0.52 M_{1/2}^2 + 0.5 M_Z^2 \cos 2\beta, %,\\
\end{split}
\label{eq_sfermionmasses}
\end{align}
where $m_{\tilde{\ell}_{R,L}}$ denotes the mass of a
right-/left-handed selectron or smuon, respectively, $m_{\tilde{\nu}}$
the mass of a left-handed electron or muon sneutrino, and $\theta_W$
the electroweak mixing angle. $M_Z$ is the mass of the $Z$
boson.

\medskip

For sfermions of the third generation, the mixing between left- and
right-handed gauge-current eigenstates has to be taken into
account.  The stau mass matrix squared
$\mathfrak{M}^2_{\tilde\tau}$ is given by \cite{Gunion:1984yn}
\begin{align}
\mathfrak{M}^2_{\tilde\tau} &= \left( {m_{\tau}^2 + A_{LL} 
\qquad m_{\tau} B_{LR}} 
\atop {m_{\tau} B_{LR}\qquad m_{\tau}^2 + C_{RR} } \right),
\label{eq_staumassmatrix}
\end{align}
with $m_{\tau}$ denoting the tau lepton mass and, expressed in terms
of left- and right-handed third generation softbreaking parameters
$m_{\tilde L_3}$ and $m_{\tilde E_3}$, respectively,
\begin{align}
\begin{split}
A_{LL}  &=  m^2_{\tilde L_3} - (0.5 - \sin^2\theta_W) M_Z^2 \cos 2\beta\,, \\
   B_{LR}  &= A_{\tau} - \mu \tan{\beta}\,, \\
   C_{RR} &=   m^2_{\tilde E_3} - \sin^2\theta_W M_Z^2 \cos 2\beta,
\end{split}
\end{align}
where $A_{\tau}$ is the trilinear coupling of the left- and
right-handed stau to the Higgs. In mSUGRA, $A_{\tau}=A_0$
at the GUT scale.  The softbreaking parameters depend on the
mSUGRA parameters as follows \cite{Drees:1995hj},
\begin{align}
% \begin{split}
m_{\tilde E_3}^2 &= M_0^2 + 0.15 M_{1/2}^2 - \frac{2}{3} X_\tau, \nn \\
m_{\tilde L_3}^2 &= M_0^2 + 0.52 M_{1/2}^2 - \frac{1}{3} X_\tau,
\label{eq_xtaudef}\\
X_\tau &\equiv 10^{-4}(1+\tan^2 \beta) \left( M_0^2 + 0.15 M_{1/2}^2 + 
0.33A_0^2 \right),
\nn
% \end{split}
 \end{align} where $X_{\tau}$ parameterizes the influence of the tau
 Yukawa coupling.  Note, that $X_\tau$ can have a strong impact on the
 stau masses due to its $\tan^2\beta$ dependence, even though $X_\tau$
 is suppressed by a factor $10^{-4}$.  We will investigate this effect
 on the $\stau_1$ decay branching ratios in the next section.

\medskip

The stau mass eigenstates $\stau_{1,2}$ are obtained
from the gauge eigenstates by a unitary rotation $U$ such that 
$U$ diagonalizes the mass matrix, $U \mathfrak{M}^2_{\tilde\tau} U^{\dagger} \, = \, \text{diag}
\left(m_{\tilde{\tau}_1}^2, m_{\tilde{\tau}_2}^2\right)$, yielding for
the masses $m_{\tilde{\tau}_{1,2}}$
\begin{align} 
\begin{split}
m_{\tilde{\tau}_{1,2}}^2   = & \,  m_{\tau}^2 +  \frac{1}{2} (A_{LL} + C_{RR})
 \\
	& \mp \frac{1}{2} \sqrt{(A_{LL}-C_{RR})^2
                + 4 m_{\tau}^2 B_{LR}^2}\,.
\end{split}
\label{eq_staumass}
\end{align}

\medskip

The gaugino masses can be approximated in terms of the universal
gaugino mass $M_{1/2}$ \cite{Drees:1995hj},
\begin{align}
\begin{split}
	m_{\tilde{\chi}_1^0} &\simeq M_1 =  0.41 M_{1/2}, 
\\
	m_{\tilde{\chi}_2^0} &\simeq M_2 =  0.84 M_{1/2}.
\label{neutralino_masses}
\end{split}
\end{align}
Here it has been used that the lightest neutralino $\tilde{\chi}_1^0$
is bino-like in many mSUGRA models and that its mass can be
approximated by the bino mass parameter $M_1$ at the weak scale.
Accordingly, the second lightest neutralino $\tilde{\chi}_2^0$ is
mainly wino-like and its mass governed by the wino mass parameter
$M_2$.

\subsection{Reference Scenarios with a $\stau_1$~LSP}

For the purpose of numerical studies and as future reference points,
we define two specific sets of $\rpv$ mSUGRA scenarios with a
$\stau_1$~LSP: 
\begin{align}
\begin{split}
\textnormal{\bf Set A:\,\,}& M_0 = 0 \textnormal{\,GeV},\,  M_{1/2}=500 
\textnormal{\,GeV},
	\\& A_0=600 \textnormal{\,GeV},\, \tan\beta=13, \textnormal{sgn}(\mu) = +1,
	\\& \textnormal{a single\,} \lamp_{ijk} \neq 0\lvert_{\rm GUT}, 
	% \\& \textnormal{quark mixing in the up- or down sector as specified};
\\[2ex]
\textnormal{\bf Set B:\,\,}& M_0=0 \textnormal{\,GeV},\, M_{1/2}=700 \textnormal{\,GeV},
	\\& A_0=1150 \textnormal{\,GeV}\,, \tan\beta=26, \textnormal{sgn}(\mu) = +1,
	\\& \textnormal{a single\,} \lamp_{ijk} \neq 0\lvert_{\rm GUT}.
\label{eq_sets}
\end{split}
\end{align}

They are chosen in accordance with the following bounds \footnote{In
  Ref.~\cite{Allanach:2006st}, specific benchmark scenarios
  with a $\stau_1$~LSP where proposed. We do not consider them here
  because even the weakest bounds on $\lamp$ assuming down-type quark
  mixing are at the order of $\mathcal{O}(10^{-3})$ for which the rate
  of resonant slepton production is suppressed.}:
\begin{itemize}
\item ${\rm BR}(B_s \rightarrow \mu^+ \mu^-) < 5.8 \times 10^{-8}$ 
at the $95\%$ C.L. obtained by the CDF collaboration \cite{:2007kv}.
\item $2.76 \times 10^{-4} < {\rm BR}(b \rightarrow s \gamma) < 4.34 \times 10^{-4}$
which is the central theoretical value at $2\sigma$ \cite{Allanach:2006st} 
using the experimental value of \cite{Barberio:2007cr}.
\item The discrepancy between experiment and the SM prediction of the
anomalous magnetic moment of the muon is $\delta a_\mu=a_\mu^{\rm exp}-a_
\mu^{\rm SM} = (29.5 \pm 8.8) \times 10^{-10}$, {\it i.e.} $3.4\sigma$ 
\cite{Bennett:2006fi,Miller:2007kk,Stockinger:2007pe}. The sets 
(\ref{eq_sets}) are chosen such that $\delta a_\mu^{\rm SUSY}=a_
\mu^{\rm MSSM} - a_\mu^{\rm SM}$ agrees with $\delta a_\mu$ within $2 \sigma$.
\item Higgs mass $m_{h^0}\geq 112.4$ GeV. This value corresponds
to the LEPII bound of $m_{h^0}\geq 114.4$ GeV at $95\%$ C.L. \cite{Barate:2003sz} 
assuming a numerical error of 2 GeV for the mass prediction.
\item A non-vanishing coupling $\lambda'_{ijk}$ at the GUT scale
  will generate a tree-level neutrino mass
  \cite{Hall:1983id,Ellis:1984gi,deCarlos:1996du,Nardi:1996iy,Allanach:2003eb}.
  All couplings $\lambda'_{ijk}$ in the following are chosen such
  that the tree-level neutrino mass is smaller than the cosmological
  bound on the sum of neutrino masses from WMAP \cite{Spergel:2003cb}
  combined with 2dGRFS data \cite{Colless:2003wz}: $\sum_im_{\nu_i}<
  0.71$ eV. A corresponding comprehensive set of bounds for the mSUGRA
  parameter set SPS1a \cite{Allanach:2002nj} with one non-vanishing
  coupling $\lambda'_{ijk}$ is given in Ref.~\cite{Allanach:2003eb}.
  Note, that the generated tree-level neutrino mass depends on all
  mSUGRA parameters (\ref{eq_rpvmsugra}). The neutrino mass bounds on
  $\lambda'_{ijk}$ for Set A and Set B are weaker compared to those
  for SPS1a.
\end{itemize}
We use the computer programs provided by
\cite{onlinecalc,Allanach:2003jw,Belanger:2005jk} to calculate ${\rm BR}(B_s
\rightarrow \mu^+ \mu^-)$, ${\rm BR}(b \rightarrow s \gamma)$, and $\delta
a_\mu^{\rm SUSY}$. These programs do not include the B$_3$~couplings. But
the corresponding effects are negligible for $\lambda'_{ijk}
\lsim \mathcal{O}(10^ {-2})$ \cite{Allanach:2006st}.
\begin{table}
\begin{ruledtabular}
\begin{tabular}{c|cc||c|cc}
\hspace*{7ex}& \multicolumn{2}{c||}{  masses [GeV]} &
\hspace*{5ex}& \multicolumn{2}{c}{  masses [GeV]} \\
 & Set A  & Set B  && Set A  & Set B \\
\hline
$\stau_1$ & 179 & 146 & $\neut{1}$ & 203 & 290\\
$\tilde{e}_R$ & 193 & 266 &$\neut{2}$ & 380 & 544 \\
$\stau_2$ & 340 & 453 & $\neut{3}$ & 571 & 754\\
$\tilde{e}_L$ & 340 & 471 & $\neut{4}$ & 587 & 765\\
$\tilde{\nu}_{\tau}$ & 326 & 437 & $\tilde{\chi}_{1}^{\pm}$ & 383 & 549\\
$\tilde{\nu}_{e}$ & 329 & 461 & $\tilde{\chi}_{2}^{\pm}$ & 583 & 761\\
\hline
$\tilde{t}_1$ & 841 & 1160 & $h^0$ & 113 & 115\\
$\tilde{b}_1$ & 970 & 1300 & $H^0$ & 643 & 795 \\
$\tilde{u}_R$ & 1010 & 1370 & $A^0$ & 642 &  795 \\
$\tilde{t}_2$ & 1010 & 1340 &  $H^+$ & 648 & 799 \\
$\tilde{b}_2$ & 995 & 1340 & &&\\
$\tilde{u}_L$ & 1040 & 1410 & $\tilde{g}$ & 1150 & 1560
\end{tabular}
\caption{\label{tab_masses} Sparticle masses for the $\rpv$ mSUGRA
  sets A and B as defined in \eqref{eq_sets}, evaluated for a
  renormalization scale $Q_{\rm susy} = \sqrt{m_{\tilde{t}_1}(Q_{\rm
      susy}) \,m_{\tilde{t}_2}(Q_{\rm susy})}$ using {\tt Softsusy 2.0.10}
  \cite{Allanach:2001kg}.  The variation due to different $\lamp_{ijk}
  \neq 0 |_{\rm GUT}$ and quark mixing (see Sect.~\ref{subec_mixing}) is
  below the percent level.  The masses in the second generation
  coincide with those in the first generation.  }
\end{ruledtabular}
\end{table}

\medskip

We show in Table~\ref{tab_masses} the supersymmetric mass spectra of
the parameter sets A and B (\ref{eq_sets}).  We have neglected the
mass dependence on the different non-zero B$_3$~couplings which is
valid if $\lambda'_{ijk} \lsim\mathcal{O}(10^{-2})$
\cite{Allanach:2006st}. The main B$_3$ effect on the spectrum
  is that we allow for a $\tilde\tau_1$~LSP.

\medskip

One naturally obtains a $\tilde\tau_1$~LSP spectrum for $M_{1/2}
\gg M_0$. The large $M_{1/2}$ raises the lightest neutralino mass 
(\ref{neutralino_masses}) faster than the right-handed slepton masses
(\ref{eq_sfermionmasses}). It also drives the gluino and indirectly
via the RGEs the squark masses up. We thus see in
Table~\ref{tab_masses} squark and gluino masses $\gsim1\,$TeV, while
the slepton masses are below 500 GeV.  Another general feature of a
$\stau_1$~LSP scenario is that the second lightest neutralino and the
lightest chargino are also heavier than the sleptons. Therefore the
only conventional supersymmetric decays of the left-handed sleptons
are via the lightest neutralino. Depending on the dominant B$_3$
coupling and its size, the left-handed sleptons can also decay into
two jets.

\medskip

Nearly all sparticles in Set B ($M_{1/2}= 700$~GeV) are heavier than
in Set A ($M_{1/2}= 500$~GeV).  The most important difference for the
phenomenology at colliders arises from the different values of $\tan
\beta$ (\mbox{$\tan\beta=13$} in Set A, $\tan\beta=26$ in Set B). According
to \eqref{eq_xtaudef}, the soft breaking parameters of the stau
decrease for increasing $\tan\beta$ and thus both stau mass
eigenstates are reduced for large values of $\tan\beta$.  Furthermore,
the mass of the lighter stau is reduced due to the larger L--R-mixing,
\cf Eq.~(\ref{eq_staumassmatrix}). This effect can be seen in
Table~\ref{tab_masses}, where the mass of the $\stau_1$~LSP is 179~GeV
in Set~A but only 146~GeV in Set~B. The $\stau_1$~mass and $\tan\beta$
strongly influence the possible 2- and 4-body $\tilde\tau_1$~LSP
branching ratios.  We will investigate this topic in detail in
Sect.~\ref{stau_LSP_decays}.

% % % % % % % % % % % % % % % % % % % % % % % % % % % % % % % % 
\subsection{Fermion Mixing}
\label{subec_mixing}

Since the B$_3$ RGEs are coupled, given one non-zero B$_3$~coupling at
$M_{\rm GUT}$, we will generate many non-zero couplings at the weak scale $M_
Z$. As we will see in the next section, the size of the
dynamically generated B$_3$~couplings depends sensitively on the
composition of the quark Yukawa matrices. For this reason we prepend
here a short discussion of quark mixing in B$_3$ models.

\medskip

Initially at $M_{\rm GUT}$, all parameters are given in the
weak-current eigenstate basis.  This includes the quark and lepton
Yukawa coupling matrices ${\bf Y}_U,\, {\bf Y}_D,\, {\bf Y}_E$ and the
corresponding mass matrices $\bf{m}_u,\, \bf{m}_d,\, \bf{m}_e$.
Since, in general, these matrices are not diagonal, we need to rotate
the (charged) lepton and quark fields from the weak into the mass
eigenstate basis,
\begin{align}
	f_{L,R}^{\rm mass} & = 
		{\bf V_{f\, L,R}^{}} \,\, f_{L,R}^{\rm weak}  \,,
\end{align}
with $f_{L,R}$ denoting the left- and right-handed fermion fields,
respectively and $\bf V_{f\, L,R}$ denoting the corresponding rotation
matrices.  The mass matrices in the mass eigenstate basis are then
given by
\begin{align}
\begin{split}
	\bf{V_{u\, L}^{}} \, \bf{m}_u \, \bf{V_{u\, R}^+} &= \text{diag} (m_u, m_c, m_t),\\
	\bf{V_{d\, L}^{}} \, \bf{m}_d \, \bf{V_{d\, R}^+} &= \text{diag} (m_d, m_s, m_b),\\
	\bf{V_{e\, L}^{}} \, \bf{m}_e \, \bf{V_{e\, R}^+} &= \text{diag} (m_e, m_{\mu}, m_{\tau}),
\end{split}
\end{align}
defined at the weak scale $M_Z$. 
The rotation matrices ${\bf V_{f\, L,R}^{}}$ are not directly 
experimentally accessible but only the
CKM matrix $\bf V_{\rm CKM}$,
\begin{align}
	\bf
	V_{\rm CKM} = V_{u\,L}^{} V_{d\,L}^+.
\end{align}

In general, the rotation matrices for the left-handed fields
differ from those for the right-handed fields.  In the following,
however, for simplicity and definiteness, we assume real and
symmetric Yukawa coupling matrices, thus $\bf V_{f\,L}=
V_{f\,R}$. Furthermore we neglect neutrino masses in this context and
assume that ${\bf Y}_E$ is diagonal in the weak-current basis.
Correspondingly, ${\bf V_{e\, L,R}} = \mathbf{1}_{3\times3}$.

\medskip

To further constrain the quark Yukawa couplings, we restrict ourselves
to the extreme cases of quark mixing taking place completely in the
up- or the down-quark sector, respectively. We will refer to it
as ``up-type mixing" if
\begin{align}
	{\bf V_{u\,L,R}^{}}  = {\bf V_{\rm CKM}}, \quad {\bf V_{d\,L,R}^{}}  = {\bf 1}_{3\times3},
\end{align}
at the weak scale $M_Z$ and as ``down-type mixing" if
\begin{align}
{\bf V_{u\,L,R}^{}}  = {\bf 1}_{3\times3} , \quad {\bf V_{d\,L,R}^{}}  = {\bf V_{\rm CKM}^+}
\end{align}
at the weak scale.
Therefore, in up-type mixing scenarios, the Yukawa matrices are
\begin{align}
	{\bf Y}_U (M_Z)\times v_u &= {\bf V_{\rm CKM}^+}\cdot \text{diag} (m_u, m_c, m_t) 
\cdot{\bf V_{\rm CKM}}, \nonumber \\
	{\bf Y}_D (M_Z)\times v_d &=  \text{diag} (m_d, m_s, m_b) ,
\label{Yukawa_upmix}
\end{align} 
and in down-type mixing scenarios, the Yukawa matrices are
\begin{align}
	{\bf Y}_U (M_Z) \times v_u &= \text{diag} (m_u, m_c, m_t), \\
	{\bf Y}_D (M_Z) \times v_d &= {\bf V_{\rm CKM}}\cdot \text{diag} (m_d, m_s, m_b) 
	\cdot{\bf V_{\rm CKM}^+} \nonumber,
\end{align} 
respectively. In the following we will consider these two
extreme cases.  $v_u$ ($v_d$) is the vacuum expectation value of the
up-type (down-type) neutral CP-even Higgs with
\begin{equation}
	v_u = v \sin \beta \, , \quad \quad v_d = v \cos \beta \, ,
\label{vevs}
\end{equation}
where $v=174$ GeV is the SM vacuum expectation value 
\footnote{In $\rpv$ SUSY models, \eqref{vevs} is in
general modified by additional sneutrino vacuum expectation values
$v_i$. But $v_i \ll v$ in order to be consistent with neutrino
masses. We therefore neglect $v_i$ in Eq.~(\ref{vevs}).}.

\medskip

As a consequence of the non-trivial quark rotation matrices, the
$\lamp_{ijk}$~coupling in \eqref{eq_rpvmsugra} also has to be
rotated from the weak basis into the quark mass basis for a comparison
with experimental data. In case of up-type mixing, the $L_i Q_j \bar
D_k$ interactions of the superpotential (\ref{superpot}) in the quark
mass basis are in terms of SU(2) component superfields
\begin{align}
	\lamp_{ijk}[N_i D^m_j 
		- E_i {\bf (V_{\rm CKM}^+})_{jl} U^m_l ] \bar D^m_k \, .
\end{align}
In the case of down-mixing they are 
\begin{align}
	\lamp_{ijk}[N_i {\bf (V_{\rm CKM}})_{jl} D^m_l 
		- E_i U^m_j ] {\bf (V_{\rm CKM}^+})_{nk} \bar D^m_n \, .
\end{align}
See also Ref.~\cite{Agashe:1995qm}. However for slepton production
cross sections, we do not take into account these CKM effects.  If
needed, the corresponding rescaling of the $\lamp$~coupling can be
done easily. Furthermore the sub-dominant interactions, which include
non-diagonal matrix elements of ${\bf V_{\rm CKM}}$, do not allow for
large production cross sections since $\lamp$ enters only quadratically.

%%%%%%%%%%%%%%%%%%%%%%%%%%%%%%%%%%%%%%%%%%%%% 
\subsection{Renormalization Group Equations}
\label{sec_rges}

One of the most important consequence of including $\rpv$ effects in
SUSY models is that the LSP is no longer stable. This is of
special interest for phenomological studies if the LSP couples
directly to the dominant $\rpv$ operator. This leads to large
LSP decay widths and to distinctive final state signatures.

\medskip

In the scenarios considered in this work, {\it cf.}
\eqref{eq_rpvmsugra}, the dominant coupling is a $\lamp_{ijk}$;
for $i\neq3$ it does not couple to the $\stau_1$~LSP. However,
the RGEs of the $\rpv$~couplings are
coupled via non-diagonal entries of Higgs-Yukawa matrices and a
$\lamp_{ijk}$ generates dynamically other $\rpv$~couplings. Among
those, we want to focus on the $\lam_{i33}$ which do couple directly
to the $\stau_1$~LSP.

\medskip 

The aim of the next two sections is to study the RGEs of the dominant
$\lamp_{ijk}$ and to quantitatively determine the generated
  $\lam_{i33}$. We then use these results to predict the low energy
spectrum of $\rpv$ mSUGRA scenarios given by \eqref{eq_rpvmsugra}.  We
will also derive approximate formul{\ae} that allow for a
numerical implementation of the running of the couplings.

\medskip

The full renormalization group equations for the $\rpv$~couplings 
$\lamp_{ijk}$ and $\lam_{i33}$  
are \cite{Martin:1993zk, Allanach:2003eb,Allanach:1999mh},
\begin{align}
\begin{split}
	16\pi^2 \frac{d}{dt} \lambda'_{ijk} =\,&
		\lambda'_{ijl} \, \gamma_{D_l}^{D_k}
	+	\lambda'_{ilk} \, \gamma_{Q_l}^{Q_j}
	+	\lambda'_{ljk} \, \gamma_{L_l}^{L_i} 
\\
	&-	\Yd_{jk} \, \gamma_{H_1}^{L_i},
\label{RGElambdap}
\end{split}
\\
\begin{split}
	16\pi^2 \frac{d}{dt} \lambda_{i33} =\,&
		\lambda_{i3l} \,\gamma_{E_l}^{E_3}
	+	\lambda_{il3} \, \gamma_{L_l}^{L_3}
	+	\lambda_{l33} \, \gamma_{L_l}^{L_i} 
\\
	&-	\Ye_{33} \, \gamma_{H_1}^{L_i}
	+ 	\Ye_{i3} \, \gamma_{H_1}^{L_3},
\label{RGElambda}
\end{split}
\end{align}
with $t = \ln Q$, $Q$ being the renormalization scale.  The anomalous
dimensions $\gamma$ are listed in \cite{Allanach:2003eb} at one-loop
level and in \cite{Allanach:1999mh} at two-loop level. The RGEs
simplify considerably under the assumption of the single $\rpv$
coupling dominance hypothesis \cite{Dimopoulos:1988jw, Barger:1989rk}.
Products of two or more
$\rpv$~couplings including quadratic contributions of the
dominant coupling can be neglected for $\lambda'\lsim\mathcal{O}(10^
{-2})$. In this limit, the one-loop anomalous dimensions read
\begin{align}
\begin{split}
	\gamma_{Q_j}^{Q_i} =\,&  \big( \mathbf{Y}_D \mathbf{Y}_D^+ \big)_{ij}
		+	 \big( \mathbf{Y}_U \mathbf{Y}_U^+ \big)_{ij}  
\\
		&- 	\delta_{j}^{i} \Big( \frac{1}{30} g_1^2 
				+ \frac{3}{2} g_2^2 + \frac{8}{3} g_3^2 \Big),
\\
	\gamma_{D_j}^{D_i} =\,& 2 \big( \mathbf{Y}_D^+ \mathbf{Y}_D \big)_{ji}
		- \delta_{j}^{i} \Big(\frac{2}{15} g_1^2 + \frac{8}{3} g_3^2\Big),
\\
	\gamma_{L_j}^{L_i} =\,& \big( \mathbf{Y}_E \mathbf{Y}_E^+ \big)_{ij}
		- 	\delta_{j}^{i} \Big( \frac{3}{10} g_1^2 + \frac{3}{2} g_2^2 \Big),
\\
	\gamma_{E_j}^{E_i} =\,& 2 \big( \mathbf{Y}_E^+ \mathbf{Y}_E \big)_{ji}
		- 	\delta_{j}^{i} \Big( \frac{6}{5} g_1^2 \Big),
\\
	\gamma_{H_1}^{L_i} =\,& -3\lambda'_{iaq} \big(\mathbf{Y}_D \big)_{aq}
		- 	\lambda_{ibq} \big(\mathbf{Y}_E \big)_{bq}\,,
\label{eq_anomaldim}
\end{split}
\end{align}
where $g_1,\,g_2,\,g_3$ are the three gauge couplings.

\medskip

From Eqs.~(\ref{RGElambda}) and (\ref{eq_anomaldim}), we see that the
terms related to $\gamma_{H_1}^{L_i}$ allow for the dynamical
generation of $\lambda_{i33}$ by a non-zero $\lamp_{iaq}$~coupling
[and \textit{vice versa} for \eqref{RGElambdap}].  All other terms in
\eqref{RGElambda} only alter the running of $\lambda_{i33}$ once it is
generated.  The RGEs can be further simplified. At one-loop level,
all $\rpv$~couplings but the dominant $\lamp_{ijk}$ and the generated
$\lam_{i33}$ can be neglected in the RGEs since they must be
generated first by $\lamp$ and thus contribute at two-loop level only.

\medskip

Since we work in a diagonal charged lepton Yukawa basis, the last term
in \eqref{RGElambda}, proportional to $\Ye_{i3}$ does not contribute
to the running of $\lam_{i33}$. It is only non-zero if $i=3$, but
owing to the $ij$-antisymmetry of $\lam_{ijk}$ no coupling is
generated in this case ($\lam_{333} = 0$).

\medskip

Next, a general ordering of the parameters in the anomalous dimensions
is \footnote{$\tan \beta$ will increase $\Yd$ and $\Ye$, \cf
Eqs.~(\ref{Yukawa_upmix})-(\ref{vevs}). Thus for $\tan\beta\gsim30$, 
the ordering of the parameters can change to $\Yd_{33}^2>\Ye_{33}^2>
g_1^2$.}
\begin{align}
	g_3^2 >\Yu_{33}^2 > g_2^2 > g_1^2 > \Yd_{33}^2 > \Ye_{33}^2,
\label{eq_y2}
\end{align}
and all other entries of the $\mathbf{Y}$ matrices are smaller by at
least one order of magnitude {\footnote{Note that the charm Yukawa
coupling $\Yu_{22}$ is roughly equal to the tau Yukawa coupling
$\Ye_{33}$ if $\tan\beta=\mathcal{O}(1)$. We have neglected the charm
Yukawa coupling, because we will assume that $\tan\beta=\mathcal{O}
(10)$.}}.  The contributions to the RGEs are thus largest for diagonal
anomalous dimensions.

\medskip

As a result, the RGEs for a non-zero $\lamp_{ijk}$ at the GUT scale
and a generated $\lam_{i33}$ reduce to
\begin{align}
\begin{split}
16\pi^2 \frac{d}{dt}\lambda'_{ijk}  =\,&
\lambda'_{ijk} \,\Big[ -  \frac{7}{15}g_1^2 -3 g_2^2 -\frac{16}{3}g_3^2 
\\
	&+ \Yd_{33}^2 \big( 2 \delta_{k3} + \delta_{j3}
				+ 3 \delta_{j3} \delta_{k3} \big)
\\
	&+ \Yu_{33}^2 \delta_{j3}	+ \Ye_{33}^2 \delta_{i3}\Big],
\label{RGElambdaPX}
\end{split} \\
\begin{split}
	16\pi^2 \frac{d}{dt}\lambda_{i33}  =\,&
		\lambda_{i33} \,\Big[ - \frac{9}{5}g_1^2 - 3 g_2^2 
			+ 4 \Ye_{33}^2 \Big] \\
	&+	3 \lambda'_{ijk} \Ye_{33} \big(\mathbf{Y}_D\big)_{jk}\, .
\label{RGElambdaX}
\end{split}
\end{align}
A similar analytical approximation for the generation of $\lam$ is
derived in \cite{steve}. But the effect of the gauge
couplings is neglected there. See also Ref.~\cite{deCarlos:1996du}.

\medskip

\begin{figure}
	\includegraphics[width=.6\columnwidth]{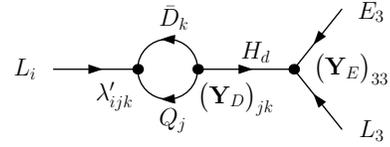}
\caption{\label{fig_lamErzeugung}
Superfield diagram for the dynamical generation of $\lam_{i33}$ by 
$\lamp_{ijk}$ at one loop order, see \eqref{RGElambdaX}.}
\end{figure}

The last term in \eqref{RGElambdaX} induces the dynamical generation
of $\lam_{i33}$.  Diagrammatically, this process can be understood as
shown in Fig.~\ref{fig_lamErzeugung}. We see that at one-loop the
lepton-doublet superfield mixes with the Higgs doublet superfield
$H_d$ via the $\rpv$~coupling $\lam'_{ijk}$ and the standard down
quark Yukawa coupling. $H_d$ then couples via the tau Yukawa coupling
$\Ye_{33}$ purely leptonically. The resulting effective interaction is
of the $\lam_{i33}$-type.

\medskip

It is important to notice that the generation is related to
$\Yd_{jk}$. Whether a given $\lamp_{ijk}$ can generate $\lam_{i33}$ or
not depends on whether $\Yd_{jk}\not=0$. For $j\not=k$ it thus depends
crucially on the origin of the CKM mixing: is it dominantly down-type
or up-type mixing. In case of down-type mixing, all entries of the
$\Ydsim$ matrix are non-zero and all $\lamp_{ijk}$ can therefore
generate a $\lam_{i33}$. In contrast, if the quark mixing takes place
in the up-sector, only the diagonal entries of $\Ydsim$ are non-zero
and $j=k$ is required. The flavor and size of the generated coupling
depends on $\tan \beta$ and on the precise $j,k$ configuration.  A
strong ordering is expected that goes along with the ordering of the
entries of the $\Ydsim$ matrix.

\medskip 

In order to study the running of the $\rpv$~couplings, the RGEs for
the Yukawa matrix elements $\Yd_{jk}$, $\Yu_{33}$, and $\Ye_{33}$ and
the gauge couplings are also needed. The full RGEs for the Yukawa
couplings are given in \cite{Martin:1993zk,Allanach:2003eb}.
Applying the single coupling dominance hypothesis, neglecting
quadratic terms in $\lamp_{ijk}$, and considering only the dominant
terms \eqref{eq_y2}, they read
\begin{align}
	16\pi^2 \frac{d}{dt} \Yu_{33}=\,& \Yu_{33} \Big[
		- \frac{13}{15} g_1^2 - 3 g_2^2  -\frac{16}{3} g_3^2
\nn\\
		&+ 6 \Yu_{33}^2  + \Yd_{33}^2 \Big],
\label{eq_RGEyu33X}
\\
	16\pi^2 \frac{d}{dt}\Ye_{33} =\,& \Ye_{33} \Big[
		- \frac{9}{5} g_1^2 - 3 g_2^2 
\nn\\
	&+ 4 \Ye_{33}^2 + 3 \Yd_{33}^2
		 \Big],
\label{eq_RGEye33X}
\\
	16\pi^2 \frac{d}{dt}\Yd_{jk}  =\,& \Yd_{jk} \Big[
		- \frac{7}{15} g_1^2 - 3 g_2^2  -\frac{16}{3} g_3^2 
\nn\\
	&+ \Yd_{33}^2 \big( 3 + \delta_{j3} + 2\delta_{k3}\big)
\nn\\
	&+ \Yu_{33}^2 \delta_{j3}  + \Ye_{33}^2 \Big].
\label{eq_RGEydjkX}
\end{align}

The one-loop order RGEs for the three gauge couplings 
within the MSSM are given by \cite{Martin:1993zk}
\begin{align}
	16\pi^2 \frac{d}{dt} g_i =\,& b_i\, g_i^3,
\label{eq_RGEgauge}
\end{align}
with $b_i = \lbrace 33/5,\, 1,\, -3\rbrace$ for $i=1,\,2,\,3$. Thus in
total, a set of nine coupled differential equations,
Eqs.~(\ref{RGElambdaPX}) - (\ref{eq_RGEgauge}), has to be solved
\footnote{In case of $j=k=3$ only 8 equations need to be solved. But
  this implies that the slepton has to be produced by parton quarks of
  the third generation which is strongly suppressed due to their
  negligible parton density.}.

%%%%%%%%%%%%%%%%%%%%%%%%%%%%%%%%%%%%%
\subsection{Numerical Results}
\label{subsec_numres}

For the numerical implementation of the RGEs we start from the
framework provided by {\tt Softsusy 2.0.10}
\cite{Allanach:2001kg}.
First, {\tt Softsusy} evaluates all necessary parameters at the SUSY scale 
\begin{equation}
Q_{\rm susy} = \sqrt{m_{\tilde{t}_1}(Q_{\rm susy}) \,m_{\tilde{t}_2}
(Q_{\rm susy})}\,\,.
\label{SUSY_scale}
\end{equation}
In a second step, we apply the (R-parity conserving) RGEs
(\ref{eq_RGEyu33X})-(\ref{eq_RGEgauge}) to run the Yukawa couplings
and gauge couplings up to the GUT scale.  Here we add the $\rpv$
couplings $\lamp_{ijk}\neq0|_{\mathrm{GUT}}$ and $\lam_{i33}=0|_{
\mathrm{GUT}}$ and evolve these  
couplings down to the scale $Q$ using the above given $\rpv$ RGEs 
(\ref{RGElambdaPX}) and (\ref{RGElambdaX}). We have implemented the 
RGEs using a standard Runge Kutta formalism \cite{NumRecipies}.

\medskip

\begin{figure}
	\includegraphics[width=\columnwidth]{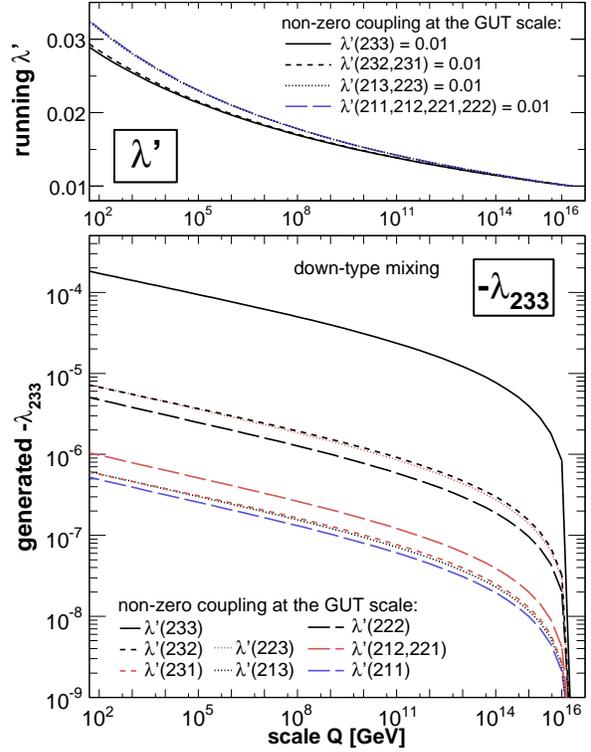}
\caption{\label{fig_runningcouplingsD}%          
  Running of $\rpv$~couplings assuming a single non-zero $\lamp=0.01$
  coupling at the GUT scale (upper panel) leading to a non-zero
  $\lam_{233}$~coupling (lower panel) at lower scales within the
  $\rpv$ mSUGRA scenario Set A for down-type mixing.}
\end{figure}

\begin{figure}
	\includegraphics[width=\columnwidth]{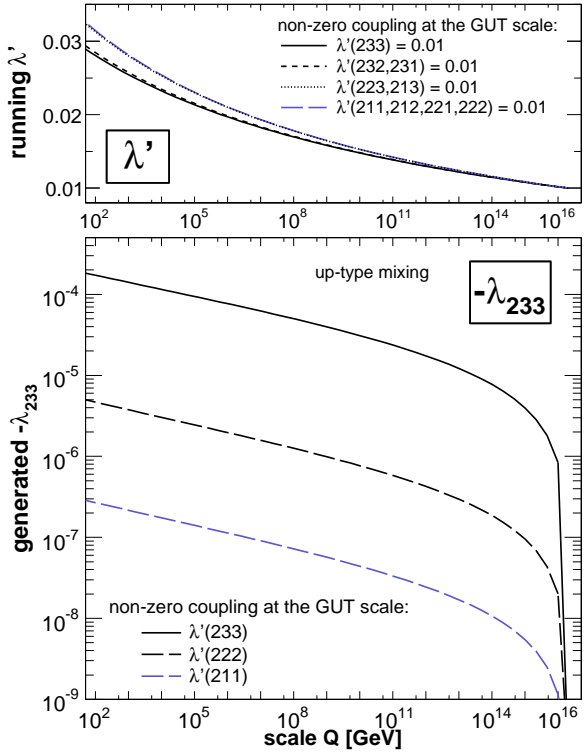}
  \caption{\label{fig_runningcouplingsU}%
Same as Fig.~\ref{fig_runningcouplingsD}, but for quark mixing in the up-sector.}
\end{figure}

In Figs.~\ref{fig_runningcouplingsD}, \ref{fig_runningcouplingsU}, we
show the running of different $\lam'_{2jk}$~couplings, starting with
$\lamp_{ijk} = 0.01|_{\mathrm{GUT}}$, for the case of down- and
up-mixing respectively. In the corresponding lower panel, we show the
scale dependence of the generated $\lam_{323} = -\lam _{233}$~coupling.
Here, we use the mSUGRA parameters of Set A ($\tan \beta=13$).

\medskip

We see that the dominant $\lamp_{ijk}$~coupling grows by about a
factor of 3, running from the GUT scale to the weak scale. This effect
is mainly due to the gauge couplings, see Ref.~\cite{deCarlos:1996du},
where the Yukawa couplings were omitted.  Including the Yukawa
couplings reduces this effect, maximally for $j=k=3$. The generated
$\lam_{233}$~coupling is at least two orders of magnitude smaller than
the original $\lamp$~coupling. Furthermore it depends sensitively on
the flavor structure $(ijk)$ of the original $\lamp$~coupling. This
reflects the dependence on the Yukawa matrix $\Yd_{jk}$. In case of
down-type mixing, the ordering of the corresponding entries is
\begin{align}
\begin{split}
	 \Yd_{33} &> \Yd_{23,32} > \Yd_{22} > 
\\ 
&> \Yd_{12,21} > \Yd_{13,31} > \Yd_{11},
\label{eq_yukordering}
\end{split}
\end{align}
reflecting precisely the ordering of the generated couplings in
Fig.~\ref{fig_runningcouplingsD}.  Small differences between the
couplings generated by $\lamp_{i23}$ ($\lamp_{i13}$) or $\lamp_{i32}$
($\lamp_{i31}$) are related to the different running of the respective
$\lamp$ and $\Yd_{jk}$~coupling, depending in turn on whether $j$ or
$k$ equals 3.

\medskip

In the case of up-type mixing, Fig.~\ref{fig_runningcouplingsU}, not
all $\lamp$~couplings can generate a $\lam$. Since the down Yukawa
coupling is diagonal, $j=k$ is required.  Other couplings can generate
$\lam_{i33}$ at higher loop levels only and are not included in our
approximations. 

\medskip

Our results can easily be translated to other scenarios: The running
of the dominant coupling $\lambda'$ is mainly driven by gauge
interactions, Eq.~(\ref{RGElambdaPX}), and thus depends only weakly on
the specific SUSY parameters. The dependence of the generated coupling
$\lam$ on SUSY parameters is more involved  but we expect
$\tan\beta$ to have the largest impact. In general, the generated
$\lam$~coupling scales with $\tan^2 \beta$,
\begin{equation}
	\lam_{i33} \propto \tan^2 \beta \, ,
\label{lam_proptp_tanb2}
\end{equation} 
if $\tan^2 \beta \gg 1$. This is because the down-quark Yukawa
couplings $\Yd_{jk}$ [and the tau Yukawa coupling $\Ye_{33}$] are
proportional to $1/\cos\beta=\sqrt{1+\tan^2 \beta}$, which directly
follows from Eqs.~(\ref{Yukawa_upmix})-(\ref{vevs}). Therefore the
magnitude of the generated $\lambda$~coupling for other scenarios can
be estimated by rescaling $\lambda$ of Fig.~\ref{fig_runningcouplingsD}
and Fig.~\ref{fig_runningcouplingsU} according to 
Eq.~(\ref{lam_proptp_tanb2}).

%%%%%%%%%%%%%%%%%%%%%%%%%%%%%%%%%%%%%
\subsection{Comparison with the Program {\tt Softsusy}}

In this section, we compare our results for $\lambda'_{ijk}$ and the
generated coupling $\lambda_{i33}$ at the SUSY scale,
Eq.~(\ref{SUSY_scale}), with an unpublished version of {\tt Softsusy}
\cite{Markus:2008}. This version of {\tt Softsusy} contains the complete
 one loop RGEs for
$\lambda'_{ijk}$ (\ref{RGElambdap}) and $\lambda_{i33}$
(\ref{RGElambda}), without our approximations.

\medskip

\begin{table*}
\begin{ruledtabular}
\begin{tabular}{c|cc|cc|cc}
\hspace*{3ex}Set A\hspace*{3ex} &\multicolumn{2}{c|}{$\lamp_{ijk}$} & 
\multicolumn{2}{c|}{$\lam_{i33}$ (down-type mixing)} 
& \multicolumn{2}{c}{$\lam_{i33}$ (up-type mixing)}\\
& \eqref{RGElambdaPX} & {\tt Softsusy}\hspace*{2ex}& \eqref{RGElambdaX} 
& {\tt Softsusy}& \eqref{RGElambdaX} & {\tt Softsusy}\\
\hline
$\lambda'_{211}$ & $2.82 \times 10^{-2}$& $2.85 \times 10^{-2}$  & $-3.96 \times 10^{-7}$ & $-3.89 \times 10^{-7}$ &  $-2.17\times 10^{-7}$ & $-2.13\times 10^{-7}$ \\
$\lambda'_{231}$ & $2.58 \times 10^{-2}$ & $2.61 \times 10^{-2}$& $-4.65\times 10^{-7}$  & $-4.80\times 10^{-7}$ &  $0$  & $\,\,+2.06 \times 10^{-12}$ \\
$\lambda'_{223}$ & $2.81 \times 10^{-2}$& $2.83 \times 10^{-2}$  & $-5.55 \times 10^{-6}$ & $-5.73\times 10^{-6}$& $0$   & $-8.45 \times 10^{-9}$  \\
$\lambda'_{233}$ & $2.55 \times 10^{-2}$ & $2.58 \times 10^{-2}$ & $-1.41 \times 10^{-4}$  & $-1.42 \times 10^{-4}$ & $-1.42 \times 10^{-4}$  & $-1.43 \times 10^{-4}$\\
$\lambda'_{311}$ & $2.81 \times 10^{-2}$ & $2.84 \times 10^{-2}$ & $0$ & $0$ & $0$   & $0$  \\
\end{tabular}
\caption{\label{comparison_softsusy} Comparison between our results,
\eqref{RGElambdaPX} and \eqref{RGElambdaX}, and the results 
of an unpublished version of {\tt Softsusy} \cite{Markus:2008} for
$\lambda'_{ijk}$ and the generated coupling $\lambda_{i33}$ at the
SUSY scale, \eqref{SUSY_scale}. We choose different couplings
$\lambda'_{ijk}=0.01$ at the GUT scale as given in the first column of
the table. The running of $\lamp_{ijk}$ is the same for down- and
up-type quark mixing. The generation of $\lam_{i33}$ depends on the
quark mixing assumptions and the values at the SUSY scale are given
separately.  The remaining mSUGRA parameters are these of Set A
(\ref{eq_sets}).}
\end{ruledtabular}
\end{table*}

We show in Table~\ref{comparison_softsusy} our results and the
results of {\tt Softsusy} for the case of down-type mixing and up-type
mixing assuming different couplings $\lambda'_{ijk}=0.01$ at the GUT
scale. For the other parameters, we consider the Set A of
Eq.~(\ref{eq_sets}).

\medskip

At the SUSY scale, the differences between our results and {\tt
Softsusy} for the case of down-type mixing,
% Table~\ref{comparison_softsusy}, 
are less than $2\%$ for all $\lambda'
_{ijk}$~couplings and less than $4\%$ for the $\lambda_{i33}$,
respectively. In case of up-type mixing, we find the same for
the couplings $\lambda'_{ijk}$ with $j=k$. However for $j
\not=k$ and up-type mixing, we observe a discrepancy between our
results and {\tt Softsusy} for the coupling $\lambda_{233}$ generated
by $\lambda'_{223}\neq 0|_{\mathrm{GUT}}$ and $\lambda'_{231}\neq0|
_{\mathrm{GUT}}$, respectively. This behavior can easily be
understood.

\medskip

The off-diagonal Yukawa matrix elements $({\bf Y}_D)_{jk}$ are equal
to zero at the weak  scale for up-type mixing. Running from
the weak scale to the GUT scale generates Yukawa couplings $({\bf
Y}_D)_{jk},\, j\neq k$, at the one loop level
\cite{Martin:1993zk,Allanach:2003eb}. The generation of
$\lambda_{233}$ via \eqref{RGElambdaX} occurs therefore formally at
two-loop level and has been neglected in our approximation. In {\tt
Softsusy} this two-loop effect is taken into account and small
couplings are generated also for $j \not =k$ and up-type mixing.
Compared to the case of down-type mixing, see
Table~\ref{comparison_softsusy}, the $\lambda_{233}$~couplings
are suppressed by five (with $\lambda'_{231}=0.01\lvert_{\rm GUT}$) and
three (with $\lambda'_{223}=0.01\lvert_{\rm GUT}$) orders of
magnitude. Note that the generation of $({\bf Y}_D)_{jk}$ is not the
only two loop effect that enters the full RGEs
\cite{Martin:1993zk,Allanach:2003eb,Allanach:1999mh}. 

\medskip

Therefore, our approximation for the generation of $\lambda_{i33}$ by
a non-zero $\lambda'_{ijk}$ at the GUT scale (\ref{RGElambdaX}) breaks
down in the case of up-type mixing and $j\not=k$.  But concerning $\stau_1$~LSP decays,
the corresponding 2-body decay branching ratio for $\lambda_{i33}$ is
negligible compared to the 4-body decay branching ratio via
$\lambda'_{ijk}$ and our approximations are applicable for such
phenomenological studies. For example, the 2-body decay branching
ratio for up-type mixing and $\lambda'_{231}=0.01|_{\rm GUT}$ or
$\lambda'_{223}=0.01|_{\rm GUT}$  is less than $10^{-4}$ in Set A.

\medskip

We conclude that our approximations are valid for the signal and decay
rates that we study in this work.  We also note that we have provided
an independent check of the yet-to-be published version of {\tt
Softsusy} \cite{Markus:2008}. Using a different set of mSUGRA
parameters leads to a similar level of agreement.

%% file: Chapter3-LSPdecays.tex
\section{$\stau_1$ LSP decays in $\rpv$ mSUGRA}
\label{stau_LSP_decays}
\input{axodraw.sty}

\subsection{General LSP Decay Modes}
\label{subsec_LSPdecays}

As we showed in Sect.~\ref{sec_model}, a non-vanishing coupling
$\lamp_{ijk}$ at the GUT scale generates an additional coupling
$\lam_{i33}$ at the weak scale which is roughly at least two orders of
magnitude smaller than $\lamp_{ijk}$, \cf
Figs.~\ref{fig_runningcouplingsD} and \ref{fig_runningcouplingsU}. In
this section, we compare the possible decay modes of the LSP via these
two couplings for different $\rpv$ scenarios.

\medskip

First, let us discuss $\neut{1}$~LSP scenarios.  The leading
order decay modes of the $\neut{1}$~LSP via the dominant
$\lamp_{ijk}$ and the generated $\lam_{i33}$~couplings are all
three body decays,
\begin{eqnarray}
	\tilde{\chi}_1^0 \overset{\lam'_{ijk}}{\longrightarrow}
	\left\{ \begin{array}{l}
		\ell_i^+ \,\overline{u}_j\, d_k    \\[.5ex]
		\ell_i^-\, u_j\, \overline{d}_k   
	\end{array}\right.
,\quad 
	\tilde{\chi}_1^0 \overset{\lam'_{ijk}}{\longrightarrow}
	\left\{ \begin{array}{l}
		\bar{\nu}_i \, \bar d_j \,d_k  \\[.5ex]
		\nu_i \,d_j\,\overline{d}_k    
	\end{array}\right. \!, \quad 
\label{3body_neut_decays_lamp}
\end{eqnarray}
and
\begin{eqnarray}
	\tilde{\chi}_1^0 \overset{\lam_{i33}}{\longrightarrow}
	\left\{ \begin{array}{l}
		\ell_i^+\,\bar{\nu}_{\tau} \,   \tau^-	\\[.5ex]
		\ell_i^- \,\nu_{\tau} \,  \tau^+
	\end{array}
	\right. 
,\quad 
	\tilde{\chi}_1^0 \overset{\lam_{i33}}{\longrightarrow}
	\left\{ \begin{array}{l}
		\bar{\nu}_i\, \tau^+ \, \tau^-	\\[.5ex]
		\nu_i \,\tau^- \, \tau^+		
	\end{array}\right.  \!. \quad 
\label{3body_neut_decays_lam}
\end{eqnarray}
The corresponding partial widths depend quadratically on $\lamp_{ijk}$ 
and $\lam_{i33}$, respectively
\cite{Butterworth:1992tc,Dreiner:1994tj,Baltz:1997gd,Richardson:2000nt}.
Therefore, the $\neut{1}$ decay via $\lam_{i33}$ is heavily suppressed
and a $\neut{1}$~LSP decays predominantly via $\lamp_{ijk}$ into SM
particles.

\medskip

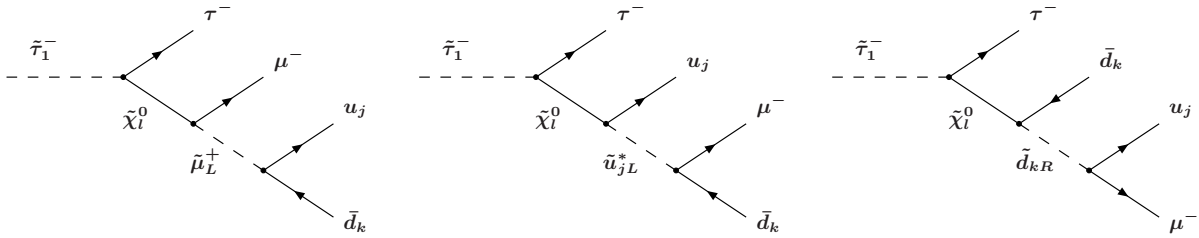
\begin{figure*}
  \begin{center}
    \scalebox{0.88}{
      \begin{picture}(190,140)(5,10)
	\DashLine(20,100)(70,100)5                       
	\ArrowLine(70,100)(100,120)                   
	\ArrowLine(100,80)(130,100)                     	
	\ArrowLine(130,60)(160,80)                      
	\ArrowLine(160,40)(130,60)                      
	\Line(100,80)(70,100)                      		
	\DashLine(100,80)(130,60)5                      
	\GCirc(70,100){1}{0}
	\GCirc(100,80){1}{0}	
	\GCirc(130,60){1}{0}
	\put(30,110){$\boldsymbol{\tilde{\tau}_1^-}$}      
	\put(105,125){$\boldsymbol{\tau^-}$}      
	\put(135,105){$\boldsymbol{\mu^-}$}        
	\put(165,85){$\boldsymbol{u_j}$}           
	\put(165,35){$\boldsymbol{\bar d_k}$}     
	\put(70,80){$\boldsymbol{\tilde{\chi}_l^0}$}   	  
	\put(99,61){$\boldsymbol{\tilde{\mu}_L^+}$}        
    \end{picture}}
\hspace{-0.7cm}
    \scalebox{0.88}{
      \begin{picture}(190,140)(5,10)
	\DashLine(20,100)(70,100)5                        
	\ArrowLine(70,100)(100,120)                     
	\ArrowLine(100,80)(130,100)                     	
	\ArrowLine(130,60)(160,80)                      
	\ArrowLine(160,40)(130,60)                     
	\Line(100,80)(70,100)                      		
	\DashLine(100,80)(130,60)5                      
	\GCirc(70,100){1}{0}
	\GCirc(100,80){1}{0}	
	\GCirc(130,60){1}{0}
	\put(30,110){$\boldsymbol{\tilde{\tau}_1^-}$}    
	\put(105,125){$\boldsymbol{\tau^-}$}      
	\put(135,105){$\boldsymbol{u_j}$}         
	\put(165,85){$\boldsymbol{\mu^-}$}            
	\put(165,35){$\boldsymbol{\bar d_k}$}      
	\put(70,80){$\boldsymbol{\tilde{\chi}_l^0}$}   	  
	\put(99,61){$\boldsymbol{\tilde u^*_{jL}}$}        
    \end{picture}}
\hspace{-0.7cm}
    \scalebox{0.88}{
      \begin{picture}(190,140)(5,10)
	\DashLine(20,100)(70,100)5                       
	\ArrowLine(70,100)(100,120)                    
	\ArrowLine(130,100)(100,80)                   	
	\ArrowLine(130,60)(160,80)                      	
	\ArrowLine(130,60)(160,40)                     	
	\Line(100,80)(70,100)                      		
	\DashLine(100,80)(130,60)5                      
	\GCirc(70,100){1}{0}
	\GCirc(100,80){1}{0}	
	\GCirc(130,60){1}{0}
	\put(30,110){$\boldsymbol{\tilde{\tau}_1^-}$}        
	\put(105,125){$\boldsymbol{\tau^-}$}     
	\put(135,105){$\boldsymbol{\bar d_k}$}         
	\put(165,85){$\boldsymbol{u_j}$}           
	\put(165,35){$\boldsymbol{\mu^-}$}     
	\put(70,80){$\boldsymbol{\tilde{\chi}_l^0}$}   	  
	\put(99,61){$\boldsymbol{\tilde d_{kR}}$}        
    \end{picture}}
  \end{center}
  \vspace{-1.0cm}
\caption{Feynman diagrams contributing to the 4-body decay  
	$\tilde{\tau}_1^-\rightarrow \tau^- \mu^- u_j \bar d_k$
	of the $\stau_1$~LSP  via $\lamp_{2jk}$. 
	In this example the $\tilde\tau_1$ decays via a virtual neutralino
	$\tilde{\chi}_l^0$ ($l=1,2,3,4$) into a tau $\tau^-$, 
	a muon $\mu^-$, an up-type quark $u_j$ of generation $j$ and 
	a down-type anti-quark $\bar d_k$ of generation $k$.
        \label{four_body_decay}}
\end{figure*}

\begin{figure}
  \begin{center}
    \scalebox{0.88}{
      \begin{picture}(135,70)(10,65)
	\DashLine(20,100)(70,100)5                      
	\ArrowLine(70,100)(100,120)                     
	\ArrowLine(70,100)(100,80)                        		
	\GCirc(70,100){1}{0}
	\put(30,110){$\boldsymbol{\tilde{\tau}_1^-}$}        
	\put(105,125){$\mu^-$}            
	\put(105,75){$\nu_\tau$}           
    \end{picture}}
\hspace{-0.7cm}
    \scalebox{0.88}{
      \begin{picture}(135,70)(10,65)
	\DashLine(20,100)(70,100)5                       
	\ArrowLine(70,100)(100,120)                     
	\ArrowLine(70,100)(100,80)                        		
	\GCirc(70,100){1}{0}
	\put(30,110){$\boldsymbol{\tilde{\tau}_1^-}$}        
	\put(105,125){$\tau^-$}             
	\put(105,75){$\nu_\mu$}           
    \end{picture}}
    \scalebox{0.88}{
      \begin{picture}(135,70)(10,65)
	\DashLine(20,100)(70,100)5                     
	\ArrowLine(70,100)(100,120)                   
	\ArrowLine(100,80)(70,100)                      		
	\GCirc(70,100){1}{0}
	\put(30,110){$\boldsymbol{\tilde{\tau}_1^-}$}      
	\put(105,125){$\tau^-$}             
	\put(105,75){$\bar \nu_\mu$}        	  
    \end{picture}}
  \end{center}
\vspace{-0.5cm}
\caption{Feynman diagrams leading to the 2-body decays of the $\stau_1$~LSP 
	via the generated coupling $\lam_{233}$. 
	The $\stau_1$ decays either into a muon $\mu^-$ and 
	a neutrino or into a $\tau^-$ and a neutrino.
\label{two_body_decay}}
\end{figure}
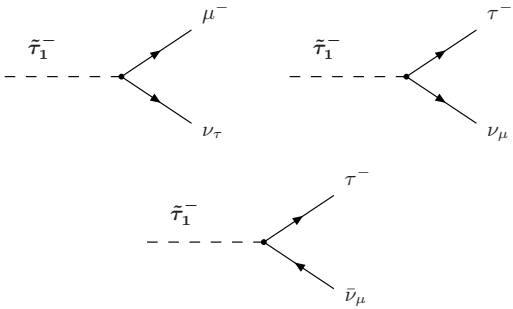

The situation changes if one considers $\rpv$ mSUGRA scenarios 
with a $\tilde{\tau}_1$~LSP, where the
$\stau_1$ couples not directly to the $L_i Q_j \bar D_k$ operator, i.e.
$i=1,2$. In this case, the $\stau_1$ must first couple to a virtual
gaugino.  The gaugino then couples to a virtual sfermion which
then decays via $\lamp_{ijk}$, resulting in a 4-body
decay of the $\stau_1$~LSP.
%, \textit{cf} Fig.~\ref{four_body_decay}.
The possible decay modes via a virtual neutralino are
\begin{align}
\begin{split}
        \tilde{\tau}_1^- \overset{\lam'_{ijk}}{\longrightarrow}
        \left\{ 
                \begin{array}{lcc}
        \tau^- \,\ell_i^+ \,\overline{u}_j\, d_k  & 	\\[.3ex]
        \tau^- \,\ell_i^- \,u_j\, \overline{d}_k & 	\\[.3ex]
        \tau^- \,\bar \nu_i \,\bar d_j \,d_k     & 	\\[.3ex]
        \tau^- \,\nu_i \,d_j \,\overline{d}_k     &
                \end{array}
        \right.\,.
\label{4body_stau_decays}
\end{split}
\end{align} 
4-body decays via a virtual chargino are also possible but they are
suppressed due to the higher chargino mass in comparison to the
lightest neutralino mass, $m(\tilde{\chi}_1^{\pm}) >
m(\tilde{\chi}_{1}^0)$. Furthermore, the (mainly right-handed)
$\stau_1$~LSP couples stronger to the (bino-like) lightest neutralino
than to the (wino-like) lightest chargino.
 
\medskip

On the other hand, the $\stau_1$ can directly decay via $\lam_{i33}$
into only two SM particles
\begin{align}
\begin{split}
        \tilde{\tau}_1^- \overset{\lam_{i33}}{\longrightarrow}
        \left\{ 
                \begin{array}{lcc}
        \tau^-\, \bar \nu_i    & \\[.3ex]
        \tau^- \,\nu_i 		& \\[.3ex]
        \ell^-_i \,\nu_\tau    	& 
                \end{array}
        \right.\,.
\label{2body_stau_decays}
\end{split}
\end{align} 
We show in Fig.~\ref{four_body_decay} (Fig.~\ref{two_body_decay}),
example diagrams for the 4-body (2-body) decay of a $\stau_1$~LSP via
$\lambda'_{2jk}$ ($\lambda_{233}$). Although the 2-body decay suffers
from the small coupling, the 4-body decay is phase space suppressed as
well as by heavy propagators.  Which decay mode dominates depends
strongly on the parameters at the GUT scale. We will discuss in detail
this topic in the next section.

\medskip

As a third type of $\rpv$ mSUGRA scenarios we want to mention 
$\stau_1$~LSP scenarios with a dominant $\lamp_{3jk}$~coupling.
Here, the dominant $\rpv$ operator couples directly to the $\stau_1$~LSP
and allows for a 2-body decay of the $\stau_1$ into two jets,
\begin{equation}
	\stau_1^- \overset{\lam'_{3jk}}{\longrightarrow} \bar u_j d_k \, .
\label{lamp3jk_decayI}
\end{equation}
$\lamp_{3jk}$ can not generate $\lam_{333}$ via the RGEs, because
$\lam_{ijk}$ has to be anti-symmetric in the indices $i,j$.
$\lam_{3nn}$ with $n \not =3$ will be generated by the muon ($n=2$) or
electron ($n=1$) Higgs Yukawa coupling, \cf
Eq.~(\ref{RGElambdaX}). But since these Yukawa couplings are
so small, the decay via $\lam_{3nn}$ is too small to be
seen.  

\medskip

For $j=3$, the up-type quark in \eqref{lamp3jk_decayI} is a
top quark and hence the decay \eqref{lamp3jk_decayI} is
kinematically forbidden for $m_{\tilde{\tau}_1} < m_t$.
The $\stau_1$~LSP than decays in a 3-body decay mode via a virtual top
quark into a $W$~boson and two jets, where at least one jet is a
$b$~jet,
\begin{equation}
\stau_1^- \overset{\lam'_{33k}}{-\!\!\!\longrightarrow} W^- \,\bar b\, d_k \, .
\label{lamp3jk_decayII}
\end{equation}
We present the squared matrix element and the partial width of this
process in Appendix~\ref{app_3body}, which to our knowledge
has not been given in the literature so far.

%%%%%%%%%%%%%%%%%%%%%%%%%%%%%%%%%%%%%%%%%%%%%%%%%%%%%%%%%%%%%
\subsection{Dependence of $\tilde\tau_1$ Decays on mSUGRA Parameters}

In this section, we investigate the conditions at the GUT scale that
lead to 2-body decays of the $\stau_1$~LSP.  We assume a non-vanishing
$\lamp_{2jk}$~coupling at the GUT scale.  This can easily be
generalized to $\lamp_{1jk}$. We point out that the branching ratios
of the $\stau_1$~LSP do not depend on the magnitude of $\lamp_{ijk}$,
since they cancel in the ratio. The following discussion is therefore
also applicable to scenarios where the couplings are too small to
produce a significant number of single slepton events at the LHC but
where the $\stau_1$~LSP is produced in cascade decays of pair produced
SUSY particles.

\medskip

For the numerical implementation we use {\tt Softsusy 2.0.10} 
\cite{Allanach:2001kg} to calculate the mass spectrum at the SUSY
scale, Eq.~(\ref{SUSY_scale}). In addition, we use our own program to
calculate $\lamp_{ijk}$ and $\lam_{i33}$ at the SUSY scale as
described in Sect.~\ref{subsec_numres}.  We than pipe the mass
spectrum and the couplings through {\tt Isawig 1.200}, which is linked
to {\tt Isajet 7.75} \cite{Paige:2003mg}. {\tt Isajet} calculates the
2-body partial width of the SUSY particles and produces an output for
{\tt Herwig} \cite{Corcella:2000bw,Corcella:2002jc,Moretti:2002eu}. We
use a special version of {\tt Herwig 6.510} which also calculates the
4-body decays of the $\stau_1$~LSP \footnote{The version of {\tt Herwig} used in
this paper was written by Peter Richardson and is available on
request.}.  As an output, we consider the total 2-body decay branching
ratio of the $\stau_1$~LSP, $\text{BR}_2$. It is defined as
\begin{equation}
	\text{BR}_2 = \frac{1}{1+\Gamma_4/\Gamma_2},
\label{eq_2body_BR}
\end{equation}
where $\Gamma_2$ and $\Gamma_4$ denote the sums of the partial widths for the 2- and 4-body decays, respectively.

\begin{figure}
	\includegraphics[scale=0.45, bb = 45 45
    510 540, clip=true]{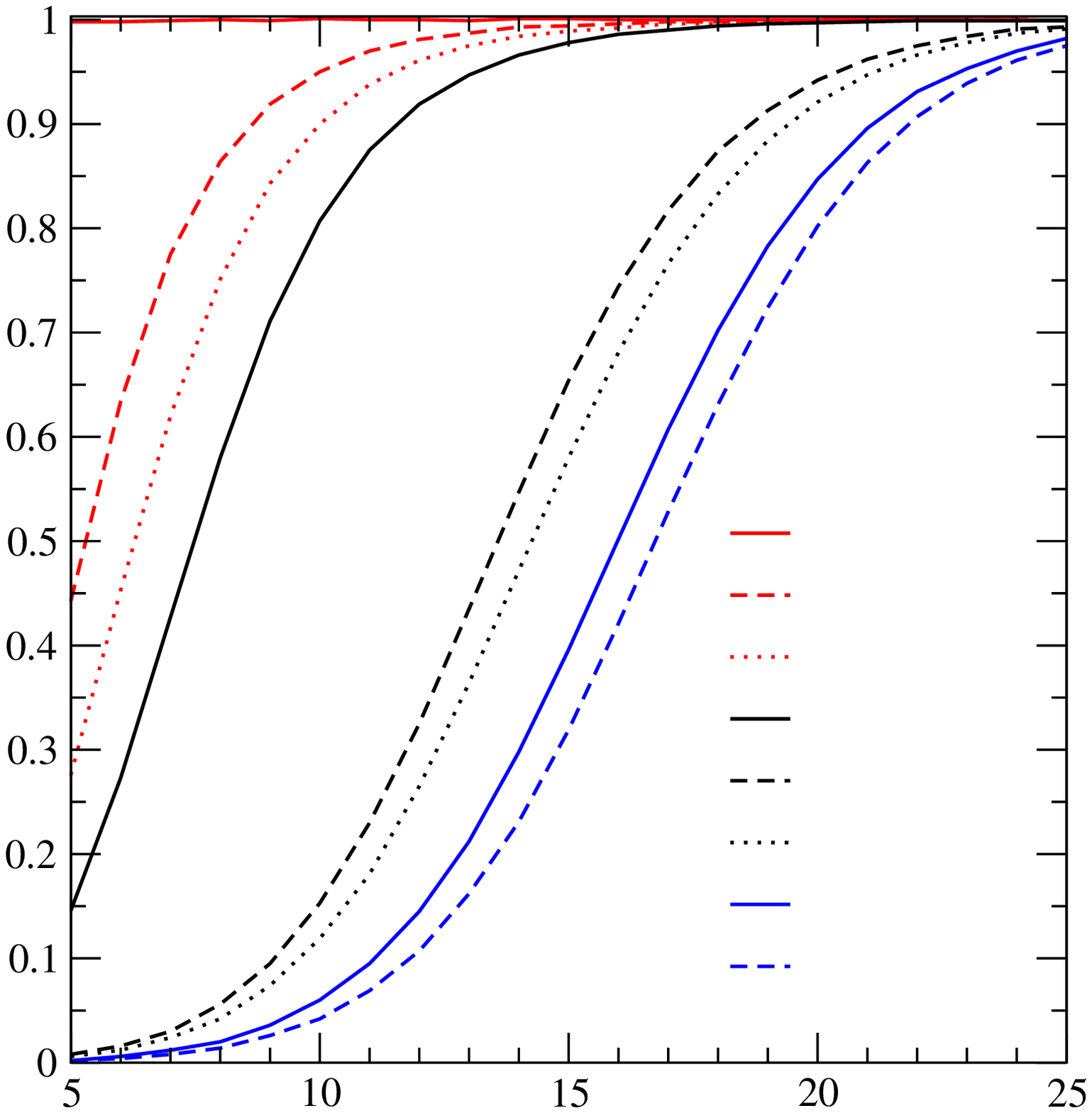}
    \put(-48.0,118.0){$\lam'_{233}$}
    \put(-48.0,106.0){$\lam'_{232}$}
    \put(-48.0,94.0){$\lam'_{223}$}
    \put(-48.0,82.0){$\lam'_{222}$}
    \put(-48.0,70.0){$\lam'_{221/212}$}
    \put(-48.0,58.0){$\lam'_{231}$}
    \put(-48.0,46.0){$\lam'_{213}$}
    \put(-48.0,34.0){$\lam'_{211}$}
    \put(-107.0,3.0){$\tan \beta$}
    \put(-215.0,80.0){\rotatebox{90}
	{BR(2-body decay)}}
 	\caption{2-body decay branching ratio as a function of $\tan \beta$ 
	for different dominating $\lambda'_{2jk}$~couplings at the GUT scale. 
    The quark mixing is in the down sector and the mSUGRA parameters are 
	$M_0=0$~GeV, $M_{1/2}=500$~GeV, $A_0=600$~GeV, $\text{sgn}(\mu)=+1$. 
	\label{tanb_dmixing}}
\end{figure}

\begin{figure}
	\includegraphics[scale=0.45, bb = 45 45
    510 540, clip=true]{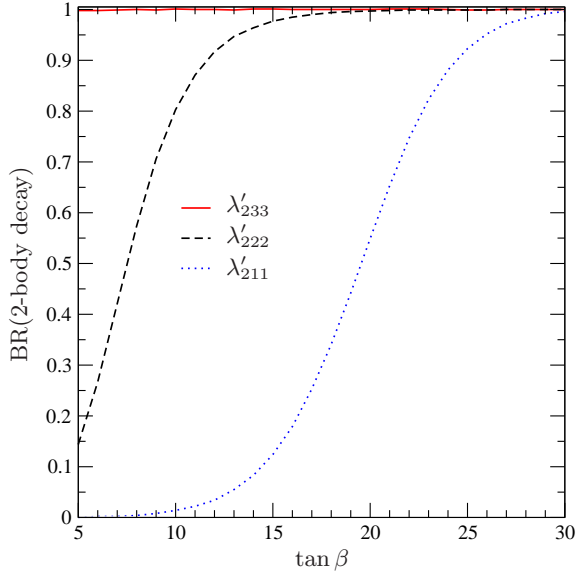}
	\put(-133.0,137.0){$\lam'_{233}$}
	\put(-133.0,125.0){$\lam'_{222}$}
	\put(-133.0,113.0){$\lam'_{211}$}
%	\put(-133.0,101.0){$\lam'_{223}$}
%	\put(-133.0,89.0){$\lam'_{213}$}
%	\put(-133.0,77.0){$\lam'_{232}$}
	\put(-107.0,3.0){$\tan \beta$}
	\put(-215.0,80.0){\rotatebox{90}
	{BR(2-body decay)}}
       \caption{2-body decay branching ratio as a function of $\tan
        \beta$ for different dominating $\lambda'_{2jk}$~couplings at
        the GUT scale.  The quark mixing is in the up sector and the
        mSUGRA parameters are $M_0=0$~GeV, $M_{1/2}=500$~GeV,
        $A_0=600$~GeV, $\text{sgn}(\mu)=+1$.  Couplings
        $\lambda'_{2jk}$ for which the 2-body decay branching ratio
        nearly vanishes are not shown. \label{tanb_upmixing}}
\end{figure}

\medskip
We first show in \figref{tanb_dmixing} (\figref{tanb_upmixing}) 
the $\tan \beta$ dependence of the 2-body decay branching ratio.
We give values for different non-vanishing couplings $\lamp_{2jk}$ 
at the GUT scale and we assume quark mixing in the down (up) sector. 

\medskip

Nearly all $\stau_1$~LSPs will decay via a 2-body decay for large values of
$\tan \beta$, {\it i.e.} $\tan \beta \gsim 30$, and down-type mixing.
In the case of up-type mixing this is also true for $\lamp_{211}$,
$\lamp_{222}$ and $\lamp_{233}$. This behavior can be easily
explained with the help of \eqref{eq_2body_BR}. The partial
widths $\Gamma_2$, $\Gamma_4$ can be approximated by
\cite{Allanach:2003eb}
\begin{eqnarray}
	\Gamma_2 & \propto & \lambda_{233}^2\,
	 m_{\tilde{\tau}_1}\,,\label{Gamma4} 
\\ 
	\Gamma_4 & \propto & \lambda^{\prime2}_{2jk} \,
	 \frac{m_{\tilde{\tau}_1}^7} {m_{\tilde{\chi}}^2
	 m_{\tilde{f}}^4}\,.
\end{eqnarray} 
$m_{\tilde{\chi}}$ denotes the mass of the relevant gaugino
and $m_{\tilde{f}}$ denotes the mass of the virtual sfermion which
couples directly to $L_2 Q_j \bar D_k$, \cf Fig.~\ref{four_body_decay}.

\medskip

As we argued in Sect.~\ref{subsec_numres}, the generated coupling
$\lambda_{233}$ scales roughly with $\tan^2\!\beta$, \cf
\eqref{lam_proptp_tanb2}.  Therefore, $\Gamma_2$ scales with $\tan^4\!
\beta$. At the same time, $\lamp_{211}$ is hardly affected by $\tan
\beta$. This is the main effect that enhances $\text{BR}_2$ for large
$\tan\beta$.

\medskip

Furthermore, increasing $\tan \beta$ increases the contribution from
the tau Yukawa couplings to the various RGEs. This is encoded in the
function $X_\tau$, \eqref{eq_xtaudef} which is proportional to
$(1+\tan ^2\beta)$. As can be seen in \eqref{eq_xtaudef}, increasing
$\tan\beta$ and $X_{\tau}$ reduces the mass of the right- and
left-handed stau and therefore, with \eqref{eq_staumass}, the mass of
the $\stau_1$~LSP, $m_{\tilde{\tau}_1}$. Furthermore, the off-diagonal
matrix elements of the stau mass matrix \eqref{eq_staumassmatrix} also
increase with $\tan\beta$. This leads to a stronger mixing between the
right- and left-handed stau and lowers the mass of the $\stau_1$, 
\cf \eqref{eq_staumass}.

\medskip

Note that $\Gamma_4/\Gamma_2$ is proportional to $m_{\tilde{\tau}_1}
^6$. According to \eqref{eq_2body_BR}, the 2-body decay branching
ratio therefore strongly increases for decreasing
$m_{\stau_1}$.

\medskip

We observe in \figref{tanb_dmixing} also a large hierarchy between the
different couplings $\lamp_{2jk}$. For example, a dominant $\lamp_{233
}$~coupling leads to $\text{BR}_2 \approx 100 \%$ for any value of
$\tan \beta$, whereas for $\lamp_{211}$ this is only the case for
$\tan \beta \gsim 25$.  This hierarchy reflects the hierarchy of the
down quark Yukawa matrix elements, \eqref{eq_yukordering}, which enter
as the dominant term in the RGE of $\lam_{233}$, \eqref{RGElambdaX}.

\medskip

For up-type quark mixing, \figref{tanb_upmixing}, and $j\not=k$ the
down-quark Yukawa matrix elements and therefore $\text{BR}_2$ are
nearly vanishing. 

\begin{figure}
	\includegraphics[scale=0.45, bb = 45 45
    515 540, clip=true]{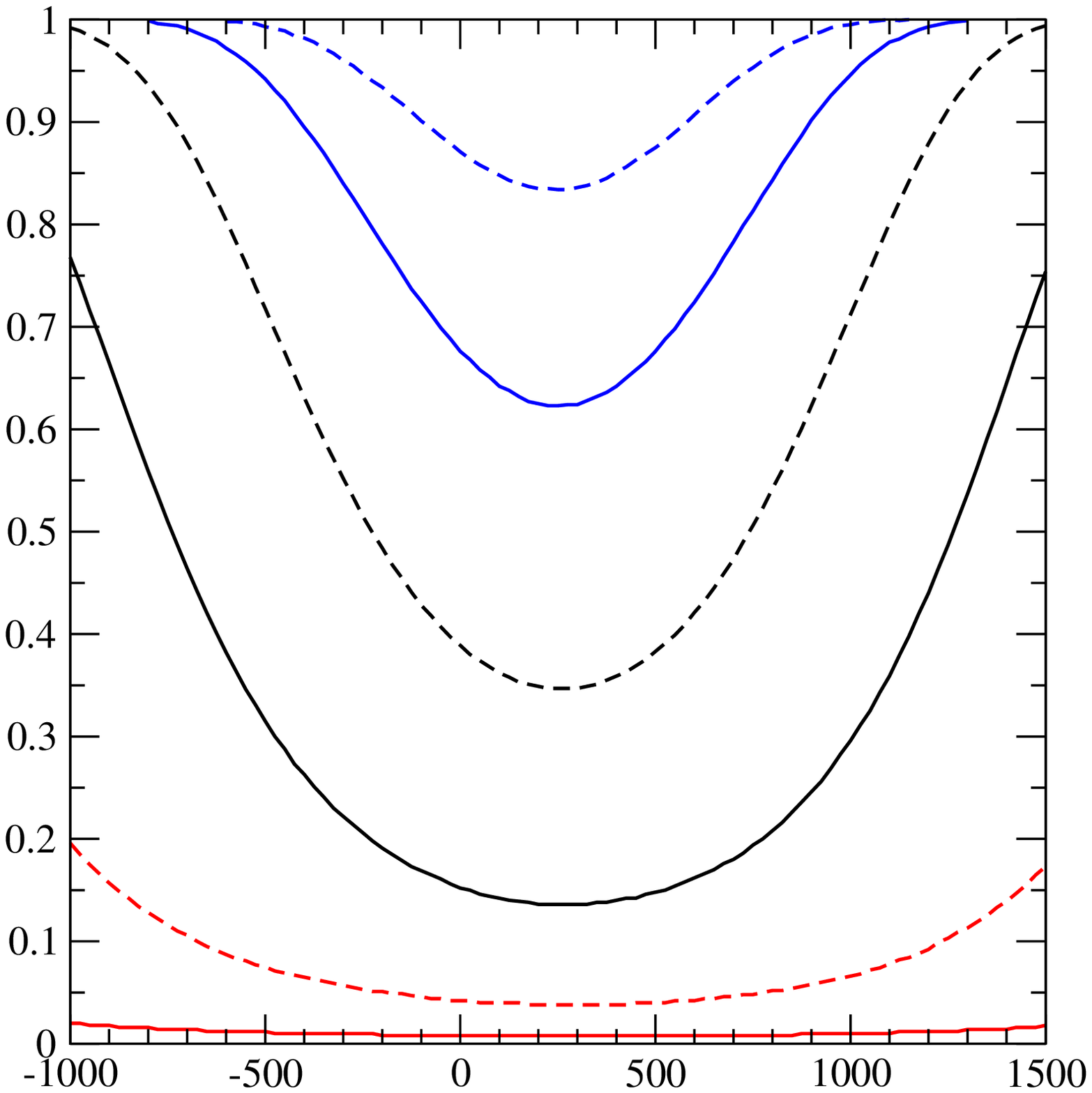}
    \put(-119.0,35.0){$\tan \beta = 10$}
    \put(-119.0,60.0){$\tan \beta = 13$}
    \put(-119.0,109.0){$\tan \beta = 16$}
    \put(-119.0,162.0){$\tan \beta = 19$}
    \put(-119.0,200.0){$\tan \beta = 22$}
    \put(-117.0,3.0){$A_0$ [GeV]}
    \put(-217.0,80.0){\rotatebox{90}
	{BR(2-body decay)}}
      \caption{\label{A0_dependence} 2-body decay branching ratio as a
        function of $A_0$ for non vanishing $\lamp_{211}$ at the GUT
        scale and different $\tan \beta$. We assume down-type quark
        mixing.  The other mSUGRA parameters are $M_0=0$~GeV,
        $M_{1/2}=500$~GeV, $\text{sgn}(\mu)=+1$. The solid red curve
        corresponds to $\tan \beta=7$.}
\end{figure}

\medskip

We investigate the dependence of $\text{BR}_2$ on $A_0$ in
\figref{A0_dependence}, for a dominant coupling $\lamp_{211}$ and
down-type mixing. We see a minimum at $A_0 \approx 250\,$GeV.
Here, $\text{BR}_2$ is reduced by up to 70\% compared to $A_0=\pm 1\,
\text{TeV}$. The minimum and the position of the minimum is dominated
by the following two effects.

\medskip

The right-handed stau couples to a left-handed stau (tau sneutrino)
and a neutral Higgs (charged Higgs) via a trilinear scalar interaction
$(\mathbf{h}_E)_{33}$ \cite{Allanach:2003eb}.  The coupling
$(\mathbf{h}_ E)_{33}$ has dimension one and in mSUGRA models it is
equal to $A_0 \times (\mathbf{Y}_E)_{33}$ at the GUT scale.  The RGE
of the right-handed scalar tau mass, $m_{\tilde{\tau}_R}$, depends in
the following way on $(\mathbf {h}_E)_{33}^2$ \cite{Allanach:2003eb}:
\begin{equation}
	\frac{d m_{\tilde{\tau}_R^2}}{dt} = + 4 (\mathbf{h}_E)_{33}^2 
+ \dots \, .
\end{equation}
This term decreases $m_{\tilde{\tau}_R}$ when we go from the GUT scale
to the SUSY scale (\ref{SUSY_scale}) due to the plus sign. The
(negative) contribution of this term to $m^2_{\tilde{\tau}_R}$ is
proportional to the integral of $(\mathbf{h}_E)_{33}^2$ from
$t_{\rm min}=\ln({M_{\rm GUT}})$ to $t_{\rm max}=\ln({M_Z})$.  For the mSUGRA
parameters given in \figref{A0_dependence}, $M_0=0$~GeV, $M_{1/2}=500$~GeV,
 $\text{sgn}(\mu)=+1$, the integral of $(\mathbf{h}_E)_{33}^2$ is
minimal at $A_0 \approx 180$~GeV and, therefore, $m_{\tilde{\tau}_R}$
is maximal.  For $m_{\tilde{\tau}_1}=m_{\tilde{\tau}_R}$ this also
leads to a maximum of $\Gamma_4/\Gamma_2 \sim m_{\tilde{\tau}_1}^6$
and hence to a minimum of $\text{BR}_2$.

\medskip

But the lightest stau is an admixture of the right- and left-handed
stau.  The off-diagonal mass matrix elements $B_{LR}$,
\eqref{eq_staumassmatrix}, depend also on the value of $(\mathbf{h}_E)
_{33}$ at the SUSY scale, Eq.~(\ref{SUSY_scale}), through $A_
\tau=(\mathbf{h}_E)_{33}/(\mathbf{Y}_E)_{33}$.  For $A_0=180$~GeV we find
$A_\tau \approx  -110$~GeV.  A negative value of $A_\tau$ enhances the effect
of L--R-mixing which decreases $m_{\tilde{\tau}_1}$.  Therefore,
the maximum of $m_{\tilde{\tau}_1}$ as a function of $A_0$ is shifted
to $A_0 \approx 250$~GeV compared to $m_{\tilde{\tau} _R}$. Note
however that the $A_\tau$ dependence of stau L--R-mixing is
sub-dominant around the minimum because of $\mu \tan\beta \gg A_\tau$.

\begin{figure}
	\includegraphics[scale=0.45, bb = 45 45
    510 540, clip=true]{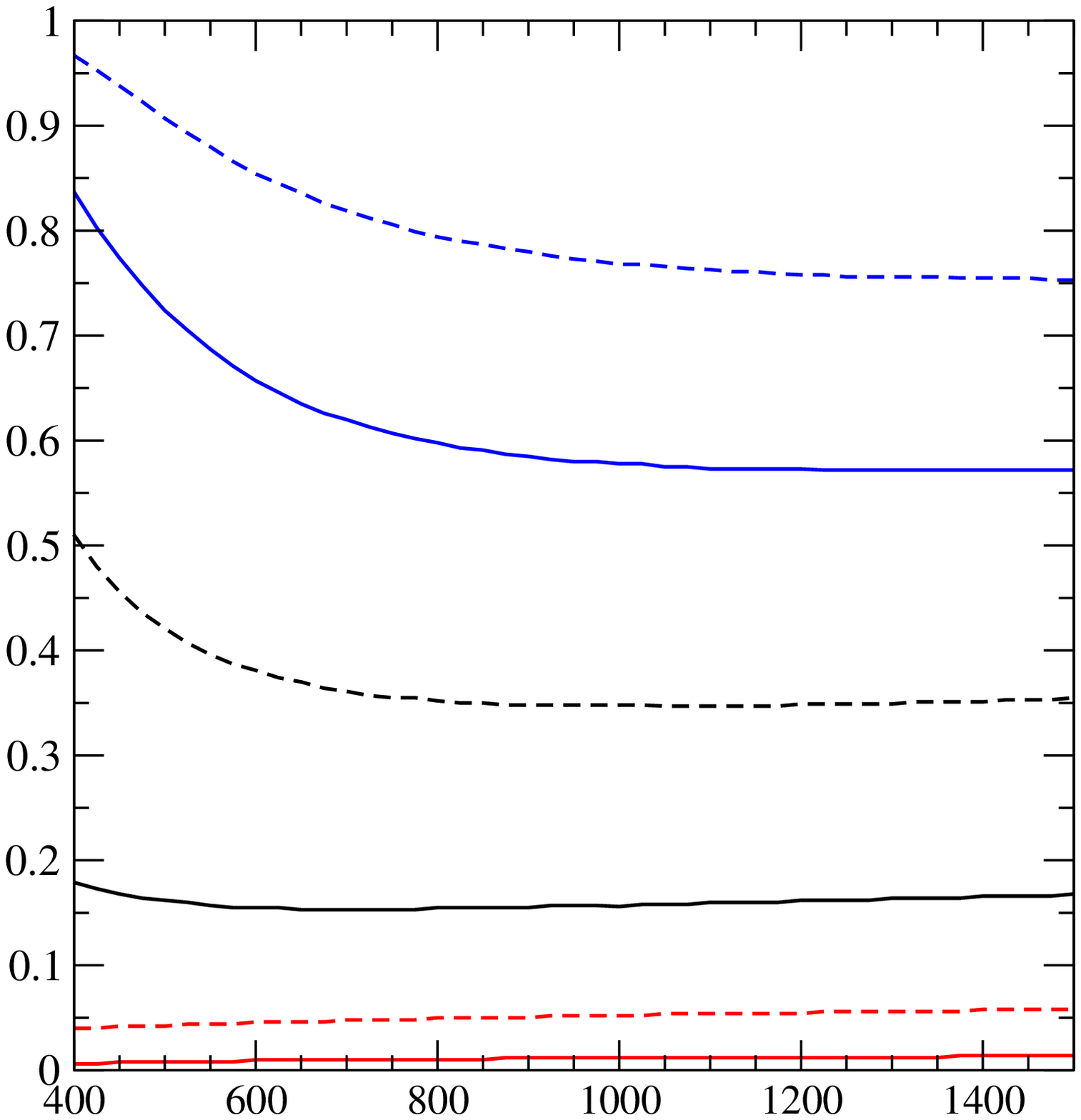}
    \put(-130.0,36.0){$\tan \beta = 10$}
    \put(-130.0,60.0){$\tan \beta = 13$}
    \put(-130.0,98.0){$\tan \beta = 16$}
    \put(-130.0,145.0){$\tan \beta = 19$}
    \put(-130.0,183.0){$\tan \beta = 22$}
    \put(-117.0,3.0){$M_{1/2}$ [GeV]}
    \put(-215.0,80.0){\rotatebox{90}
	{BR(2-body decay)}}
	\caption{\label{M12_dependence}
	2-body decay branching ratio as a function of $M_{1/2}$
	for non vanishing $\lamp_{211}$ at the GUT scale and
	different $\tan \beta$. We assume quark mixing in the down sector. 
	The other mSUGRA parameters are $M_0=0$~GeV, $A_{0}=600$~GeV, $\text{sgn}(\mu)=+1$. 
	The solid red curve corresponds to $\tan \beta=7$.}
\end{figure}

\medskip

Next, we study the dependence of $\text{BR}_2$ on the universal
gaugino mass $M_{1/2}$.  We show this behavior in
Fig.~\ref{M12_dependence}, again for a dominant $\lamp_{211}$ and
down-type mixing. The 2-body decay branching ratios approach a
constant value for increasing $M_{1/2}$. Both, the squared mass of the
gauginos, \cf \eqref{neutralino_masses}, and the squared masses
of the sfermions, \cf \eqref{eq_sfermionmasses}, depend linearly
on $M_{1/2}^2$.  Therefore,
\begin{equation}
	\lim_{M_{1/2}\rightarrow \infty} \Gamma_4/\Gamma_2 \propto 
	\frac{m_{\tilde{\tau}_1}^6}{m_{\tilde{\chi}}^2 m_{\tilde{f}}^4} 
	= \text{constant} \, .
\end{equation}  
 
The dependence of $\text{BR}_2$ on $M_{1/2}$ for $M_{1/2} \lsim 1$ TeV
is more involved, because the ratio $\Gamma_4/\Gamma_2$ depends also
on the other mSUGRA parameters, mainly through the running sfermion
masses, \cf \eqref{eq_sfermionmasses}. For example, we observe in
Fig.~\ref{M12_dependence} that the slope of $\text{BR}_2$ for $M_{1/2}
\lsim 1\,$TeV strongly depends on $\tan \beta$. For $\tan \beta=10$, 
the slope is small and positive whereas for $\tan\beta\gsim 13$ the 
slope is negative. The magnitude of the slope also increases when we
consider larger values of $\tan \beta$. This behavior is again related
to the tau Yukawa coupling $(\mathbf{Y}_E)_{33}$ and its effects on the
$\stau_1$ mass described by the function $X_\tau $, \eqref{eq_xtaudef}.
For large values of $M_{1/2}$, the influence of $X_\tau$ on the $\stau_1$
mass nearly vanishes. But as we go to smaller values of $M_{1/2}$ the
(negative) contributions due to $(\mathbf{Y}_E)_{33}$ become more and
more important. For example, for $\tan \beta=22$ and $M_{1/2}=1$ TeV
($M_{1/2}=400$~GeV) the $X_\tau$ term reduces the mass of the
right-handed stau by $3\%$ ($10\%$) compared to vanishing $(\mathbf{Y}
_E)_{33}$. This reduction of $m_{\tilde{\tau}_ 1}$ will also reduce 
$\Gamma_4/\Gamma_2$ resulting in an increase of $\text{BR}_2$. This 
effect is more pronounced for large $\tan\beta$ because $X_\tau$ is
proportional to $(1+\tan^2\beta)$. If we neglect the effect of $(
\mathbf{Y}_E)_{33}$, the $\text{BR}_2$ curves in
Fig.~\ref{M12_dependence} all get a small positive slope.

\begin{figure}
	\includegraphics[scale=0.45, bb = 45 45
    515 540, clip=true]{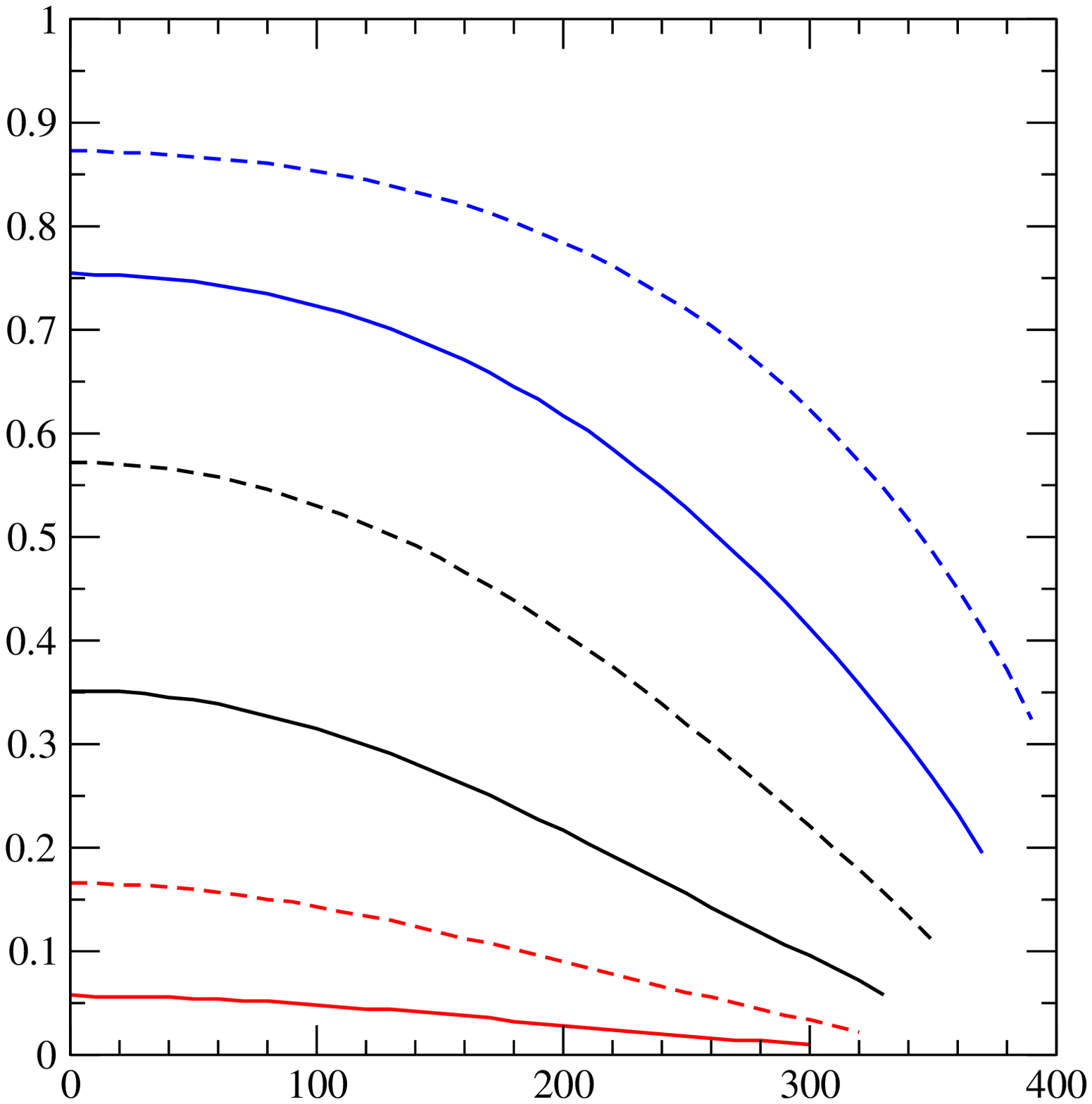}
    \put(-175.0,37.0){$\tan \beta = 10$}
    \put(-175.0,58.0){$\tan \beta = 13$}
    \put(-175.0,93.0){$\tan \beta = 16$}
    \put(-175.0,136.0){$\tan \beta = 19$}
    \put(-175.0,171.0){$\tan \beta = 22$}
    \put(-175.0,195.0){$\tan \beta = 25$}
    \put(-117.0,3.0){$M_{0}$ [GeV]}
    \put(-215.0,80.0){\rotatebox{90}
	{BR(2-body decay)}}
	\caption{\label{M0_dependence} 2-body decay branching ratio as a function of $M_{0}$
	for non vanishing $\lamp_{211}$ at the GUT scale and
	different $\tan \beta$. We assume quark mixing in the down sector. The other mSUGRA
	parameters are $M_{1/2}=1400$~GeV, $A_{0}=600$~GeV, $\text{sgn}(\mu)=+1$.}
\end{figure}

\medskip

Finally, we show in Fig.~\ref{M0_dependence} the dependence of
$\text{BR}_2$ on the universal softbreaking scalar mass $M_0$.  Here,
we have chosen a rather large value of $M_{1/2}$, $M_{1/2}= 1400$~GeV,
because otherwise a $\stau_1$~LSP would exist only in a small interval of
$M_0$. 

\medskip

The behavior of $\text{BR}_2$ can easily be understood.  Increasing
$M_0$ increases the mass of the sfermions, \eqref{eq_sfermionmasses},
but not the mass of the gauginos.  Therefore, the nominator of
$\Gamma_4/\Gamma_2\propto m_{\tilde{\tau}_1}^6/(m_{\tilde{\chi}}^2
m_{\tilde{f}}^4)$ is a polynomial of order ${\cal{O}}(M_0^6)$, whereas
the denominator is only a polynomial of order ${\cal{O}}(M_0^4)$.
Therefore, the 2-body decay branching ratios fall off for increasing
$M_0$ as shown in Fig.~\ref{M0_dependence}. The lines in the
  Figure terminate at values of $M_0$ above which the $\stau_1$ is no
  longer the~LSP.

%% file: Chapter4-SingleSlepton.tex
\section{Resonant single slepton production in $\stau_1$~LSP scenarios}
\label{single_slep_signatures}

\begin{figure}
	\scalebox{.75}{\epsffile{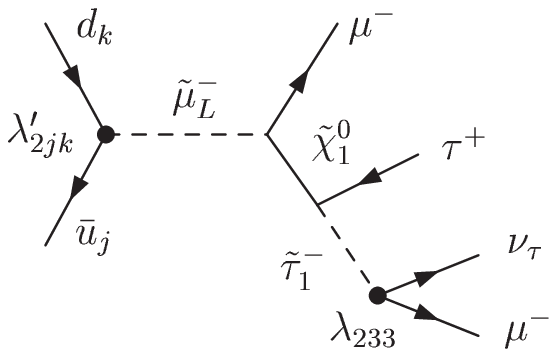}}
	\caption{\label{fig_2bodysleptondecay} Example Feynman graph
	for single slepton production in $\stau_1$~LSP scenarios where
	the slepton decay proceeds via the generated $\lam_{233}$
	coupling (2-body decay mode).}
\end{figure}

\begin{figure}
 	\scalebox{.75}{\epsffile{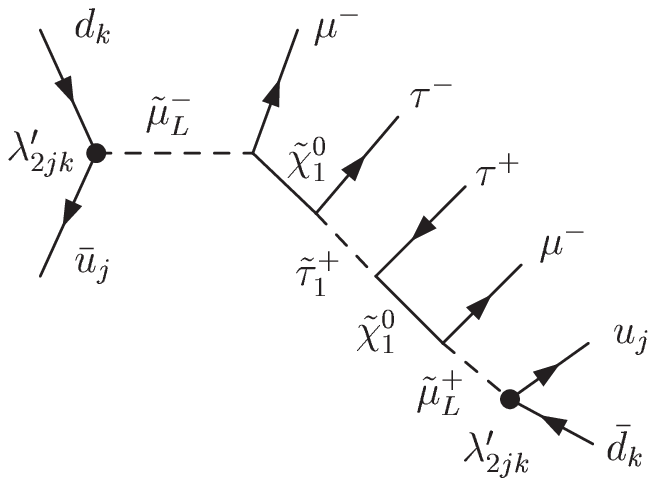}}
 	\caption{\label{fig_4bodysleptondecay} Example Feynman
 	graph for single slepton production in $\stau_1$~LSP
 	scenarios where the slepton decay proceeds via the dominant
 	$\lamp_{2jk}$~coupling (4-body decay mode).}
\end{figure}

We now apply the previous discussion to resonant single slepton
production in $\rpv$~mSUGRA scenarios with a $\stau_1$~LSP. Charged sleptons $
\tilde{\ell}_{Li}$ and sneutrinos $\tilde{\nu}_i$ can be produced 
singly on resonance at the LHC via $q_k\,\bar{q}_j$~annihilation
processes. The production cross section is proportional to $|\lamp_
{ijk}|^2$ and therefore large slepton production rates are
expected in scenarios with a dominant $\lamp_{ijk}$~coupling. The
RGE generation of $\lam_{i33}$ is important for the subsequent
slepton decay in $\stau_1$~LSP scenarios. As discussed in the previous
section, a non-vanishing $\lam_{i33}$ introduces new 
2-body decay channels for the $\stau_1$~LSP. The interplay of these
2-body decays and the 4-body decays via $\lamp_{ijk}$ determines the
final state signatures. In Figs.~\ref{fig_2bodysleptondecay} and
\ref{fig_4bodysleptondecay}, example Feynman graphs for single slepton
production and the subsequent decay in $\stau_1$~LSP scenarios
are shown.

\medskip

It is the aim of this section to first give a general overview
of the possible final states for these reactions and second to
discuss the special cases $\lamp_{2jk} \neq 0|_{\rm GUT}$ and $\lamp_{3jk}
\neq 0|_{\rm GUT}$ in more detail (Sects.~\ref{subsec_lamp2jk} and
\ref{subsec_lamp3jk}).

\subsection{General Signatures}

In the last section, the ratio of 2- to 4-body $\stau_1$~LSP decay
rates and its dependence on various SUSY parameters has been
studied. Now, we focus on single slepton production in $\stau_1$~LSP
scenarios and are interested in the general decay patterns,
independent of the precise SUSY parameters. We first give an
overview over all possible final states and signatures which could be
used as the starting point for an experimental analysis.

\medskip

A (left-handed) charged slepton or sneutrino can be
produced directly via $\lamp_{ijk}$ and has several decay modes:
\begin{align}
\begin{split}
        {\bar u}_j\, d_k \ra{\tilde\ell}_{Li}^-&\ra
        \left\{ 
                \begin{array}{ccc}
        {\bar u}_j d_k,          & \\[.3ex]
        \ell^-_i \, \tilde\chi^0_m, & \\[.3ex]
        \nu_i \, \tilde\chi^-_n,    & 
                \end{array}
        \right.\,,
\label{eq_slepdecays}
\end{split}
\\
\begin{split}
        {\bar d}_j \, d_k\ra{\tilde\nu}_i&\ra
        \left\{ 
                \begin{array}{cc}
        {\bar d}_j \, d_k,          & \\[.3ex]
        \nu_i\,\tilde\chi^0_m,    & \\[.3ex]
        \ell^-_i\,\tilde\chi^+_n, &
                \end{array}
        \right.\,.
\label{eq_sneutdecays}
\end{split}
\end{align}
Both can decay via the $\rpv$~coupling, which is the inverse
production process. It is however suppressed by $\lvert\lamp_{ijk}\rvert^
2$. If $\lamp_{ijk}\leq \mathcal{O}(10^{-2})$, it contributes
typically at the percent level. The dominant decay channels are 
2-body decays into a lepton-gaugino pair. Further 3- and
more-body decays are expected to be negligible, due to phase space
suppression.

\medskip

In case of $j=3$, the hadronic production of a charged slepton cannot
proceed via two quarks as given in \eqref{eq_slepdecays}, due to the
vanishing top-quark parton density inside a proton. Instead, the
slepton can for example be produced via a $g \bar{d}_k$ initiated
Compton process in association with a single top quark.  Furthermore,
the decay into $t\bar{d}_k$ may be kinematically forbidden. In this
case, the slepton decays via a virtual top. The corresponding decay
width is given in Appendix \ref{app_3body}. Sneutrino production for $j=3$
is possible, \eqref{eq_sneutdecays}, but due to the low bottom-quark
density small cross sections are expected.  We do not consider $j=3$
any further here and refer the reader to
\cite{Bernhardt:2008mz,Borzumati:1999th,Accomando:2006ga,Belyaev:2004qp}
for a detailed investigation of this topic. 

\medskip

For the following discussion, we assume that the produced slepton
predominantly decays into a lepton and the lightest neutralino. This
assumption is motivated by the fact that we consider $\stau_1$~LSP
scenarios. In these scenarios, sleptons are light compared to gauginos
and decays into heavier neutralinos or charginos will be kinematically
excluded or strongly suppressed. See also the computed
branching ratios in explicit SUSY models in \cite{Allanach:2006st}.

\medskip
 
The produced $\neut{1}$ is not the lightest SUSY particle and will
decay further into the $\stau_1$~LSP,
\begin{align}
	\neut{1} \ra  \tau^{\mp}\, \stau_1^{\pm}.
\label{eq_neutstau}
\end{align}
Since the neutralino is a Majorana fermion, both charge
conjugated decays are possible. In most $\stau_1$~LSP scenarios this is
the only possible decay mode of the neutralino.  However, in some
scenarios, the right-handed sleptons $\tilde{\mu}_R$ and $\tilde{e}_R$
are lighter than the $\neut{1}$ and the additional channels
$\neut{1}\rightarrow \tilde{\ell}_R^{\pm}\ell^{\mp}$ are open (for
$\ell = \mu, e$). The $\tilde{\ell}_R$ subsequently decays into the
$\stau_1$~LSP, a $\tau$, and a lepton via a virtual neutralino
\begin{align}
\begin{split}
	\neut{1} \ra  \ell^{\mp} \, \tilde{\ell}_R^{\pm}, \quad
	\tilde{\ell}_R^{\pm} \ra 
	\left\{ 
                \begin{array}{cc} 
	\ell^{\pm}\,  \tau^{\mp} \,\stau_1^{\pm}& \\[.3ex]
	\ell^{\pm} \,  \tau^{\pm} \, \stau_1^{\mp}&	
                \end{array}
        \right.\,.
\end{split}
\label{eq_neutlept}
\end{align}
These decay chains have smaller BRs than the decays in
\eqref{eq_neutstau}. However, they lead to an additional 
lepton pair in the final state and could be, therefore, of special
interest for experimental analyses.

%%%%%%%%%%%%%%%%%%%%%%%%%%%%%%%%%%%%%%%%%%%%%%%%5

\subsection{$\lamp_{2jk} \neq 0 |_{\rm GUT}$, $\lam_{233} \ll \lamp_{2jk}$}
\label{subsec_lamp2jk}

Let us now study more detailed the final state signatures in a
scenario with $\lamp_{2jk} \neq 0 |_{\rm GUT}$ and a generated $\lam_{233}
$~coupling which is small but non-zero at lower scales. In these
scenarios, resonant single $\tilde{\mu}_L$~production and resonant
single $\tilde{\nu}_{\mu}$~production at hadron colliders is possible,
\begin{align}
\begin{split}
   \bar{u}_j\, d_k   \ra \tilde{\mu}^-_L &\ra  \bar{u}_j\, d_k /  \mu^- \, \neut{1},
\\[3mm]
   \bar{d}_j\, d_k   \ra \tilde{\nu}_{\mu} &\ra \bar{d}_j\, d_k /  \nu_{\mu}\, 
\neut{1}.
\end{split}
\end{align}
As explained above, a small fraction of the sleptons decay via
the inverse production process. Predominantly they decay into a lepton
and the lightest neutralino, $\neut{1}$. The decays involving heavier
neutralinos or charginos are typically not accessible.

\medskip

\begin{table}
\begin{ruledtabular}
\begin{tabular}{c|c|c}
% \rule[-3mm]{0mm}{8mm}
&\multicolumn{2}{c}{$\bar{u}_j \,d_k \rRPV \tilde{\mu}_L^- \longrightarrow 
\bar{u}_j\, d_k / \mu^- \neut{1}$}
\\
&\multicolumn{2}{c}{ or }
\\
&\multicolumn{2}{c}{$\bar{d}_j \,d_k \rRPV \tilde{\nu}_{\mu} \longrightarrow
\bar{d}_j\, d_k / \nu_{\mu}  \neut{1}$}
\\[1ex]
\cline{2-3}\rule[-2mm]{0mm}{6mm}
&\hspace*{5.5ex}$\neut{1} \rightarrow  \tau^+\, \stau_1^-$ \hspace*{5.5ex}&
\hspace*{5.5ex} $\neut{1} \rightarrow \tau^-\, \stau_1^+$ \hspace*{5.5ex}
\\
&
$\Big[ \neut{1} \rightarrow  \tau^+\, \stau_1^- \, \ell^+ \ell^- \Big]$ &
 $\Big[ \neut{1} \rightarrow \tau^-\, \stau_1^+\, \ell^- \ell^+ \Big]$
\\[1ex]
\hline
\hspace*{2ex}$\lamp_{2jk}$\hspace*{2ex}
&$\stau_1^- \rightarrow \tau^- \mu^- \,u_j \, \bar{d}_k$ &
$\stau_1^+ \rightarrow \tau^+ \mu^+ \,\bar{u}_j \,d_k $
 \\
&$\stau_1^- \rightarrow \tau^- \mu^+ \bar{u}_j d_k $ &
$\stau_1^+ \rightarrow  \tau^+ \mu^- \,u_j \,\bar{d}_k$
\\
&$\stau_1^- \rightarrow \tau^- \nu_{\mu} \,d_j \,\bar{d}_k$ &
$\stau_1^+ \rightarrow  \tau^+ \bar{\nu}_{\mu} \,\bar{d}_j \,d_k $
\\
&$\stau_1^- \rightarrow \tau^- \bar{\nu}_{\mu} \,\bar{d}_j \,d_k$ &
$\stau_1^+ \rightarrow \tau^+ \nu_{\mu} \,d_j \,\bar{d}_k$
% 
%  DECAYS VIA A VIRTUAL CHARGINO
% 
% \\[1ex]
% &$\stau_1^- \rightarrow  \nu_{\tau}\, \mu^+ d_j\, \bar{d}_j $&
% $\stau_1^+ \rightarrow  \bar{\nu}_{\tau}\, \mu^+ d_j\, \bar{d}_j $
% \\
% &$\stau_1^- \rightarrow  \nu_{\tau} \,\mu^- \bar{d}_j \,d_j $&
% $\stau_1^+ \rightarrow  \bar{\nu}_{\tau} \,\mu^- \bar{d}_j \,d_j $
% \\
% &$\stau_1^- \rightarrow \nu_{\tau} \,\nu_{\mu} \,u_j \,\bar{d}_j $&
% $\stau_1^+ \rightarrow \bar{\nu}_{\tau} \,\nu_{\mu} \,u_j \,\bar{d}_j $
% \\
% &$\stau_1^- \rightarrow \nu_{\tau} \,\bar{\nu}_{\mu} \,\bar{u}_j\, d_j $&
% $\stau_1^+ \rightarrow \bar{\nu}_{\tau} \,\bar{\nu}_{\mu} \,\bar{u}_j\, d_j $
\\[.5ex]  
\hline
$\lam_{233}$&$\stau_1^- \rightarrow  \tau^- \nu_{\mu}$&
$\stau_1^+ \rightarrow \tau^+ \bar{\nu}_{\mu}$
\\
&$\stau_1^- \rightarrow \tau^- \bar{\nu}_{\mu}$&
$\stau_1^+ \rightarrow \tau^+ \nu_{\mu}$
\\
&$\stau_1^- \rightarrow \mu^- \nu_{\tau}$&
$\stau_1^+ \rightarrow \mu^+ \bar{\nu}_{\tau}$
\end{tabular}
\end{ruledtabular}
 \caption{\label{tab_singleslep-zerfaelle} Slepton decay chains with
 all possible final states for single $\tilde{\mu}_L^-$ and single
 $\tilde{\nu}_{\mu}$~production via $\lamp_{2jk}$, respectively.  The
 charge conjugated processes are not shown explicitly.  Slepton
 decays into heavier neutralinos or charginos are neglected.  The
 $\neut{1}$~decays predominantly into a $\stau_1$~LSP and a $\tau$. In
 some scenarios, decays as in \eqref{eq_neutlept} are 
 possible, they are cited in brackets.  Owing to the Majorana
 type nature of the neutralino two charge conjugated decays of the
 neutralino are possible (second and third column). In the first
 column the $\rpv$~coupling involved in the subsequent 4- or 2-body
 $\stau_1$ decays are given.}
\end{table}

The difference between $\tilde{\mu}_L$ and $\tilde{\nu}_{\mu}$
production concerns the flavor of the initial quarks involved (which
is related to different parton density functions and is thus important
for the hadronic cross sections), and the nature of the lepton
resulting from the slepton decay. In both processes a neutralino is
produced in the predominant decay, which in turn
decays into the $\stau_1$~LSP, as given in \eqref{eq_neutstau} and 
\eqref{eq_neutlept}. Finally, the $\stau_1$ decays either via the 
dominant $\lamp_{2jk}$~coupling (4-body decay) or via the generated
$\lam_{233}$~coupling (2-body decay). For the 4-body decays, only the
decays via virtual neutralinos have to be considered. Decay modes via
virtual charginos are suppressed due to the larger mass and
their weaker couplings to the predominantly right-handed
$\stau_1$~LSP.  The complete cascade decay chains are listed in
Tab.~\ref{tab_singleslep-zerfaelle}.

\medskip

A classification of all possible final state signatures is
given in Tab.~\ref{tab_summaryfinalstates2}, for $\tilde{\mu}_L$ and
for $\tilde{\nu}_{\mu}$~production.  For completeness, we include here
the direct B$_3$~decays via $\lamp_{2jk}$, which usually
contribute at the percent level for couplings at the order of
$\mathcal{O}(10^{-2})$.  Neutrinos do not give a signal in a detector
and are denoted as missing transverse energy, $\met$. Final state
quarks are treated as indistinguishable jets, $j$.

\medskip

The 4-body decays via $\lamp_{2jk}$ and the 2-body decays via
the inverse production process lead to two jets in the final state.
In contrast, the 2-body decays via $\lam_{233}$ are purely leptonic.
Many cascade decay chains provide missing transverse energy. 
Furthermore, since we are considering $\stau_1$~LSP scenarios, 
there is always at least one $\tau$ among the final state particles.
The experimentally most promising signatures are most likely
those involving a large number of muons, for example
like-sign dimuons and three or four final state muons.
If the $\neut{1}$ decays only into
$\stau_1\tau$, there are two signatures including like-sign dimuons for
$\tilde{\mu}_L$~production. For $\tilde{\nu}_{\mu}$~production, muons
can be produced singly only. But if the decays \eqref{eq_neutlept} are
open, both slepton production processes allow for dimuon and
trimuon production. In case of $\tilde{\mu}_L$~production, even four
final state muons are possible. Additionally,
depending on how easily taus will be identified, an analysis could be
based on like-sign $\mu\tau$-pairs.

\medskip

The final state signatures depend sensitively on which particle is the
LSP. Compared to slepton production in the $\neut{1}$~LSP scenarios
\cite{Dimopoulos:1988jw,Dimopoulos:1988fr,Dreiner:2000vf,Dreiner:2000qf,Dreiner:1998gz,Moreau:2000bs,Deliot:2000mf,Abazov:2002es,Abazov:2006ii}, there are three main differences here.  
First, for a $\stau_1$~LSP we have always one or two
taus in the final state, which in $\neut{1}$~LSP scenarios is only
possible for smuon production if heavier neutralinos are involved in
the decay chain. These heavy neutralinos then decay into the lightest
neutralino and possibly taus.
Second, the generation of a $\lam$~coupling can be neglected in
$\neut{1}$~LSP scenarios. As argued above, $\lam$ only allows
for additional 3-body decays which are thus not phase-space enhanced
compared to the 3-body decays via the dominant $\lamp$~coupling. As a
consequence, purely leptonic final state signatures are absent in
$\neut{1}$~LSP scenarios. 
Third, due to the modified spectra in
$\neut{1}$~LSP scenarios, also $\tilde{\nu}_\mu$~production can provide
like-sign dimuon events. In this case, $\tilde{\nu}_{\mu}$ can often
decay into a $\mu$ and a chargino. Like-sign dimuons arise either if
the chargino directly decays via $\lamp$ into a $\mu$ and two quarks,
or if the chargino first decays into the $\neut{1}$~LSP and then the
$\neut{1}$~LSP decays via $\lamp$ into a $\mu$ and two quarks.

\medskip

This discussion can easily be translated to scenarios with
$\lamp_{1jk}\neq 0$ by replacing the muons by electrons (and
{\it vice versa}). Since there is typically no difference in mass between
sleptons of the first and second generation, respectively, the
kinematics are the same. Note however that the bounds on the $\rpv$
couplings are stronger for $\lamp_{1jk}$ than for $\lamp_{2jk}$ for
example due to the non-observation of neutrinoless double beta decays.

\begin{table}
\begin{ruledtabular}
\begin{tabular}{c|lllll} 
&\multicolumn{5}{c}{$\tilde{\mu}^-_L$ production}
\\[1ex]
\hline
\hspace*{3ex}$\lamp_{2jk}$\hspace*{3ex} & $\hspace*{1ex} \tau^+\,  \tau^-$&$ \mu^- \, \mu^{\pm}$&&&$ jj$
\\
 & $\hspace*{1ex} \tau^+\, \tau^-$&$ \mu^- $&&$ \met $&$ jj$\hspace*{1ex}
\\
& $ [\, \tau^+\,  \tau^-$&$ \mu^- \, \mu^-\, \mu^{\pm}\,\mu^+$&&&$ jj \,]$
\\
 & $[\, \tau^+\, \tau^-$&$ \mu^-\,\mu^-\,\mu^+ $&&$ \met $&$ jj \,]$\hspace*{1ex}
\\
& $ [\, \tau^+\,  \tau^-$&$ \mu^- \, \mu^{\pm}$&$e^+e^-$&&$ jj \,]$
\\
 & $[\, \tau^+\, \tau^-$&$ \mu^- $&$e^+\,e^-$&$ \met $&$ jj \,]$\hspace*{1ex}
\\
\hline
\hspace*{3ex}$\lam_{233}$\hspace*{3ex} & $\hspace*{1ex} \tau^{\pm} $&$ \mu^-\, \mu^{\mp} $&&$ \met $ &
\\
& $\hspace*{1ex} \tau^+\,  \tau^-$&$ \mu^- $&&$ \met$ &
\\
 & $[\, \tau^{\pm} $&$ \mu^-\,\mu^-\, \mu^{\mp}\,\mu^+ $&&$ \met $ & $\hspace*{3ex}]$
\\
 & $[\, \tau^+\,  \tau^-$&$ \mu^- \, \mu^- \mu^+$&&$ \met $ & $\hspace*{3ex}]$
\\
 & $[\, \tau^{\pm} $&$ \mu^-\, \mu^{\mp} $&$e^+\,e^-$&$ \met$ & $\hspace*{3ex}]$
\\
 & $[\, \tau^+\,  \tau^-$&$ \mu^- $&$e^+\,e^-$&$ \met $ & $\hspace*{3ex}]$
\\
\hline
inv. prod. & &&& & $jj$
\\
\hline\hline
& \multicolumn{5}{c}{$\tilde{\nu}_{\mu}$ production} 
\\[1ex]
\hline
\hspace*{3ex}$\lamp_{2jk}$\hspace*{3ex} & 
	$\hspace*{1ex} \tau^+ \, \tau^-$&$ \mu^{\pm}$&&$\met $&$jj$
\\
  & $\hspace*{1ex} \tau^+ \, \tau^-$&&&$ \met $&$jj$ 
\\
 & $[\, \tau^+ \, \tau^-$&$\mu^-\, \mu^{\pm}\,\mu^+$&&$\met $&$jj \,]$
\\
  & $[\, \tau^+ \, \tau^-$&$\mu^-\,\mu^+$&&$ \met $&$jj \,]$
\\
& $[\, \tau^+ \, \tau^-$&$ \mu^{\pm}$&$e^+\,e^-$&$\met $&$jj \,]$
\\
  & $[\, \tau^+ \, \tau^-$&&$e^+\,e^-$&$ \met $&$jj \,]$
\\
\hline
\hspace*{3ex}$\lam_{233}$\hspace*{3ex} &  $\hspace*{1ex} \tau^{\pm} $&$ \mu^{\mp}$&&$ \met $ &
\\
 & $\hspace*{1ex} \tau^+\,  \tau^-$&&&$ \met $ &
\\
 &  $[\, \tau^{\pm} $&$ \mu^-\, \mu^{\mp}\,\mu^+$&&$ \met$ & $\hspace*{3ex}]$
\\
& $[\, \tau^+\,  \tau^-$&$\mu^-\,\mu^+$&&$ \met  $ &$\hspace*{3ex}]$
\\
 &  $[\, \tau^{\pm} $&$ \mu^{\mp}$&$e^+\,e^-$&$ \met$ &$\hspace*{3ex}]$
\\
 & $[\, \tau^+\,  \tau^-$&&$e^+\,e^-$&$ \met $ &$\hspace*{3ex}]$
\\
\hline
inv. prod. & &&& & $jj$
\end{tabular}
\end{ruledtabular}
\caption{\label{tab_summaryfinalstates2}%
  Summary of all possible final states for single slepton production via
  $\lamp_{2jk}$. Decays involving the dominant $\lamp_{2jk}$~coupling 
  and involving the generated $\lam_{233}$~coupling are listed
  separately, \cf Tab.~\ref{tab_singleslep-zerfaelle}. If 
  kinematically allowed, the $\neut{1}$ may also decay into a light-flavor
  lepton-slepton pair which gives rise to an additional $\mu^+\mu^-$ or
  $e^+e^-$ pair in the final state. The corresponding signatures are
  given in brackets. The decay via the inverse production process is
  also listed.}
\end{table}

\subsection{$\lamp_{3jk} \neq 0 |_{\rm GUT}$}
\label{subsec_lamp3jk}

Some additional remarks are in order for a dominant $\lamp_{3jk}$ $\rpv$~coupling.
 These couplings allow for resonant single $\tilde{\nu}_{\tau}$ production 
 and, owing to the L-R-mixing in the stau-sector, also both
resonant $\stau_1$ and $\stau_2$ production ($j\not=3$).

\medskip

For $\stau_1$~production, we refer to the discussion of LSP decay
modes in Sect.~\ref{subsec_LSPdecays}.  Here the LSP couples
directly to the $\rpv$~operator and the inverse production
process dominates the decay rate,
\begin{align}
   \bar{u}_j\, d_k   \ra \stau_1^- &\ra  \bar{u}_j\, d_k \,.
\end{align}
This decay is kinematically accessible if $j\neq3$. For $j=3$ 
the stau decays via a virtual top-quark, \textit{cf.}
\eqref{lamp3jk_decayII}, for $m_{\stau_1} < m_t$.
Note that $j=3$ requires associated
production, \textit{e.g.} $g\,d_k \rightarrow \stau \, t$, due to the
absence of top quarks inside the proton
\cite{Bernhardt:2008mz,Borzumati:1999th,Accomando:2006ga,Belyaev:2004qp}.

\medskip

For $\stau_2$ and $\tilde{\nu}_{\tau}$~production, there are
the following 2-body decay modes:
\begin{align}
\begin{split}
   \bar{u}_j\, d_k   \ra \stau_2^- & \ra 
        \left\{ 
                \begin{array}{lll}
        \bar{u}_j \, d_k,           \\
	\tau^-\, \neut{1} 	   \\
	\stau_1^-\, h^0/Z^0   \\
                \end{array}
        \right.\,,
\end{split}
\end{align}
\begin{align}
\begin{split}
   \bar{d}_j\, d_k   \ra \tilde{\nu}_{\tau} & \ra 
        \left\{ 
                \begin{array}{lll}
        \bar{d}_j \, d_k,           \\
	\nu_{\tau} \, \neut{1} \\
	\stau_1^-  \, W 		    \\
                \end{array}
        \right.\,.
\end{split}
\end{align}
The inverse production process 
contributes and leads to a $jj$ final state. The decay into a lepton
and a neutralino often dominates for small $\tan \beta$
($\tan\beta\lsim10$). The neutralino decays further into the $\stau_1$
LSP which directly decays into two quarks: 
\begin{align}
\begin{split}
	\neut{1} \ra \, \tau^\pm \, &\stau_1^\mp, \quad
	\stau_1^- \ra \bar{u}_j \, d_k\,,
\end{split}
\end{align}
where we have included the two charge conjugated decays of the
neutralino. The final states of these decay modes are 
$\tau^-\tau^{\pm} jj$, and there is the possibility of like-sign tau
events. If the $\neut{1}$~decay (\ref{eq_neutlept}) is 
kinematically allowed, we can have an additional pair of 
electrons or muons in the final state.

\medskip

The singly produced slepton can also decay into the $\stau_1$~LSP and
a SM particle, $Z^0$, $h^0$, or $W$, respectively (final states:
$h^0/Z^0/W\, jj$). This decay mode is special
for singly produced sleptons of the third generation because they are
L-R mixed eigenstates. It can be the dominant decay mode of the
$\stau_2$ and $\tilde{\nu}_\tau$, depending on the parameters.

\medskip

The branching ratios for all $\rpv$~conserving $\stau_2$ and $\tilde
{\nu}_{\tau}$ 2-body decay modes are given in
Tab.~\ref{tab_allgBRs} in Appendix \ref{app_3body}, for the SUSY
parameter sets A and B.

%% file: Chapter5-ExampleCalc.tex
\section{Single smuon production: An Explicit Numerical Example}
\label{numerical_example}

In this section, we present explicit calculations of promising signal
rates for resonant slepton production at the LHC in the $\rpv$~mSUGRA
 model with a $\stau_1$~LSP, focussing on parameter sets A and B,
\textit{cf.}~\eqref{eq_sets}.  First, we consider in
Sect.~\ref{sec_example} (exclusive) like-sign dimuon events,
\ie events with exactly two muons of the same charge in the final
state. An analysis of SM and SUSY backgrounds for the like-sign dimuon
signature is given in Sect.~\ref{background}.  Second, in
Sect.~\ref{subsec_34muonen}, we present event rates for single smuon
production leading to three or four muons in the final states, which
are kinematically accessible within sets A and B.

\subsection{Like-Sign Dimuon Events}
\label{sec_example}

Following Refs.~\cite{Dreiner:2000qf,Dreiner:2000vf}, we first
concentrate on events with exclusive like-sign dimuons. 
Here events with more than two muons are rejected. In this sense, in $\stau_1$~LSP
scenarios, only single smuon production leads to exclusive like-sign
dimuon pairs, \cf Tab.~\ref{tab_summaryfinalstates2}. It has been
shown in Refs.~\cite{Dreiner:2000qf,Dreiner:2000vf} that this
selection criterion enhances the signal to background ratio
considerably. In Refs.~\cite{Dreiner:2000qf,Dreiner:2000vf} it was
shown that using a set of cuts, the SM background rate at the LHC,
$\Gamma_B|_{\rm SM}$, can be reduced to
\begin{equation}
	\Gamma_B|_{\rm SM} = 4.9 \pm 1.6 \; \text{events/10 fb}^{-1}.
\label{SM_bg}
\end{equation}
At the same time the cut efficiency, {\it i.e.} the number of signal
events which pass the cuts, lies roughly between $20\%$ and $30\%$.
Note that Refs.~\cite{Dreiner:2000qf,Dreiner:2000vf} assume a
$\neut{1}$~LSP. As we will argue in Sect.~\ref{background}, similar
cuts are also applicable in $\stau_1$~LSP scenarios. For the numbers
presented in this section, however, no cuts are applied and full cross
sections and event rates are given.

\medskip

The total cross section for like-sign dimuon events is given by the
resonant $\tilde{\mu}_L^+$ or $\tilde{\mu}_L^-$ production cross
section multiplied by the respective branching ratios leading to
like-sign dimuon final states. Both decays via the dominant
$\lamp_{2jk}$~coupling and a generated $\lam_{233}$~coupling
contribute. For a negatively charged smuon they are:
\begin{align}
\begin{split}
	\bar{u}_j \,d_k \rRPV \tilde{\mu}_L^- \rightarrow \mu^- 
	 &\neut{1}, 
\\
	& \hr \tau^+\, \stau_1^- 
\\
	& \qquad\quad \hrRPV \tau^-  \mu^- \,u_j \, \bar{d}_k  \,,
\\
	& \qquad\quad \hrRPVlam \nu_{\tau}\,\mu^- \,,
\\
	 & \hr \tau^-\, \stau_1^+ 
\\
	& \qquad\quad \hrRPV \tau^+ \mu^- \,u_j \,\bar{d}_k  \,,
\label{eq_contribdecaychains}
\end{split}
\end{align}
plus the analogous decay chains where the neutralino decays first into
an $\tilde{e}_R^\pm$-$e^\mp$ pair, \cf \eqref{eq_neutlept}.  The
couplings depicted on the arrows indicate the employed $\rpv$
coupling. The decay chain for a positively charged smuon can be
obtained by charge conjugation.  However, one should keep in mind that
the production cross sections for $\tilde{\mu}^+_L$ and $\tilde{\mu}^
-_L$ differ at $pp$ colliders, since charge conjugated quarks (and
corresponding parton densities) are involved.

\medskip

The cross sections for the exclusive like-sign dimuon final states are
presented in Tab.~\ref{tab_numbersA} for Set A and in
Tab.~\ref{tab_numbersB} for Set B. The smuon production cross
sections, $\sigma_{\rm prod.}( \tilde{\mu}_L^{\mp})$ (see also
Tabs.~\ref{xsections_setA} and \ref{xsections_setB}), include NLO QCD
and SUSY-QCD corrections \cite{Dreiner:2006sv}, see
Appendix~\ref{xsections_and_BRs}. For the numerical analysis, we only
consider couplings $\lambda'_{2jk}$ that involve partons of the first
generation leading to large production cross sections at the LHC.

\medskip

As already discussed, the $\stau_1$~LSP can either decay via $\lamp$
(4-body decay) or via $\lam$ (2-body decay). A list of the respective
branching ratios is given in Appendix A, Tabs.~\ref{stau_BRs_setA} and
\ref{stau_BRs_setB}, for sets A and B and for several $\lamp_{2jk}$ 
couplings.  Here we show the resulting cross section times branching
ratio, $\sigma_{\rm prod.}\times \text{BR}_{\lamp}$ and $\sigma_{\rm prod.}
\times \text{BR}_{\lam}$, for like-sign dimuon events involving $\stau_1$ 
decays via $\lamp$ and $\lam$, respectively, as described in
\eqref{eq_contribdecaychains}.

\begin{table*}
\begin{ruledtabular}
\begin{tabular}{rlcc|cccc}
\multicolumn{2}{c}{}&\multicolumn{2}{c|}{}&\multicolumn{2}{c}{up-type mixing\hspace*{3ex}} 
& \multicolumn{2}{c}{down-type mixing\hspace*{3ex}}
 \\[1.5ex]
\multicolumn{2}{c}{\rb{Set A}}&
& $\sigma_{\rm prod.}(\tilde{\mu}^{\mp}_L)$ [fb]\hspace*{2ex}
& $\sigma_{\rm prod.} \times \text{BR}_{\lamp}$ 
& \hspace*{3ex}$\sigma_{\rm prod.} \times \text{BR}_{\lam}$ \hspace*{3ex}
& $\sigma_{\rm prod.} \times \text{BR}_{\lamp}$
& \hspace*{3ex}$\sigma_{\rm prod.} \times \text{BR}_{\lam}$ \hspace*{3ex}
\\[1.5ex]
\hline
			& & $\mu^-\, \mu^-$ & 61.6 & 11.1 & 0.71 & 9.81 & 2.09 \\
\hspace*{3ex}\rb{$\lambda'_{211}$}&\rb{$= 2 \times 10^{-3}|_{\rm GUT}$}\hspace*{3ex}& $\mu^+\, \mu^+$ & 108 & 19.4 & 1.25 & 17.2 & 3.66
			\\[1ex]
\hline
			&& $\mu^-\, \mu^-$ & 42.0 & 7.84 & $-$ & 4.51 & 3.88  \\
\hspace*{3ex}\rb{$\lambda'_{221}$}&\rb{$=2 \times 10^{-3}|_{\rm GUT}$}\hspace*{3ex}& $\mu^+\, \mu^+$ & 16.2 & 3.03 & $-$ & 1.74 & 1.50
			\\[1ex]
\hline
			&& $\mu^-\, \mu^-$ & 18.6 & 3.46 & $-$ & 1.99 & 1.71  \\
\hspace*{3ex}\rb{$\lambda'_{212}$}&\rb{$=2 \times 10^{-3}|_{\rm GUT}$}\hspace*{3ex}& $\mu^+\, \mu^+$ & 86.0 & 16.1 & $-$ & 9.23 & 7.94
			\\[1ex]
\hline
			&& $\mu^-\, \mu^-$ & 8.80 & 1.67 & $-$  & 1.32  & 0.40   \\
\hspace*{3ex}\rb{$\lambda'_{213}$}&\rb{$=2 \times 10^{-3}|_{\rm GUT}$}\hspace*{3ex}& $\mu^+\, \mu^+$ & 49.8 & 9.43 & $-$  & 7.43  & 2.24
% 			\\[1ex]
\end{tabular}
\caption{\label{tab_numbersA}%
  Cross sections for exclusive like-sign dimuon ($\mu^-\mu^-$ or $\mu^
  +\mu^+$) final states at the LHC within Set A. In the left column,
  we present the single-smuon production cross sections, $\sigma_{\text
  {prod.}}(\tilde{\mu}_L^{\mp})$, see also Tabs.~\ref{xsections_setA}
  and \ref{xsections_setB}. In the right column, we have folded in the
  relevant decay branching ratios, in order to obtain like-sign
  dimuons. All cross sections are given in fb. Where they exist, we
  have assumed always a cascade of 2-body decays. We consider in turn
  quark mixing in the up- and down-sector, when determining the dominant
  $\stau_1$ decay mode.  The $\stau_1$~LSP can either decay via $\lamp$
  (4-body decay) or via $\lam$ (2-body decay), \cf
  Tab.~\ref{tab_singleslep-zerfaelle}, which leads to different
  like-sign dimuon cross sections, $\sigma_{\rm prod.}\times \text{BR}_{
  \lamp}$ and $\sigma_{\rm prod.}\times \text{BR}_{\lam}$, respectively.  The
  $\lambda'_{2jk}$~couplings are in accordance with neutrino mass
  bounds \cite{Allanach:2003eb, Markus:2008}.  In case of up-type
  mixing, larger values of $\lambda'_{2jk}$ for the four considered
  couplings are allowed by the neutrino mass bounds.  The cross
  sections scale with $|\lamp|^2$ and the corresponding rescaling can
  easily be performed.}
\end{ruledtabular}
\end{table*}

\begin{table*}
\begin{ruledtabular}
\begin{tabular}{rlcc|cccc}
\multicolumn{2}{c}{}&\multicolumn{2}{c|}{}&\multicolumn{2}{c}{up-type mixing\hspace*{3ex}}
 & \multicolumn{2}{c}{down-type mixing\hspace*{3ex}}
 \\[1.5ex]
\multicolumn{2}{c}{\rb{Set B}}&
& $\sigma_{\rm prod.}(\tilde{\mu}^{\mp}_L)$ [fb]\hspace*{2ex}
& $\sigma_{\rm prod.} \times \text{BR}_{\lamp}$ 
& \hspace*{3ex}$\sigma_{\rm prod.} \times \text{BR}_{\lam}$ \hspace*{3ex}
& $\sigma_{\rm prod.} \times \text{BR}_{\lamp}$
& \hspace*{3ex}$\sigma_{\rm prod.} \times \text{BR}_{\lam}$ \hspace*{3ex}
\\[1.5ex]
\hline
			& & $\mu^-\, \mu^-$ & 476 & 1.04 & 101 & 0.21 & 102  \\
\hspace*{3ex}\rb{$\lambda'_{211}$}&\rb{$= 1 \times 10^{-2}|_{\rm GUT}$}\hspace*{3ex}& $\mu^+\, \mu^+$ & 885 & 1.93 & 188 &0.39 & 189
			\\[1ex]
\hline
			&& $\mu^-\, \mu^-$ & 309 & 62.8 & $-$ & $-$ & 66.2  \\
\hspace*{3ex}\rb{$\lambda'_{221}$}&\rb{$= 1 \times 10^{-2}|_{\rm GUT}$}\hspace*{3ex}& $\mu^+\, \mu^+$ & 105 & 21.4 & $-$ & $-$ & 22.5
			\\[1ex]
\hline
			&& $\mu^-\, \mu^-$ & 123 & 25.1 & $-$ & $-$ & 26.3  \\
\hspace*{3ex}\rb{$\lambda'_{212}$}&\rb{$= 1 \times 10^{-2}|_{\rm GUT}$}\hspace*{3ex}& $\mu^+\, \mu^+$ & 681 & 139 & $-$ & $-$ & 146
			\\[1ex]
\hline
			&& $\mu^-\, \mu^-$ & 54.6  & 11.2 & $-$  & 0.02   &11.7    \\
\hspace*{3ex}\rb{$\lambda'_{213}$}&\rb{$= 1 \times 10^{-2}|_{\rm GUT}$}\hspace*{3ex}& $\mu^+\, \mu^+$ & 370  & 75.6  & $-$  &0.16 &  79.4
% 			\\[1ex]
\end{tabular}
\caption{\label{tab_numbersB}%
  Same as Tab.~\ref{tab_numbersA} but for single slepton production
  within Set B. The neutrino mass bounds are less restrictive in the
  case of Set B and $\lamp_{2jk} =0.01|_{\rm GUT}$ are considered for both
  up- and down-type quark mixing. All cross sections are given in fb.}
\end{ruledtabular}
\end{table*}

\medskip
 
The total number of exclusive like-sign dimuon events is given by the
integrated luminosity multiplied by the total cross section.  In Set A
with up-type (down-type) quark mixing, we obtain per 10\,fb$^ {-1}$
\begin{align}
\begin{split}
	&N(\mu^- \mu^- + \mu^+ \mu^+)/10\,\text{fb}^{-1} =
\\[.5ex]
	&\Big[ \sigma_{\rm prod.}(\tilde{\mu}^-_L) 
	      + \sigma_{\rm prod.}(\tilde{\mu}^+_L) \Big] 
	\times \Big[ \text{BR}_{\lamp} +\text{BR}_{\lam}\Big] \times 10
\\
	&\approx \begin{cases} 325 \;\;(330) \\110 \;\;(115)\\ 195 \;\;(210)\\110 \;\;(115) \end{cases} \!\!\!
	/10\,\text{fb}^{-1} \;\;\;\textnormal{for}\;
		 \begin{cases} \lamp_{211} \\ \lamp_{221}\\ \lamp_{212}\\ \lamp_{213} \end{cases}
		 \!\!\!\!= 0.002\lvert_{\rm GUT}.
\label{total-events-A}
\end{split}
 \end{align}

Note that for up-type mixing, some larger couplings may be considered.
From the neutrino mass bounds, also $\lamp_{211,\, 221,\,212,\, 213} =
0.01|_{\rm GUT}$ (and even larger) are allowed.  The cross sections are
proportional to $|\lamp|^2$ and thus a five times larger coupling
implies cross sections and event numbers multiplied by a factor of 25
compared to those of Tab.~\ref{tab_numbersA}.

\medskip

For Set B, $\lamp_{2jk} = 0.01|_{\rm GUT}$ is allowed for both up- and
down-type mixing. The numbers of like-sign dimuon events are,
\begin{align}
\begin{split}
	&N(\mu^- \mu^- + \mu^+ \mu^+)/10\,\text{fb}^{-1} =
\\
	&\approx \begin{cases} 2920\, (2920) \\840 \;\;\,(890)\\ 1640 \,(1720)\\ 870 \,\;\;(910) \end{cases} \!\!\!
	/10\,\text{fb}^{-1} \;\;\;\textnormal{for}\;
		 \begin{cases} \lamp_{211} \\ \lamp_{221}\\ \lamp_{212}\\ \lamp_{213} \end{cases}
		 \!\!\!\!= 0.01\lvert_{\rm GUT},
\label{total-events-B}
\end{split}
\end{align}
for up-type (down-type) quark mixing, respectively.

\medskip

As can be seen in Eqs.~(\ref{total-events-A}), (\ref{total-events-B}),
for each non-zero $\lamp$~coupling the total event numbers for up- and
down-mixing are of the same order.  But as Tabs.~\ref{tab_numbersA}
and \ref{tab_numbersB} show, the parts contributing to the event rate
can be quite different. In case of up-type mixing and $j\neq k$, the
4-body decays via $\lamp$ dominate and the contributions of the 2-body
decay are negligible [since the size of the necessary $\lam$~coupling
is proportional to $\Yd_{jk}$].  In contrast, for down-type mixing all
four considered couplings can generate a relatively large
$\lam_{233}$, \cf Fig.~\ref{fig_runningcouplingsD}, and the 2-body
decay modes contribute considerably. In Set B, where $\tan\beta$ is
large and where thus the fraction of 2-body decays is especially high
(see discussion of Fig.~\ref{tanb_dmixing}), reliable event numbers
are only obtained if the generation of $\lam_{233}$ is included in the
theoretical framework. Moreover, a measurement of the ratio of 2-body
to 4-body $\stau_1$ decays can reveal information about where the
quark mixing takes place.

\medskip

For $j=k$, the generation of a $\lam$~coupling is also possible in
case of up-type mixing.  In Set A, the generated $\lam_{233}$ is not
large enough to allow for large 2-body decay rates. However in Set B,
due to the large $\tan\beta$ value, the 2-body decays dominate
over the 4-body decays.  Thus, the different $\stau_1$ decay modes contain
also information about $\tan \beta$.

\medskip

We present in Tabs.~\ref{tab_numbersA} and \ref{tab_numbersB}
also the total hadronic cross sections for single smuon production,
$\sigma_{\rm prod.}(\tilde{\mu}^{\mp}_L)$.  Within one parameter set,
the cross sections vary strongly for different $\lamp_{2jk}$. This is
of course related to corresponding required parton density
  functions.  The largest cross section is obtained for
$\lamp_{211}\neq 0$, \ie for the processes $\bar{u}\,d \rightarrow
\tilde{\mu}_L^-$ and $u\,\bar{d} \rightarrow \tilde{\mu}_L^+$. Smaller
cross sections are obtained for $\lamp_{212}\neq 0$ (involving an up
quark and a strange quark) and the smallest cross section for
$\lamp_{221}\neq 0$ (charm quark and down quark) and $\lamp_{213}\neq
0$ (up quark together with bottom quark).

\medskip

Since the LHC is a $pp$ collider, there is an asymmetry between the
$\tilde\mu^+_L$ and $\tilde\mu^-_L$ production cross sections. If
experimentally a distinction between $\mu^+\mu^+$ and $\mu^-\mu^-$
event rates is found, the ratio can be used to
constrain the indices of the non-zero $\lamp_{2jk}$~coupling.  For
example, a non-vanishing coupling $\lambda'_{211}$ leads to a ratio of
$N(\mu^+ \mu^+):N(\mu^- \mu^-) \sim 2:1$ in sets A and B, whereas for
non-vanishing $\lambda'_{221}$ the ratio is $1:2.5$ in Set A and $1:3$
in Set B.  The highest event rates are obtained for processes that
involve the valence quarks $u$ and $d$.  The charge conjugated
processes, involving $\bar{u}$ or $\bar{d}$, are suppressed in
comparison.  Thus, a larger fraction of $\mu^+\mu^+$ events goes along
with $j=1$ (where the production process is $u\, \bar{d}_k \ra
\tilde{\mu}^+_L$) and a larger fraction of $\mu^-\mu^-$ events is
related to $k=1$ and $j \not= 1$ (production process $\bar{u}_j \,d
\ra \tilde{\mu}^-_L$).

%%%%%%%%%%%%%%%%%%%%%%%%%%%%%%%%%%%%%%%%%%%%%%%%%%%%%%%%%%%%
\subsection{Discussion of Background and Cuts for Like-Sign Dimuon Final States}
\label{background}

In this section, we discuss the background for like-sign dimuon events
from the SM and from SUSY particle pair production
via gauge interactions. We follow
Refs.~\cite{Dreiner:2000qf,Dreiner:2000vf} closely.  There, single
smuon production via $\lambda'_{211}$ was investigated assuming a
$\neut{1}$~LSP. A detailed signal over background analysis was
performed based on like-sign dimuon events. We argue that a similar or
even the same set of cuts might be used to suppress the background in
our case and we compare background and signal rates to determine the
discovery potential of our analysis. 
% A detailed Monte-Carlo simulation lies beyond the scope of this paper.

\medskip

The main SM background sources are $t\bar t$ production, $b \bar b$
production, single top production, and gauge boson pair production,
{\it i.e.} $WW$, $WZ$ and $ZZ$ production.  In
Refs.~\cite{Dreiner:2000qf,Dreiner:2000vf}, the dominant signature from
single smuon production including like-sign dimuon events is
\begin{equation}
	\tilde{\mu}_L^- \rightarrow 
		\mu^- \tilde{\chi}^0_1 \rightarrow 
		\mu^- (\mu^- u \bar d).
\label{chi_ls_muons}
\end{equation} 
The two muons of the signal (\ref{chi_ls_muons}) are isolated because
they stem from different decays of SUSY particles. In addition, the
muons carry large momenta since they originate from the decay of
(heavy) SUSY particles.  The following cuts were proposed to improve
the signal over SM background ratio at the LHC:
\begin{itemize}
\item The muon rapidity $|\eta|<2.0$, thus requiring all the
      leptons in the central region of the detector,
\item a cut on the transverse momentum on each muon: 
	  $p_T|_\mu \geq 40$~GeV,
\item an isolation cut on each of the muons,
\item a cut on the transverse mass of each of the muons, 60~GeV $< M_T <$ 85~GeV,
\item a veto on the presence of a muon with the opposite charge as the
      like-sign dimuons,
\item a cut on the missing transverse energy, $\met \leq 20\, \text{GeV}$ . 
\end{itemize}

These cuts reduce the SM background to $4.9\pm 1.6$ events per 10
fb$^{-1}$ at the LHC , \cf \eqref{SM_bg}. Among the above cuts, the
isolation and $p_T$ cut lead to the strongest suppression of the SM
background.

\medskip

We now investigate the case of a $\stau_1$~LSP. If the 4-body decays
(\ref{4body_stau_decays}) of the $\stau_1$~LSP dominate, the leading
signature of resonant single smuon production including like-sign
dimuon events can be written as
\begin{equation}
	\tilde{\mu}_L^- 
	\rightarrow \mu^- \tilde{\chi}^0_1 
	\rightarrow \mu^- \tau^\mp \tilde\tau^\pm 
	\rightarrow \mu^- \tau^\mp (\tau^\pm \mu^- u \bar d).
\label{tau_ls_muons_4KZ}
\end{equation}
As above, the muons originate from the decay of heavy particles
($\stau_1$ and $\tilde{\mu}_L$), are in general well isolated, and
carry large momenta. Thus, for both signals \eqref{chi_ls_muons} and
\eqref{tau_ls_muons_4KZ}, the same cuts should allow to discriminate
between the signal and the SM background.  Furthermore, the additional
pair of taus in \eqref{tau_ls_muons_4KZ} allows to require
one or two (isolated!) taus. This might additionally improve the
signal to background ratio.

\medskip

If the $\stau_1$~LSP predominantly decays via 2-body decay
modes, \eqref{2body_stau_decays}, 
the situation is a bit different. 
The like-sign dimuon signature is now 
\begin{equation}
	\tilde{\mu}_L^- 
	\rightarrow \mu^- \tilde{\chi}^0_1
	\rightarrow \mu^- \tau^+ \tilde\tau^- 
	\rightarrow \mu^- \tau^+ (\mu^- \nu_\tau).
\label{tau_ls_muons_2KZ}
\end{equation} 
We again have two isolated muons with large momenta and the same
isolation and $p_T|_{\mu}$ cuts as before should be useful to suppress
the SM background. But the neutrino of the $\stau_1$ decay leads to high
missing transverse energy $\met$ in the signal and an upper bound on
$\met$ is not appropriate anymore. Alternatively we propose a cut that
requires a minimum missing energy, \eg $\met \geq 60$~GeV.  This would
also reduce the SM background where the main source of $\met$ are
low-energetic neutrinos from $W$ decays.  Furthermore, we can again
require an additional tau in the final state. Finally, one can exploit
the fact that the 2-body decays lead to a pure leptonic final state
and a jet veto can be applied.

\medskip

In Refs.~\cite{Dreiner:2000qf,Dreiner:2000vf}, the SUSY background on
like-sign dimuon events is suppressed by vetoing all events with more
than two jets of $p_T|_{\rm jet} > 50$~GeV.  This cut will also work if
the 4-body decay mode of the $\stau_1$~LSP (\ref{4body_stau_decays})
dominates. The 2-body decay modes lead to purely leptonic final states
and even no high-$p_T$ jet may be required.

\medskip

We conclude that for $\stau_1$~LSP scenarios, the background for
like-sign dimuon events can be suppressed similarly as it has been
proposed for $\neut{1}$~LSP scenarios in
\cite{Dreiner:2000qf,Dreiner:2000vf}.

\medskip

We thus compare our signal, as given in \eqref{total-events-A} and
\eqref{total-events-B} for sets A and B respectively, to the
background, assuming that cuts as discussed above reduce the SM
background to less than 5 events per 10~fb$^{-1}$, \cf
\eqref{SM_bg}. For the signal efficiency, we assume $20\%$, \ie $20\%$
of signal events pass the cuts. We neglect systematic errors, at this
stage of the analysis.

\medskip

For Set A a more than 5$\sigma$ excess over the SM background can be
obtained for an integrated luminosity of 10~fb$^{-1}$ for all
couplings given in \eqref{total-events-A}.  For Set B, a cut
efficiency of $20\%$ for the signal corresponds to an excess between
$100\,\sigma$ and $300\,\sigma$ for the number of like-sign muon
events over the SM background! Therefore, within Set B, couplings can
be tested at the LHC down to $\lamp_{2jk}\lvert_{\rm GUT}
\sim\mathcal{O}(10^{-3})$. But a detailed Monte-Carlo based signal
over background analysis remains to be done.

%%%%%%%%%%%%%%%%%%%%%%%%%%%%%%%%%%%%%%%%%%%% 
\subsection{Final States with 3 and 4 Muons}
\label{subsec_34muonen}

\begin{table*}
\begin{ruledtabular}
\begin{tabular}{c|ccccc|cc}
Set A&
$\sigma(--+)$ & $\sigma(++-)$& $\sigma(--++)$ & 
$\sigma(+++-)$ & $\sigma(---+)$
& $\sum \sigma(--\dots)$ & $\sum \sigma(++\dots)$
\\[1.5ex]
\hline
$\lambda'_{211}= 2 \times 10^{-3}|_{\rm GUT}$\hspace*{2ex}& 9.38 (9.39) & 12.9 (13.0) & 5.32 (5.26)& 3.39 (3.35) & 1.93 (1.91)
	& 16.6 (16.6) & 21.7 (21.6)
			\\[1ex]
$\lambda'_{221}= 2 \times 10^{-3}|_{\rm GUT}$\hspace*{2ex}& 5.77 (5.77) & 3.84 (3.74) & 1.89 (1.77)  & 0.53 (0.49) & 1.36 (1.27)
	& 9.02 (8.81) & 6.26 (6.00)
			\\[1ex]
$\lambda'_{212}= 2 \times 10^{-3}|_{\rm GUT}$\hspace*{2ex}& 4.02 (3.93) & 9.05 (9.24) & 3.39 (3.17)  &2.79 (2.61) & 0.60 (0.56)
	& 8.01 (7.66) & 15.2 (15.0)
			\\[1ex]
$\lambda'_{213}= 2 \times 10^{-3}|_{\rm GUT}$\hspace*{2ex}& 2.04 (2.02) & 5.14 (5.19) & 1.85 (1.80)  & 1.57 (1.53) & 0.28 (0.27)
	& 4.17 (4.09) & 8.56 (8.52)
\end{tabular}
\caption{\label{tab_mehrals2muonenA}
Cross sections for signals with three or four final state muons within
parameter Set A, assuming down-type (up-type) quark mixing. Given are
the cross sections as defined in
Eqs.~(\ref{eq_sigmmmp})-(\ref{eq_sig3muons}) and the sums for two
negatively or positively charged muons, $\sum \sigma(--\dots)$ or $\sum
\sigma(++\dots)$, respectively. All cross sections are given in fb.
}
\end{ruledtabular}
\end{table*}

\begin{table*}
\begin{ruledtabular}
\begin{tabular}{c|ccccc|cc}
Set B&
$\sigma(--+)$ & $\sigma(++-)$& $\sigma(--++)$ & 
$\sigma(+++-)$ & $\sigma(---+)$
& $\sum \sigma(--\dots)$ & $\sum \sigma(++\dots)$
\\[1.5ex]
\hline
$\lambda'_{211}= 1 \times 10^{-2}|_{\rm GUT}$\hspace*{2ex}& 20.8 (20.8) & 29.1 (29.1) & 13.4 (13.4)& 8.73 (8.73) & 4.69 (4.69)
	& 38.9 (38.9) & 51.3 (51.3)
			\\[1ex]
$\lambda'_{221}= 1 \times 10^{-2}|_{\rm GUT}$\hspace*{2ex}& 11.9 (12.0) & 7.77 (7.59) & 4.08 (3.88)  & 1.04 (0.98) & 3.05 (2.89)
	& 19.1 (18.7) & 12.9 (12.4)
			\\[1ex]
$\lambda'_{212}= 1 \times 10^{-2}|_{\rm GUT}$\hspace*{2ex}& 8.14 (7.98) & 19.5 (19.9) & 7.93 (7.53)  & 6.72 (6.39) & 1.21 (1.15)
	& 17.3 (16.7) & 34.2 (33.8)
			\\[1ex]
$\lambda'_{213}= 1 \times 10^{-2}|_{\rm GUT}$\hspace*{2ex}& 3.94 (3.85) & 10.4 (10.6) & 4.20 (4.00)  & 3.66 (3.48) & 0.54 (0.51)
	& 8.68 (8.36) & 18.3 (18.1)
\end{tabular}
\caption{\label{tab_mehrals2muonenB}
Same as Tab.~\ref{tab_mehrals2muonenA} but for single slepton production within Set B. All cross sections are given in fb.}
\end{ruledtabular}
\end{table*}

To round off our studies, we consider in this section final states
with more than two muons.  For example, for parameter sets A and B,
the $\neut{1}$ cannot only decay into a $\stau_1$-$\tau$ pair but also into a
$\tilde {\mu} _R$-$\mu$ or $\tilde{e}_R$-$e$ pair. These are
kinematically accessible and have non-negligible branching ratios (Set
A: $7.0\%$, Set B: $2.2\%$; see Tab.~\ref{tab_allgBRs}).  As we have
shown in Tab.~\ref{tab_summaryfinalstates2}, these decays lead to
three or even four muons of mixed signs in the final state. Each of
the muons stems from the decay of a different SUSY particle.
Especially the four-muon final state cannot be found at a high rate in
$\neut{1}$~LSP scenarios and its observation could be a hint for a
$\stau_1$~LSP. Therefore, we analyze the three- and four-muon final
states in this section. All necessary branching ratios and production
cross sections are given in the Appendix, see
Tabs.~\ref{xsections_setA}-\ref{stau_BRs_setB}.

\medskip

The four--muon events may be classified into $\mu^-\mu^-\mu^-\mu^+$,
$\mu^-\mu^-\mu^+\mu^+$, and $\mu^-\mu^+\mu^+\mu^+$ signatures and we
introduce the notations $\sigma(---+)$, $\sigma(--++)$, and $\sigma
(+++-)$, for the respective cross sections. The four-muon final states
require a long decay chain and many different decays contribute at
various stages. For smuon production, summing up all contributions,
the cross sections can be written in the following compact form
\begin{align}
\begin{split}
 \sigma_{\tilde \mu}(- - - +) = & \, 
	\sigma_{\rm prod.}(\tilde{\mu}^-_L)  
	\times \text{BR}(\tilde{\mu}^-_L \ra \neut{1}\,\mu^-) \,
\\
	& \times  
	\text{BR}(\neut{1} \ra \tilde{\mu}_R^{+}\, \mu^{-})
	\times P_{\stau_1}(1\mu) \, ,
\\
\sigma_{\tilde \mu}(+ + + -) = & \, \sigma_{\tilde \mu}(- - - +)
\times \sigma_{\rm prod.}(\tilde{\mu}^+_L) / \sigma_{\rm prod.}(\tilde{\mu}^-_L) \, ,
\\
 \sigma_{\tilde \mu}(--++) = & \, \sigma_{\tilde \mu}(---+) +\sigma_{\tilde \mu}(+++-),
\end{split}
\label{eq_sigmmmp}
\end{align}
where $P_{\stau_1}(1\mu) = \text{BR}(\stau_1^- \ra \mu^-\dots) + \text{BR}(\stau_1^+ \ra
\mu^- \dots)$
denotes the probability of a negatively charged final state muon in a
$\stau_1$ decay.  The difference between $\sigma_{\tilde \mu}(---+)$
and $\sigma_{\tilde \mu}(+ + + -)$ stems from the different partons
and parton densities involved in the production cross sections.

\medskip

Smuon production can also lead to exactly three final state charged
muons, $\mu^-\mu^-\mu^+$ or $\mu^+\mu^+\mu^-$. The corresponding
cross sections now involve the probability $P_{\stau_1}(0\mu)$ for a
$\stau_1$ decay without a final state muon,
\begin{align}
\begin{split}
 \sigma_{\tilde\mu}(- - +) = & \,
	\sigma_{\rm prod.}(\tilde{\mu}^-_L)  
	\times \text{BR}(\tilde{\mu}^-_L \ra \neut{1}\,\mu^-) 
\\
	& \times  
	\text{BR}(\neut{1} \ra \tilde{\mu}_R^{+}\, \mu^{-})
	\times 2 P_{\stau_1}(0\mu)\, ,
\\
 \sigma_{\tilde\mu}(+ + -) = & \, \sigma_{\tilde\mu}(- - +) 
 \times \sigma_{\rm prod.}(\tilde{\mu}^+_L) / \sigma_{\rm prod.}(\tilde{\mu}^-_L) \, .
 \label{three_mu_from_snu}
\end{split}
\end{align}
There are 16 different decay chains of the $\tilde{\mu}_L^-$ leading
to a $\mu^- \mu^- \mu^+$ final state. The factor of 2 in
Eq.~(\ref{three_mu_from_snu}) is a consequence of summing over all
these decay chains.

\medskip

The same final state signatures (exactly three muons) can be obtained via
$\tilde{\nu}_{\mu}$ production. The decay chain is similar to that of
a produced smuon. The missing muon from the slepton decay is here
replaced by demanding a muon in the final $\stau_1$ decay,
\begin{align}
\begin{split}
  \sigma_{\tilde{\nu}}(- - +) 
=& \,
	\left[ \sigma_{\rm prod.}(\tilde{\nu}_{\mu}) +
		\sigma_{\rm prod.}(\tilde{\nu}_{\mu}^*) \right]
\\
	& \times \text{BR}(\tilde{\nu}_{\mu} \ra \neut{1}\,\nu_{\mu}) 
	\times  
	\text{BR}(\neut{1} \ra \tilde{\mu}_R^{+}\, \mu^{-})
\\
	& \times P_{\stau_1}(1\mu)\, ,
\\
\sigma_{\tilde{\nu}}(+ + -) = & \, \sigma_{\tilde{\nu}}(- - +) \, .
\end{split}
\end{align}
The total cross sections for (exactly) three final state 
muons are then given by 
\begin{align}
\begin{split}
 \sigma(\mp \mp \pm) = & 
	\sigma_{\tilde\mu} (\mp\mp \pm)+ \sigma_{\tilde{\nu}}(\mp\mp\pm).
\end{split}
\label{eq_sig3muons}
\end{align}
Tabs.~\ref{tab_mehrals2muonenA} and \ref{tab_mehrals2muonenB} give an
overview over the numerical results. The same $\lamp$~couplings as in
the previous Tabs.~\ref{tab_numbersA} and \ref{tab_numbersB} are
considered. The generation of $\lam_{233}$ has been taken into account
for the $\stau_1$ decays and the cross sections give total numbers,
including both 4- and 2-body $\stau_1$ decays.

\medskip

We see that the sum of three-- and four-muon events is in the same
order of magnitude as the results for purely like-sign
dimuons. For Set A, where $\text{BR}(\neut{1}\ra \tilde{\mu}_R\, \mu) = 7\%$,
the event numbers are even larger. In Set B, with $\text{BR}(\neut{1}\ra
\tilde{\mu}_R\, \mu) = 2\%$, the total contributions are smaller by a
factor of about three.  Depending on the experimental goals,
these channels thus give important contributions and should be
included in an analysis. On the other
hand, these events also suggest to use three or four final state muons
as a signal for slepton production since the background is
expected to be very low.

%% file: Chapter6-Appendices.tex
\section{Cross Sections and Branching Ratios
 relevant for Slepton Production and Decay}
\label{xsections_and_BRs}

%%%%%%%%%GEMEINSAM Auf DREI STELLEN%%%%%%%%
\begin{table*}
\begin{ruledtabular}
\begin{tabular}{c|cccc|cccccc}
Set & \multicolumn{4}{c|}{$\sigma_{\rm prod.}$ [fb]} 
 & \multicolumn{6}{c}{$\sigma_{\rm prod.}$ [fb]} 
\\
A & $\tilde{e}_L^+/\tilde{\mu}_L^+$ & $\tilde{e}_L^-/\tilde{\mu}_L^-$
& $\tilde{\nu}_{e/\mu}^*$ & $\tilde{\nu}_{e/\mu}$ 
& $\stau_2^+ $ & $\stau_2^-$
& $\stau_1^+ $ & $\stau_1^-$
& $\tilde{\nu}_{\tau}^*$ & $\tilde{\nu}_{\tau}$ 
\\
\hline
$\lamp_{i11} = 0.01\lvert_{\text{GUT}}$ & $2700$ & $1540$ 
& $1860$ & $1860$ 
& $2620$ & $1500$
& $434$ & $272$
& $190$ & $190$
\\
$\lamp_{i22} = 0.01\lvert_{\text{GUT}}$ & $268$ & $268$ 
& $410$ & $410$ 
& $2600$ & $2600$
& $64.5$ & $64.5$
& $421$& $421$
\\
$\lamp_{i12} = 0.01\lvert_{\text{GUT}}$ & $2150$ & $464$ 
& $1430$ & $602$ 
& $2090$ & $451$
& $360$ & $103$
& $1460$& $616$
\\
$\lamp_{i21} = 0.01\lvert_{\text{GUT}}$ & $405$ & $1050$ 
& $602$ & $1430$ 
& $393$ & $1020$
& $91.9$ & $197$
& $616$& $1460$
\\
$\lamp_{i13} = 0.01\lvert_{\text{GUT}}$ & $1240$ & $220$ 
& $788$ & $292$ 
& $1210$ & $214$
& $216$ & $51.3$
& $806$& $299$
\\
$\lamp_{i23} = 0.01\lvert_{\text{GUT}}$ & $119$ & $119$ 
& $191$ & $191$ 
& $116$ & $116$
& $30.0$ & $30.0$
& $196$& $196$
\\
$\lamp_{i31} = 0.01\lvert_{\text{GUT}}$ & $-$ & $-$ 
& $247$ & $666$ 
& $-$ & $-$
& $-$ & $-$
& $253$& $681$
\\
$\lamp_{i32} = 0.01\lvert_{\text{GUT}}$ & $-$ & $-$ 
& $161$ & $161$ 
& $-$ & $-$
& $-$ & $-$
& $166$& $166$
\\
$\lamp_{i33} = 0.01\lvert_{\text{GUT}}$ & $-$ & $-$ 
& $69.3$ & $69.3$ 
& $-$ & $-$
& $-$ & $-$
& $71.1$& $71.1$
 \\
\end{tabular}
\caption{\label{xsections_setA}Complete list of
 hadronic cross sections for resonant single slepton/sneutrino
production via $\lamp_{ijk} = 0.01\lvert_{\rm GUT}$ 
at the $pp$ collider LHC ($\sqrt{S} = 14$~TeV) 
within the parameter Set~A. 
The cross sections include QCD and SUSY-QCD corrections at NLO 
\cite{Dreiner:2006sv}.
For $\lamp_{i3k}$, sleptons cannot be produced 
because of the vanishing top-quark density in the proton.
}
\end{ruledtabular}
\end{table*}

\begin{table*}
\begin{ruledtabular}
\begin{tabular}{c|cccc|cccccc}
Set & \multicolumn{4}{c|}{$\sigma_{\rm prod.}$ [fb]} 
& \multicolumn{6}{c}{$\sigma_{\rm prod.}$ [fb]} \\
B
& $\tilde{e}_L^+/\tilde{\mu}_L^+$ & $\tilde{e}_L^-/\tilde{\mu}_L^-$
& $\tilde{\nu}_{e/\mu}^*$ & $\tilde{\nu}_{e/\mu}$ 
& $\stau_2^+ $ & $\stau_2^-$
& $\stau_1^+ $ & $\stau_1^-$
& $\tilde{\nu}_{\tau}^*$ & $\tilde{\nu}_{\tau}$ 
\\
\hline
$\lamp_{i11} = 0.01\lvert_{\text{GUT}}$ & $885$ & $476$ 
& $559$ & $559$ 
& $949$ & $515$
& $1168$ & $750$
& $657$& $657$
\\
$\lamp_{i22} = 0.01\lvert_{\text{GUT}}$ & $67.3$ & $67.3$ 
& $102$ & $102$ 
& $74.7$ & $74.7$
& $192$ & $192$
& $124$& $124$
\\
$\lamp_{i12} = 0.01\lvert_{\text{GUT}}$ & $681$ & $123$ 
& $414$ & $155$ 
& $735$ & $136$
& $976$ & $301$
& $490$& $187$
\\
$\lamp_{i21} = 0.01\lvert_{\text{GUT}}$ & $105$ & $309$ 
& $155$ & $414$ 
& $117$ & $337$
& $269$ & $548$
& $187$& $490$
\\
$\lamp_{i13} = 0.01\lvert_{\text{GUT}}$ & $370$ & $54.6$ 
& $214$ & $70.2$ 
& $401$ & $60.6$
& $572$ & $146$
& $255$& $85.4$
\\
$\lamp_{i23} = 0.01\lvert_{\text{GUT}}$ & $28.2$ & $28.2$ 
& $44.4$ & $44.4$ 
& $31.4$ & $31.4$
& $87.2$ & $87.2$
& $54.3$& $54.3$
\\
$\lamp_{i31} = 0.01\lvert_{\text{GUT}}$ & $-$ & $-$ 
& $60.4$ & $184$ 
& $-$ & $-$
& $-$ & $-$
& $73.5$& $219$
\\
$\lamp_{i32} = 0.01\lvert_{\text{GUT}}$ & $-$ & $-$ 
& $38.2$ & $38.2$ 
& $-$ & $-$
& $-$ & $-$
& $46.7$& $46.7$
\\
$\lamp_{i33} = 0.01\lvert_{\text{GUT}}$ & $-$ & $-$ 
& $14.8$ & $14.8$ 
& $-$ & $-$
& $-$ & $-$
& $18.2$& $18.2$
 \\
\end{tabular}
\caption{\label{xsections_setB} 
	Same as Tab.~\ref{xsections_setA} but for parameter Set~B.}
\end{ruledtabular}
\end{table*}

In this Appendix we give the necessary cross sections and branching
ratios to calculate rates of all possible decay signatures for single
slepton production at the LHC, within the $\rpv$ sets A and B with a
$\stau_1$~LSP, \cf \eqref{eq_sets}.

\medskip

In Tables~\ref{xsections_setA} and \ref{xsections_setB}, all hadronic
production cross sections of resonant single sleptons within parameter
Set~A and Set~B, respectively, are given.  We consider here
 $\lamp_{ijk}=0.01|_{\text{GUT}}$, but the cross section scales with 
$\lvert\lamp_{ijk}\rvert^2$.  
The running of $\lamp_{ijk}$ is taken into account according
to \eqref{RGElambdaPX}, leading to the following values at the SUSY
scale $Q_{\text{susy}}$, \cf \eqref{SUSY_scale}: 
\begin{align}
\begin{split}
\text{Set A:} \quad
\lamp_{2jk}=0.0282,\quad &\lamp_{3jk}=0.0282,
\\ 
\lamp_{23k}=0.0258,\quad &\lamp_{33k}=0.0257,
\\
\lamp_{2j3}=0.0281,\quad &\lamp_{3j3}=0.0280,
\\ 
\lamp_{233}=0.0255,\quad &\lamp_{333}=0.0254;
\end{split}
\end{align}
\begin{align}
\begin{split}
\text{Set B:} \quad
\lamp_{2jk}=0.0274,\quad & \lamp_{3jk}=0.0271,
\\
\lamp_{23k}=0.0249,\quad & \lamp_{33k}=0.0247,
\\
\lamp_{2j3}=0.0269,\quad & \lamp_{3j3}=0.0266,
\\
\lamp_{233}=0.0238,\quad & \lamp_{333}=0.0236,
\end{split}
\end{align}
where $j,k=1,2$ and $Q_{\text{susy}} = 893$~GeV for Set~A and $Q_{\text{susy}} = 1209$~GeV for Set~B.

\medskip

The production cross sections include NLO SUSY-QCD corrections
\cite{Dreiner:2006sv}. The latter depend on the trilinear
quark-squark-slepton coupling, $h_{D^k}$, defined in
Ref.~\cite{Allanach:2003eb}. Numerically, it is $h_{D^k}=-23.4$~GeV
($-21.2$~GeV) within Set~A (Set~B) at the SUSY scale. We incorporated
the running of $h_{D^k}$ by using the one-loop contributions from
gauge interactions \cite{Allanach:2003eb}.

\medskip

\begin{table}
\begin{ruledtabular}
\begin{tabular}{l|cc|cc}
 & \multicolumn{4}{c}{$\text{BRs}\, [\%]$} \\[.5ex]
\cline{2-5}
  & \multicolumn{2}{c|}{$\lamp_{2jk}=0.01|_{\text{GUT}}$}
& \multicolumn{2}{c}{$\lamp_{3jk}=0.01|_{\text{GUT}}$} \\
& Set A & Set B & Set A  & Set B
\\
\hline 
$\tilde{\mu}_L^- \rightarrow \neut{1}\,  \mu^-$& 91.1 & 91.3 & 100 & 100 \\ 
$\tilde{\mu}_L^- \rightarrow \bar{u}_j\,  d_k$& 8.9 & 8.7 & $-$ & $-$ \\ 
$\tilde{\nu}_\mu \rightarrow \neut{1}\, \nu_\mu$ & 91.7 & 91.5 & 100& 100 \\ 
$\tilde{\nu}_\mu \rightarrow \bar{d}_j \,d_k$ & 9.3 & 8.4 & $-$& $-$ \\[.5ex]
\hline 
$\neut 1 \rightarrow \tilde{\tau}_1^\pm\, \tau^\mp$ & 36.0 & 45.7 & 36.0 & 45.7 \\ 
$\neut 1 \rightarrow \tilde{\mu}_R^\pm\, \mu^\mp$ & 7.0 & 2.2& 7.0 & 2.2\\ 
$\neut 1 \rightarrow \tilde{e}_R^\pm\, e^\mp$ & 7.0 & 2.1 & 7.0 & 2.1 \\[.5ex]
\hline 
$\tilde{\mu}_R^- \rightarrow \tilde{\tau}_1^+\, \mu^-\, \tau^-$ & 54.3 & 64.1 & 54.3 & 64.1\\ 
$\tilde{\mu}_R^- \rightarrow \tilde{\tau}_1^-\, \mu^-\, \tau^+$& 45.7 & 35.9& 45.7 & 35.9\\[.5ex]
\hline\hline 
$\tilde{\tau}^-_2 \rightarrow \neut{1}\,\tau^-$ & 58.4 &14.7 & 55.5 &14.5 \\ 
$\tilde{\tau}^-_2 \rightarrow \tilde{\tau}_1^-\, h^0 $& 22.5& 41.8& 21.4& 41.2\\ 
$\tilde{\tau}^-_2 \rightarrow \tilde{\tau}_1^- \,Z^0 $& 19.1& 43.5& 18.1& 42.9\\  
$\tilde{\tau}^-_2 \rightarrow \bar{u}_j \,d_k $& $-$& $-$& 5.0& 1.3\\  
$\tilde{\nu}_\tau \rightarrow \neut{1}\, \nu_\tau$ & 62.2 &13.6 & 58.8 &13.4\\ 
$\tilde{\nu}_\tau \rightarrow \tilde{\tau}_1^-\, W^+$ & 37.8 &86.4& 35.8 &85.2\\ 
$\tilde{\nu}_\tau \rightarrow \bar{d}_j \, d_k$ & $-$ & $-$ & 5.4 & 1.4 \\
\end{tabular}
\caption{\label{tab_allgBRs}
	Table of branching ratios, $\text{BRs}$, that are relevant for single slepton production 
	and decays within the $\rpv$ mSUGRA scenarios Set~A and Set~B. Two different
	non-zero $\rpv$~couplings are considered, $\lamp_{2jk} = 0.01|_{\text{GUT}}$ for columns 2 and 3 and
	$\lamp_{3jk} = 0.01|_{\text{GUT}}$ for columns 4 and 5.
	The branching ratios for $\lamp_{1jk}\neq 0$ can be obtained from those for 
	$\lamp_{2jk} \neq 0$ by interchanging muon and electron flavor in the first four decay channels.
	The branching ratios for $\tilde{e}_L$ ($\tilde{\nu}_e,\, \tilde{e}_R$) 
	in scenarios with $\lamp_{ijk}\neq0,\, i\neq1$ 
	are equal to those of $\tilde{\mu}_L$ ($\tilde{\nu}_{\mu},\, \tilde{\mu}_R$) with $\lamp_{3jk}\neq 0$.
	The branching ratios for $\stau_1$~LSP decays are listed separately 
	in Tabs.~\ref{stau_BRs_setA}, \ref{stau_BRs_setB}.}
\end{ruledtabular}
\end{table}

Second, for the calculation of the rate for a given signature of resonant single slepton production,
the branching ratios for the slepton decay
and for the subsequent decay chains down to the $\stau_1$~LSP are needed.
For all dominant $\lamp_{ijk}$~couplings these branching ratios are
universal within parameter Set~A and Set~B, respectively, and are given in
Tab.~\ref{tab_allgBRs}.

\medskip

Finally, we show in Table~\ref{stau_BRs_setA} (Table~\ref{stau_BRs_setB}) all branching ratios of $\stau_1$~LSP
decays for different couplings $\lamp_{2jk}$ at the GUT scale.
Branching ratios within scenarios with $\lamp_{1jk}\neq 0$ are analogous and can be obtained from the tables by
replacing $\mu$ by $e$ in the final state signatures.

\medskip

In the case of a non-vanishing $\lamp_{3jk}$, the $\stau_1$~LSP directly
couples to the dominant $L_3Q_j \bar D_k$ operator and decays predominantly
via the inverse production process, see also the discussion in
Sect.~\ref{subsec_LSPdecays}.  For the special case of $\lamp_{33k}
\not = 0$ and $m_{\tilde{\tau}_1} < m_t$, however, the $\stau_1$ decays
into a $W$ boson and two jets, \cf \eqref{lamp3jk_decayII}. The
corresponding matrix element and partial width are calculated in
Appendix~\ref{app_3body}.

% % % % % % % % % % % % % % % % % % % % % % % % % % % % % % % % % % % % % % % % % % 

%%%% GEMEINSAM %%%%
%% ueberall 100 prozent oder 99.9 %%%%%

\begin{table*}
\begin{ruledtabular}
\begin{tabular}{c|cccc|cccccccc}
Set &\multicolumn{2}{c}{$\stau_1^- \stackrel{\lam}{\rightarrow} \nu_{\mu}\tau^-$}&&&\multicolumn{8}{c}{ } \\
A
&\multicolumn{2}{c}{[$=\stau_1^- \stackrel{\lam}{\rightarrow} \nu_{\tau}\mu^-$]}
&\multicolumn{2}{c|}{\rb{$\stau_1^- \stackrel{\lam}{\rightarrow} \bar{\nu}_{\mu}\tau^-$}}
&\multicolumn{2}{c}{\rb{$\stau_1^- \stackrel{\lamp}{\rightarrow} \tau^-\mu^- u_j \bar{d}_k$}}
&\multicolumn{2}{c}{\rb{$\stau_1^- \stackrel{\lamp}{\rightarrow} \tau^-\mu^+ \bar{u}_j d_k$} }
&\multicolumn{2}{c}{\rb{$\stau_1^- \stackrel{\lamp}{\rightarrow} \tau^- \nu_{\mu} d_j \bar{d}_k$}}
&\multicolumn{2}{c}{\rb{$\stau_1^- \stackrel{\lamp}{\rightarrow} \tau^- \bar{\nu}_{\mu} \bar{d}_j d_k$}}
\\[1ex]
% &&&&\\[1ex]
\hline
$\lambda'_{211}$
& $7.9\%$ &($2.7\%$) &$0.2\%$ &($0.1\%$)& $11.8\%$ &($13.3\%$) & $25.3\%$ &($28.5\%$) & $15.2\%$ &($17.1\%$)& $31.6\%$ &($35.6\%$)
\\
$\lambda'_{212}$
& $21.5\%$ &($-$) &$0.5\%$ &($-$)& $7.9\%$ &($14.2\%$) & $17.1\%$ &($29.3\%$) & $10.2\%$ &($18.1\%$)& $21.3\%$ &($38.4\%$)
\\
$\lambda'_{213}$
& $10.5\%$ &($-$) &$0.2\%$ &($-$)& $11.1\%$ &($14.1\%$) & $23.8\%$ &($30.2\%$) & $14.3\%$ &($18.1\%$)& $29.6\%$ &($37.6\%$)
\\
$\lambda'_{221}$
& $21.5\%$ &($-$) &$0.5\%$ &($-$)& $7.9\%$ &($14.2\%$) & $17.1\%$ &($29.3\%$) & $10.2\%$ &($18.1\%$)& $21.3\%$ &($38.4\%$)
\\
$\lambda'_{222}$
& $46.8\%$ &($46.8\%$) &$1.1\%$ &($1.1\%$)& $0.7\%$ &($0.8\%$) & $1.6\%$ &($1.6\%$) & $1.0\%$ &($1.0\%$)& $2.0\%$ &($2.0\%$)
\\
$\lambda'_{223}$
& $48.2\%$ &($ -$) &$1.1\%$ &($-$)& $0.4\%$ &($14.2\%$) & $0.8\%$ &($29.3\%$) & $0.5\%$ &($18.2\%$)& $1.0\%$ &($38.4\%$)
\\
$\lambda'_{231}$
& $17.9\%$ &($-$) &$0.4\%$ &($-$)& $-$ &($-$) & $-$ &($-$) & $20.7\%$ &($32.1\%$)& $43.0\%$ &($67.9\%$)
% \\
% $\lambda'_{231} M $
% & $17.9\%$ &($-$) &$0.4\%$ &($-$)& $-$ &($-$) & $-$ &($-$) & $20.7\%$ &($32.5\%$)& $43.0\%$ &($67.5\%$)
\\
$\lambda'_{232}$
& $48.8\%$ &($-$) &$1.1\%$ &($-$)& $-$ &($-$) & $-$ &($-$) & $0.4\%$ &($32.5\%$)& $0.8\%$ &($67.5\%$)
\\
$\lambda'_{233}$
& $49.4\%$ &($49.4\%$) &$1.1\%$ &($1.1\%$)& $-$ &($-$) & $-$ &($-$) & $-$ &($-$)& $-$ &($-$)\\
\end{tabular}
\caption{\label{stau_BRs_setA}
Branching ratios of the $\stau_1$~LSP for different non-zero
$\lamp_{2jk}$~couplings at the GUT scale. The branching ratios are
calculated within the mSUGRA parameter Set~A for the SUSY breaking
scale $Q_{\text{susy}} = 893$~GeV. We assume down-type (up-type)
quark mixing.  Branching ratios for non-vanishing $\lamp_{1jk}$ are
analogous, with $\mu$ replaced by $e$.}
\end{ruledtabular}
\end{table*}
%%%% GEMEINSAM %%%%

\begin{table*}
\begin{ruledtabular}
\begin{tabular}{c|cccc|cccccccc}
Set &\multicolumn{2}{c}{$\stau_1^- \stackrel{\lam}{\rightarrow} \nu_{\mu}\tau^-$}&&&\multicolumn{8}{c}{ } \\
B
&\multicolumn{2}{c}{[$=\stau_1^- \stackrel{\lam}{\rightarrow} \nu_{\tau}\mu^-$]}
&\multicolumn{2}{c|}{\rb{$\stau_1^- \stackrel{\lam}{\rightarrow} \bar{\nu}_{\mu}\tau^-$}}
&\multicolumn{2}{c}{\rb{$\stau_1^- \stackrel{\lamp}{\rightarrow} \tau^-\mu^- u_j \bar{d}_k$}}
&\multicolumn{2}{c}{\rb{$\stau_1^- \stackrel{\lamp}{\rightarrow} \tau^-\mu^+ \bar{u}_j d_k$} }
&\multicolumn{2}{c}{\rb{$\stau_1^- \stackrel{\lamp}{\rightarrow} \tau^- \nu_{\mu} d_j \bar{d}_k$}}
&\multicolumn{2}{c}{\rb{$\stau_1^- \stackrel{\lamp}{\rightarrow} \tau^- \bar{\nu}_{\mu} \bar{d}_j d_k$}}
\\[1ex]
% &&&&\\[1ex]
\hline
$\lambda'_{211}$
& $49.0\%$ &($48.6\%$) &$1.7\%$ &($1.7\%$)& $-$ &($0.1\%$) & $0.1\%$ &($0.4\%$) & $-$ &($0.1\%$)& $0.1\%$ &($0.5\%$)
\\
$\lambda'_{212}$
& $49.1\%$ &($-$) &$1.7\%$ &($-$)& $-$ &($5.6\%$) & $-$ &($41.1\%$) & $-$ &($6.3\%$)& $-$ &($46.9\%$)
\\
$\lambda'_{213}$
& $49.0\%$ &($-$) &$1.7\%$ &($-$)& $-$ &($5.7\%$) & $0.1\%$ &($41.0\%$) & $-$ &($6.4\%$)& $0.1\%$ &($46.9\%$)
\\
$\lambda'_{221}$
& $49.1\%$ &($-$) &$1.7\%$ &($-$)& $-$ &($5.6\%$) & $-$ &($41.0\%$) & $-$ &($6.3\%$)& $-$ &($47.0\%$)
\\
$\lambda'_{222}$
& $49.1\%$ &($49.1\%$) &$1.7\%$ &($1.7\%$)& $-$ &($-$) & $-$ &($-$) & $-$ &($-$)& $-$ &($-$)
\\
$\lambda'_{223}$
& $49.1\%$ &($ -$) &$1.7\%$ &($ -$)& $-$ &($5.7\%$) & $-$ &($41.0\%$) & $-$ &($ 6.4\%$)& $-$ &($ 47.0\%$)
\\
$\lambda'_{231}$
& $49.1\%$ &($-$) &$1.7\%$ &($-$)& $-$ &($-$) & $-$ &($-$) & $-$ &($12.0\%$)& $0.1\%$ &($88.0\%$)
\\
$\lambda'_{232}$
& $49.1\%$ &($ -$) &$1.7\%$ &($-$)& $-$ &($-$) & $-$ &($-$) & $-$ &($12.0\%$)& $-$ &($ 88.0\%$)
\\
$\lambda'_{233}$
& $49.1\%$ &($49.1\%$) &$1.7\%$ &($1.7\%$)& $-$ &($-$) & $-$ &($-$) & $-$ &($-$)& $-$ &($-$)\\
\end{tabular}
\caption{\label{stau_BRs_setB}
Branching ratios of the $\stau_1$~LSP for different non-zero
$\lamp_{2jk}$~couplings at the GUT scale. The branching ratios are
calculated within the mSUGRA parameter Set~B for the SUSY breaking
scale $Q_{\text{susy}} = 1209$~GeV. We assume down-type
(up-type) quark mixing.  Branching ratios for non-vanishing
$\lamp_{1jk}$ are analogous, with $\mu$ replaced by $e$.}
\end{ruledtabular}
\end{table*}

%%%%%%%%%%%%%%%%%%%%%%%%%%%%%%%%%%%%%%%%%%%%%%%%%%%%%

\section{The $\rpv$ slepton decay $\tilde\ell^-_i\rightarrow W^-\bar b d_k$}
\label{app_3body}

A non-vanishing $L_i Q_3 \bar D_k$ operator allows for slepton decay
into a top quark $t$ and a down-type quark $d_k$ of generation $k$, 
\begin{equation}
	\tilde{\ell}_i^- \rightarrow \bar t d_k \, . 
\end{equation}
However, this decay mode is kinematically only allowed if $m_{\tilde{
\ell}_i} > m_t + m_{d_k}$. For $m_{\tilde{\ell}_i} < m_t + m_{d_k}$, the
slepton decays via a virtual top quark,
\begin{align}
	\tilde{\ell}_i^- \rightarrow W^- \bar b  d_k.
\label{three_body}
\end{align}
This 3-body decay has not been considered in the literature yet and is
not implemented in the R-parity violating version of {\tt Herwig},
either.  We complete the picture by calculating the 3-body decay
(\ref{three_body}) in the following.

\medskip

The relevant parts of the supersymmetric Lagrangian are \cite{Richardson:2000nt}
\begin{align}
\begin{split}
	\mathcal{L}_{L_i Q_3 \bar{D}_k} &=
\lamp_{i3k} L_{1\beta} \tilde{\ell}_{i\beta}^-\, \bar{d}_k\, P_L\, t + h.c. \; ,
 \\
	\mathcal{L}_{bWt} &=
-\frac{g}{\sqrt{2}}\, W_\mu^+\, \bar{t}\, \gamma^\mu\, P_L\, b + h.c. \;,
\label{3body_Lag}
\end{split}
\end{align}
where $L_{\alpha\beta}$ is the slepton mixing matrix, $\alpha$ the
left/right eigenstate, and $\beta$ the mass eigenstate. From
\eqref{3body_Lag}, the squared matrix element (summed over final state
polarizations and colors) can be derived,
\begin{align}
\begin{split}
	\Big\lvert \overline{\mathcal{M}} & \big(\tilde{\ell}_{i\beta}^- \rightarrow W^- \bar b  d_k\big) \Big\rvert^2 =
% \\&
	 \frac{3}{2} 
	\frac{\lambda^{'2}_{i3k} L_{1\beta}^2 g^2}
		{[(W+ b)^2-m_t^2]^2+m_t^2 \Gamma_t^2} 
 \\
 	&  \times \Bigg \{   2 (d_k\!\cdot\! b) \left[m_b^2 - m_W^2 +4 (W\!\cdot\! b) 
 		+ \frac{4 (W\!\cdot\! b)^2}{m_W^2}\right] \, 
  \\
 	& \quad + 4(d_k\!\cdot\! W) \left[m_b^2+2(W\!\cdot\! b) 
        - m_b^2\, \frac{(W\! \cdot\! b)}{m_W^2}\right] \Bigg \}
 	 \label{M2_three_body_decay}.
\end{split}
\end{align}
We denote the particle four-momenta by the particle letter, and $m_t$,
$m_b$, and $m_W$, are the top, bottom and $W$ mass, respectively.
$\Gamma_t$ is the total width of the top quark.

\medskip

\begin{figure}[tb]
	\includegraphics[scale=0.43,bb = 0 45 540 527, clip=true]{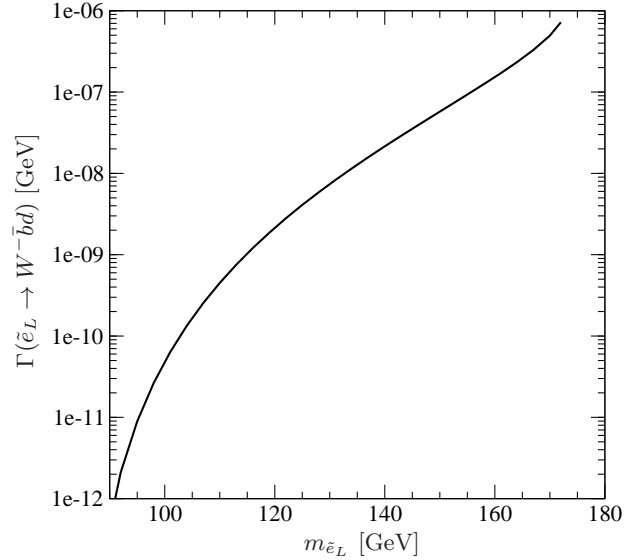}
\put(-120.0,1.0){$m_{\tilde{e}_L}$ [GeV]}
\put(-230.0,70.0){\rotatebox{90}{$\Gamma(\tilde{e}_L \rightarrow W^-
\bar b  d)$ [GeV]}}
\caption{\label{3body_width}Partial width in GeV for the 3-body decay 
$\tilde{e}_L \rightarrow W^- \bar b d$ as a function of the selectron
mass $m_{\tilde{e}_L}$. We take $\lamp_{131}=0.01$ and $L_{11}=1$,
in Eq.~(\ref{M2_three_body_decay}).}
\end{figure}

From the squared matrix element (\ref{M2_three_body_decay}) we obtain
easily the partial width for the 3-body decay (\ref{three_body}), see
\eg \cite{Richardson:2000nt}.  We show in \figref{3body_width} the
partial width $\Gamma(\tilde{e}_L \rightarrow W^- \bar b d)$ as a
function of the left-handed selectron mass $m_{\tilde{e}_L}$. Here we
take $\lamp_{131}=0.01$ and $L_{11}=1$, in
Eq.~(\ref{M2_three_body_decay}).

\medskip

In comparison to the 3-body decay (\ref{three_body}), the possible
4-body decays via $\lambda'_{i3k}$ are negligible. For example for the
parameter Set~B with non-vanishing $\lambda'_{331}$, the branching
ratio of the 3-body $\stau_1$~LSP decay (\ref{three_body}) is
larger by five orders of magnitude than the branching ratio of
the 4-body $\stau_1$~LSP decays.

